\documentclass[final,3p,times]{elsarticle}

\pdfoutput=1

\usepackage[utf8]{inputenc}
\usepackage[T1]{fontenc}
\usepackage{booktabs}
\usepackage{amssymb}
\usepackage{nomencl}
\usepackage{tcolorbox}
\usepackage{multicol}
\makenomenclature
\usepackage{geometry}
\usepackage{etoolbox}
\renewcommand\nomgroup[1]{%
  \item[\bfseries
  \ifstrequal{#1}{P}{\textbf{Physical and Mathematical Symbols}}{%
  \ifstrequal{#1}{M}{\textbf{Machine Learning Symbols}}{%
  {}}}%
]}
\usepackage[font=small,labelfont=bf,labelsep=none]{caption}
\usepackage{multirow}
\usepackage{setspace}
\usepackage{subfig}
\usepackage{amssymb}
\usepackage{algorithm}
\usepackage{algorithmic}
\usepackage{graphicx}
\usepackage{epstopdf}
\usepackage{ctable}
\usepackage{caption}
\usepackage{amssymb}
\usepackage{algorithm}
\usepackage{bm} 
\usepackage{amsmath}
\usepackage{float} 
\usepackage{chngcntr}
\usepackage{subcaption}
\usepackage{multicol}
\usepackage{hyperref}
\usepackage{tcolorbox}
\biboptions{sort&compress}
\DeclareMathAlphabet{\mathcal}{OMS}{cmsy}{m}{n}
\DeclareSymbolFont{largesymbols}{OMX}{cmex}{m}{n}
\usepackage{array, caption, threeparttable}
\usepackage{nomencl}
\makenomenclature
\usepackage{hyperref}
\hypersetup{
colorlinks=true,
linkcolor=blue,
anchorcolor=blue,
citecolor=blue}
\journal{Journal of Computational Physics}

\begin{document}
\mbox{}
\captionsetup[figure]{labelfont={bf},labelformat={default},labelsep=period,name={Fig.}}

\begin{frontmatter}

\title{LESnets (Large-Eddy Simulation nets): Physics-informed neural operator for large-eddy simulation of turbulence}

\author[inst1,inst2,inst3]{Sunan Zhao} 
\author[inst1,inst4]{Zhijie Li} 
\author[inst1,inst2,inst3]{Boyu Fan} 
\author[inst1,inst2,inst3]{Yunpeng Wang} 
\author[inst1,inst2,inst3]{Huiyu Yang} 
\author[inst1,inst2,inst3]{Jianchun Wang\corref{cor1}} 
\cortext[cor1]{Corresponding author at: Department of Mechanics and Aerospace Engineering, Southern University of Science and Technology, Shenzhen 518055, China.\\
E-mail address: \href{wangjc@sustech.edu.cn} {wangjc@sustech.edu.cn} (J. Wang).
}  

\affiliation[inst1]{organization={Department of Mechanics and Aerospace Engineering},
            addressline={Southern University of Science and Technology}, 
            city={Shenzhen},
            postcode={518055},
            country={China}}
\affiliation[inst2]{organization={Guangdong Provincial Key Laboratory of Turbulence Research and Applications},
            addressline={Southern University of Science and Technology}, 
            city={Shenzhen},
            postcode={518055}, 
            country={China}}
\affiliation[inst3]{organization={Guangdong-Hong Kong-Macao Joint Laboratory for Data-Driven Fluid Mechanics and Engineering
Applications},
            addressline={Southern University of Science and Technology}, 
            city={Shenzhen},
            postcode={518055}, 
            country={China}}
\affiliation[inst4]{organization={Department of Biomedical Engineering
Applications},
            addressline={National University of Singapore}, 
            city={Singapore},
            postcode={117583}, 
            country={Singapore}}
\begin{abstract}
{Acquisition of large datasets for three-dimensional (3D) partial differential equations (PDE) is usually very expensive.} Physics-informed neural operator (PINO) eliminates the high costs associated with generation of training datasets, and shows great potential in a variety of partial differential equations. In this work, we employ physics-informed neural operator, encoding the large-eddy simulation (LES) equations directly into the neural operator for simulating three-dimensional incompressible turbulent flows. {We develop the LESnets (Large-Eddy Simulation nets) by adding large-eddy simulation equations to two different data-driven models, including Fourier neural operator (FNO) and implicit Fourier neural operator (IFNO) without using label data. Notably, by leveraging only PDE constraints to learn the spatio-temporal dynamics, LESnets models retain the computational efficiency of data-driven approaches while obviating the necessity for data. Meanwhile, using LES equations as PDE constraints makes it possible to efficiently predict complex turbulence at coarse grids. We investigate the performance of the LESnets models with two standard three-dimensional turbulent flows: decaying homogeneous isotropic turbulence and temporally evolving turbulent mixing layer. In the numerical experiments, the LESnets models show similar accuracy as compared to traditional large-eddy simulation and data-driven models including FNO and IFNO, and exhibits a robust generalization ability to unseen regime of flow fields. By integrating a single set of flow data, the LESnets models can automatically learn the coefficient of the subgrid scale (SGS) model during the training of the neural operator. Moreover, the well-trained LESnets models are significantly faster than traditional LES, and exhibits comparable computational efficiency to the data-driven FNO and IFNO models. Thus, physics-informed neural operators have a strong potential for 3D nonlinear engineering applications.}

\end{abstract}

\begin{highlights}

\item {Large-eddy simulation (LES) equations are incorporated into physics-informed neural operators to enable efficient coarse-grid predictions of turbulence, eliminating the need for time series of flow data during neural operator training.}
\item {A well-trained physics-informed neural operator can accurately predict statistics and structures of statistically unsteady three-dimensional turbulent flows with a robust generalization ability to unseen regime of flow fields, while the speed of neural operator is 30-40 times faster than that of traditional large eddy simulations at the same grid resolution.}
\item {Employing a single set of flow data as $a$ $priori$, the coefficient of the subgrid scale model can be automatically optimized during the training process.}
\end{highlights}



\end{frontmatter}


\section{Introduction}
\label{sec1}

{Over the past few decades, the computational fluid dynamics (CFD) methods have found widespread applications in aerospace engineering and industrial production \cite{PopeTurbulence}. Multi-scale structures of turbulent flows can result in significant computational time. Thus, Reynolds-Averaged Navier-Stokes (RANS) \cite{RANS} method and Large-Eddy Simulation (LES) \cite{LES} method have been developed for turbulence simulation to reduce computational costs. In the past decade, neural networks (NNs) have been increasingly employed in the development of advanced turbulence models and simulation methods, facilitated by rapid advances in computing power \cite{brunton2020machine,duraisamy2019turbulence}}. {Most of the early approaches used machine learning (ML) to develop accurate turbulence models for RANS \cite{ling2016reynolds,wu2019reynolds} and LES \cite{BECK2019108910,park2021toward,wang2021artificial,yang2019predictive}}. {Ling et al. \cite{ling2016reynolds} proposed a novel neural network with an invariant tensor basis to embed Galilean invariance into the predicted anisotropic component of Reynolds stress, and demonstrated a high accuracy for predicting the Reynolds stress tensor. Wu et al. \cite{wu2019reynolds} proposed a metric based on local condition number for $a$ $priori$ evaluation of the conditioning of the RANS equations, which serves as a guide for subsequent data-driven RANS models.} Beck et al. \cite{BECK2019108910} proposed convolutional neural networks (CNNs) and residual neural networks (RNNs) to construct accurate subgrid scale (SGS) models for LES. Park et al. and Wang et al. \cite{park2021toward,wang2021artificial} also applied NNs to learn accurate closures of SGS stress in LES of turbulence. {Yang et al. \cite{yang2019predictive} proposed a physics-informed machine learning model for wall modeling of turbulent channel flows at moderate and high Reynolds numbers, predicting wall-shear stress more accurately than conventional methods.}

Many recent studies have focused on approximating the Navier–Stokes equations using deep neural networks \cite{lusch2018deep,sirignano2018dgm}. Raissi et al. \cite{PIDL1,PIDL2,PINN} introduced the  physics-informed neural networks (PINNs) to solve forward and inverse problems of partial differential equations (PDEs). Subsequently, this approach has also been used to address the missing flow dynamics \cite{xu2021explore} and simulate vortex induced vibrations \cite{raissi2019deep}. {Physics-informed neural networks (PINNs) provide notable benefits by removing the reliance on large training datasets, showing considerable potential in solving various partial differential equations (PDEs). However, the optimization process can face difficulties in achieving a convergence and fine-tuning the hyperparameters related to the weights of different terms within the loss function for complex tasks \cite{wang2022and,PImsNO}.} {To alleviate the above problems, PhyGeoNet model and Bayesian approach \cite{gao2021phygeonet,zou2025uncertainty} have been developed and employed in the task of turbulence prediction \cite{cai2021review,NSFNets,zhu2024new,Chen2025ThreedimensionalSW}.} {Gao et al. \cite{gao2021phygeonet} proposed a physics-informed geometry-adaptive convolutional neural networks (PhyGeoNet) that introduces an elliptic coordinate mapping to achieve coordinate transformation between irregular physical domains and regular reference domains.} {Recently, Zou et al. \cite{zou2025uncertainty} proposed a Bayesian approach to quantify the performance of PINNs in the situation of noisy input data, offering a novel perspective for studying the stability of PINNs. Cai et al. \cite{cai2021review} applied PINNs to reconstruct the 3D wake flow fields based on a few two-dimensional and two-component (2D2C) velocity observations.} Jin et al. \cite{NSFNets} developed the Navier–Stokes flow nets (NSFnets) by embedding the governing equations, initial conditions, and boundary conditions into the loss function to simulate three-dimensional turbulent channel flow. {Zhu et al. \cite{zhu2024new} developed a PINN-based method to significantly augment state-of-the-art experimental data of 3D stratified flows. Chen et al. \cite{Chen2025ThreedimensionalSW} used a multi-scale version of PINN for 3D wind field reconstruction from LiDAR measurements.} 

{Operator learning serves as another popular paradigm for solving PDE with machine learning \cite{NeuralOperator,DeepONets,FNO,li2022fourier,yang2024implicit}. By learning a map between the input condition and the PDE solution in a data-driven manner, neural operators can effectively solve a family of PDEs \cite{NeuralOperator}.} Deep operator network (DeepOnet) and Fourier neural operator (FNO) are two of most popular operator learning methods. A DeepONet consists of two sub-networks, one for encoding the input function (branch net), and another for encoding the locations (trunk net) \cite{DeepONets}. The main idea of FNO is to use Fourier transform to map high-dimensional data into the frequency domain and approximate nonlinear operators by learning the relationships between Fourier coefficients through neural networks \cite{FNO}. The FNO model outperforms the state-of-the-art models, including U-Net \cite{Unet} TF-Net \cite{wang2020towards} and ResNet \cite{he2016deep} in two-dimensional (2D) turbulence prediction. {You et al. \cite{IFNO} developed an implicit Fourier neural operator (IFNO), to efficiently model the increment between layers as an integral operator to capture the long-range dependencies in the feature space. However, the majority of the works have been focused on one-dimensional (1D) and two-dimensional (2D) problems. Utilizing deep neural networks to model 3D turbulence poses a more significant challenge due to the substantial growth in both the size and dimension of simulation data \cite{momenifar2022dimension}.} Recently, Peng et al. \cite{peng2023linear} added attention mechanisms to FNO to predict three-dimensional turbulence. {Luo et al. \cite{luoFNO} applied the FNO to the fast prediction of a three-dimensional compressible Rayleigh-Taylor turbulent flow at coarse grids. Li et al. and Wang et al. \cite{IUFNO,wang2024prediction} developed an implicit U-Net enhanced FNO (IUFNO) using the coarse-grid filtered DNS (fDNS) data for the fast prediction of 3D isotropic turbulence, turbulent mixing layer and turbulent channel flows. Large dataset of filtered DNS were used for training the IUFNO models. Although the well-trained operator learning models are accurate and effective, they still lack physical interpretability and rely on high-resolution training data.}

In recent research, the physics-informed approach has been applied to operator learning by embedding PDEs into the loss functions in a manner similar to PINNs, including physics-informed DeepONets {\cite{PI-DeepONets}}, physics-informed Fourier neural operator {\cite{PINO}}, and physics-informed token transformer {\cite{lorsung2024physics}}. {Physics-informed neural operator (PINO) models have demonstrated high accuracy across multiple PDE problems \cite{hao2023instability,rosofsky2023applications,jiao2024solving,chen2024physics,wang2024beyond}. Nevertheless, for high-dimensional complex turbulent flows, the generalization capability, the computational efficiency, and the optimization of the physics-informed neural operator models can encounter significant challenges due to the large range of involved motion scales, flow regimes, and turbulent dynamics. In the present study, we propose the Large-Eddy Simulation nets (LESnets) models by embedding large-eddy simulation equations into two neural operator models (FNO and IFNO) to predict the 3D turbulence. The proposed LESnets models aim to effectively simulate turbulent flows without training data and maintain the efficiency of  data-driven neural operators.}

{The rest of the paper is organized as follows. In Section \ref{sec2}, governing equations of the large-eddy simulation, LESnets models, and the architectures of PINNs, FNO and PINO are presented. We then present results of the LESnets models for decaying homogeneous isotropic turbulence and temporally evolving turbulent mixing layer in Section \ref{sec3}. A method to learn the coefficient of the subgrid scale model using a single set of flow data is proposed in Section \ref{sec4}. In Section \ref{sec5}, we investigate the computational efficiency of LESnets models, analyze their sensitivity to different parameters and dataset sizes, and investigate how to integrate data loss and PDE loss to optimize the LESnets. Finally, we summarize our results in Section \ref{sec6}.}

\section{Methodology}
\label{sec2}
\subsection{Governing equations}
\label{section2-1}
The 3D incompressible turbulent flows of Newtonian fluid are governed by the 3D Navier-Stokes (NS) equations, namely \cite{PopeTurbulence,ishihara2009study}

\begin{equation}
    \frac{\partial u_i}{\partial x_i}=0,
\label{eq 1}
\end{equation}

\begin{equation}
    \frac{\partial u_i}{\partial t}+\frac{\partial u_iu_j}{\partial x_j}=-\frac{\partial p}{\partial x_i}+\nu\frac{\partial^2u_i}{\partial x_j\partial x_j}+\mathcal{F}_i,
\label{eq 2}    
\end{equation}
where ${t}$ is time, ${u_i}$ denotes the ${i}$th component of velocity, ${p}$ is the pressure divided by the constant density, ${\nu}$ represents the kinematic viscosity, and ${\mathcal{F}_i}$ stands for a large-scale forcing to the momentum of the fluid in the ${i}$th coordinate direction. {In this paper, summation convention is employed for repeated indices.}

{Even though the NS equations have been developed more than a century ago, direct numerical simulation (DNS) of turbulent flow based on these equations at high Reynolds numbers  is still impractical \cite{meneveau2000scale,moser2021statistical}. Unlike DNS, LES only solves the large-scale flow structures using a coarse grid, leaving the effects of subgrid scale motions handled by the subgrid scale (SGS) models \cite{Smagorinsky,DSM}.} The filtered incompressible Navier–Stokes equations can be derived for the resolved fields as follows: \cite{PopeTurbulence,sagaut2005large}

\begin{equation}
    \frac{\partial \bar{u}_i}{\partial x_i}=0,
\label{eq 3}
\end{equation}

\begin{equation}
    \frac{\partial \bar{u}_i}{\partial t}+\frac{\partial \bar{u}_i\bar{u}_j}{\partial x_j}=-\frac{\partial \bar{p}}{\partial x_i}-\frac{\partial \tau_{ij}}{\partial x_j}+\nu\frac{\partial^2\bar{u}_i}{\partial x_j\partial x_j}+\mathcal{\bar{F}}_i,
\label{eq 4}
\end{equation}
where ${\bar{u}}$ represents the coarse-grid filtered DNS (fDNS)
velocity, and $\tau_{ij}$ is the unclosed subgrid scale (SGS) stress defined by $\tau_{ij}= \overline{u_iu_j}-\bar{u}_i\bar{u}_j$. In order to solve the LES equations, it is crucial to model the SGS stress as a function of the filtered variables. A very well-known SGS model is the Smagorinsky model (SM) \cite{Smagorinsky}, which can be written as

\begin{equation}
    \tau_{ij}^A=\tau_{ij}-\frac{\delta_{ij}}3\tau_{kk}=-2C_{\mathrm{Smag}}^2\overline{\Delta}^2|\overline{S}|\overline{S}_{ij},
\label{eq 5}
\end{equation}
{where $\delta_{ij}$ represents the Kronecker symbol, $C_{\mathrm{Smag}}$ is the Smagorinsky coefficient, $\overline{\Delta}$ is the filter width and $\overline{S}_{ij}=\frac{1}{2}(\partial\bar{u}_i/\bar{x}_j+\partial\bar{u}_j/\bar{x}_i)$ is the filtered strain rate.} $|\overline{S}|=\sqrt{2\overline{S}_{ij}\overline{S}_{ij}}$ is the characteristic filtered strain rate.

The integral length scale $L_I$ and the large-eddy turnover time $\tau$ are respectively given by

\begin{equation}
    L_I=\frac{3\pi}{2\left(u^{rms}\right)^2}\int_0^\infty\frac{E(k)}{k}dk,\quad\tau=\frac{L_I}{u^{rms}},
\label{tau}
\end{equation}
where $u^{rms}=\sqrt{\langle u_iu_i\rangle}$ is the root mean square (rms) of the velocity, and ${E(k)}$ is the energy spectrum.

{In this study, we employ the LES equations Eq. \eqref{eq 3} and Eq. \eqref{eq 4} along with the Smagorinsky model as physical constraints. We apply the physics-informed neural operator (PINO) to learn the solution operator of 3D turbulent flow at coarse grid. Furthermore, we compare the performance of PINO with conventional LES models and data-driven neural operators. A table of nomenclature is included in \ref{Appendix B}.}
\subsection{Physics-informed neural networks (PINNs)}
\label{section2-2}

{Consider a general form of the dynamical system governing by the following PDE}

\begin{eqnarray}
    \begin{aligned}\frac{\partial{u}}{\partial{t}}&=\mathcal{R}(u), && \operatorname{in}D\times(0,\infty),\\u&=g,&& \operatorname{in}\partial D\times(0,\infty),\\u&=a,&& \operatorname{in}{D}\times\{0\},\end{aligned}
\label{eq 6}
\end{eqnarray}
{where $D$ is a bounded domain, $\partial D$ is a Dirichlet boundary, $a(x)=u(x,0)\in\mathcal A\subseteq\mathcal V$ is the initial condition, $u(x,t)\in\mathcal{U}$ for time 
${t}>0$ is an unknown function of x and t, and $\mathcal{R}$ is a non-linear partial differential operator with $\mathcal{U}$ and $\mathcal{V}$ being Banach spaces.} Here, $g(x,t)$ is a known boundary condition. We assume that $u$ is bounded for all time and for every $u(x,0)\in\mathcal{U}$.

This formulation gives rise to the solution operator $\mathcal{G}^{\dagger}:\mathcal{A}\to C\big((0,T];\mathcal{U}\big)$ defined by $a\mapsto u$. Given an instance ${a}$ and a solution operator $\mathcal{G}^{\dagger}$ defined by Eq. \eqref{eq 6}, we denote $u^\dagger=\mathcal{G}^\dagger(a)$ as the unique
ground truth \cite{PINO}. The equation solving task is to approximate $u^\dagger$.

{The PINN-type methods use a neural network $u_{\theta}$ with parameters ${\theta}$ as the ansatz to approximate the solution function $u^\dagger$ as a function of x and t.} The parameters ${\theta}$ are obtained by minimizing the physics-informed loss with exact derivatives computed using automatic-differentiation (autograd) \cite{autograd}.

\begin{figure}[ht]
\centering
\includegraphics [width=1.0\textwidth]{ 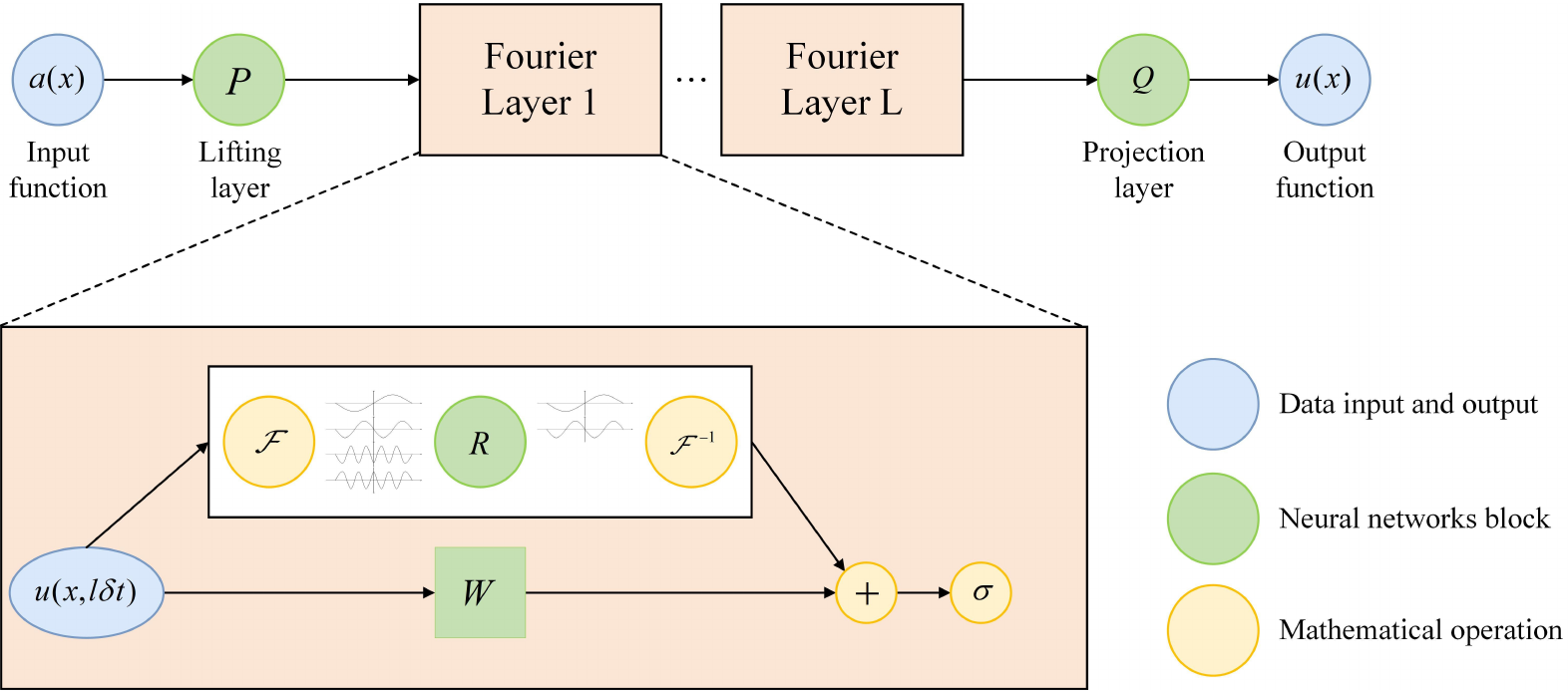}
\caption{{The architecture of the Fourier neural operator (FNO). The input function, ${a(x)}$ is transformed to a higher dimension channel space `width' $d_v$ using a fully-connected neural network ${P}$ and then apply ${L}$ Fourier layers. Each Fourier layer contains a Fourier transform $\mathcal{F}$, a linear transform ${R}$ on the lower Fourier modes, an inverse Fourier transform $\mathcal{F}^{-1}$ and a local linear transform ${W}$. At the end of the Fourier layers, a fully-connected neural network ${Q}$ is applied to convert the output to the target dimension.}}
\label{fig1}
\end{figure}

A typical choice for physics-informed loss is the $L^{2}\big((0,T];L^{2}(D)\big)$ norm, yielding \cite{PINN}
\begin{equation}
\label{eq 7}
    \begin{aligned}
\mathcal{L}_{\mathrm{pde}}(a,u_{\theta})& =\left\|\frac{du_{\theta}}{dt}-\mathcal{R}(u_{\theta})\right\|_{L^{2}(D;T)}^{2}+\alpha\left\|u_{\theta}|_{\partial D}-g\right\|_{L^{2}(\partial D;T)}^{2}+\beta\left\|u_{\theta}|_{t=0}-a\right\|_{L^{2}(D)}^{2} \\
&=\int_0^T\int_D|\frac{du_\theta}{dt}(x,t)-\mathcal{R}(u_\theta)(x,t)|^2\mathrm{d}x\mathrm{d}t \\
&+\alpha\int_0^T\int_{\partial D}|u_\theta(x,t)-g(x,t)|^2\mathrm{d}x\mathrm{d}t \\
&+\beta\int_D|u_\theta(0,x)-a(x)|^2\mathrm{d}x.
    \end{aligned}
\end{equation}
The PDE loss consists of the physics loss in the interior and the loss on the boundary and initial conditions, with hyper-parameters $\alpha,\beta>0$.

{PINNs face significant challenges in solving PDEs of multi-scale problems \cite{wang2022and,PINNlimitations}. This limitation is particularly evident in three-dimensional turbulent flows \cite{NSFNets}. Moreover, existing approaches often rely on additional observational flow data \cite{raissi2020hidden}. However, the requirement of flow data presents significant challenges for practical implementation, especially in real-world scenarios where experimental measurements and numerical simulations are either scarce or expensive \cite{cai2021review}.}

\subsection{Neural operator}
\label{section2-3}

In general, the neural operators can approximate the operator in Eq. \eqref{eq 6} which is a non-linear mapping between infinite-dimensional spaces \cite{NeuralOperator}

\begin{equation}
\label{eq 8}
    G:a=u(x,0)\in\mathcal{A}\times\Theta\mapsto u(x,t)\in\mathcal{U},
\end{equation}
where $\mathcal{A} = \mathcal{A}(D;\mathbb{R}^{d_a} )$ and $\mathcal{U}=\mathcal{U}(D;\mathbb{R}^{d_u})$ are separable Banach spaces of function taking values in $\mathbb{R}^{d_a}$ and $\mathbb{R}^{d_u}$ respectively, and $D\subset\mathbb{R}^d$ is a bounded, open set. For some finite-dimensional parameter space ${\Theta}$ by choosing $\mathcal{\theta}^{\dagger}\in \Theta$ so that $\mathcal{G}(\cdot,\mathcal{\theta}^{\dagger})=\mathcal{G}_\mathcal{\theta} \approx \mathcal{G}^{\dagger}$. {This mapping is inherently resolution-invariant. There exist numerous effective operator learning models that have demonstrated outstanding performances across a wide range of problems \cite{PImsNO,IUFNO}.} One can use a neural operator $\mathcal{G_{\theta}}$ with parameters $\theta$ as a surrogate model to approximate $\mathcal{G}^{\dagger}$. {Using a dataset $\{a_j,u_j\}_{j=1}^N$ as a ground truth, neural operator can minimize the empirical error on a given pair of data:}

\begin{figure}[ht]
\centering
\includegraphics [width=1.0\textwidth]{ 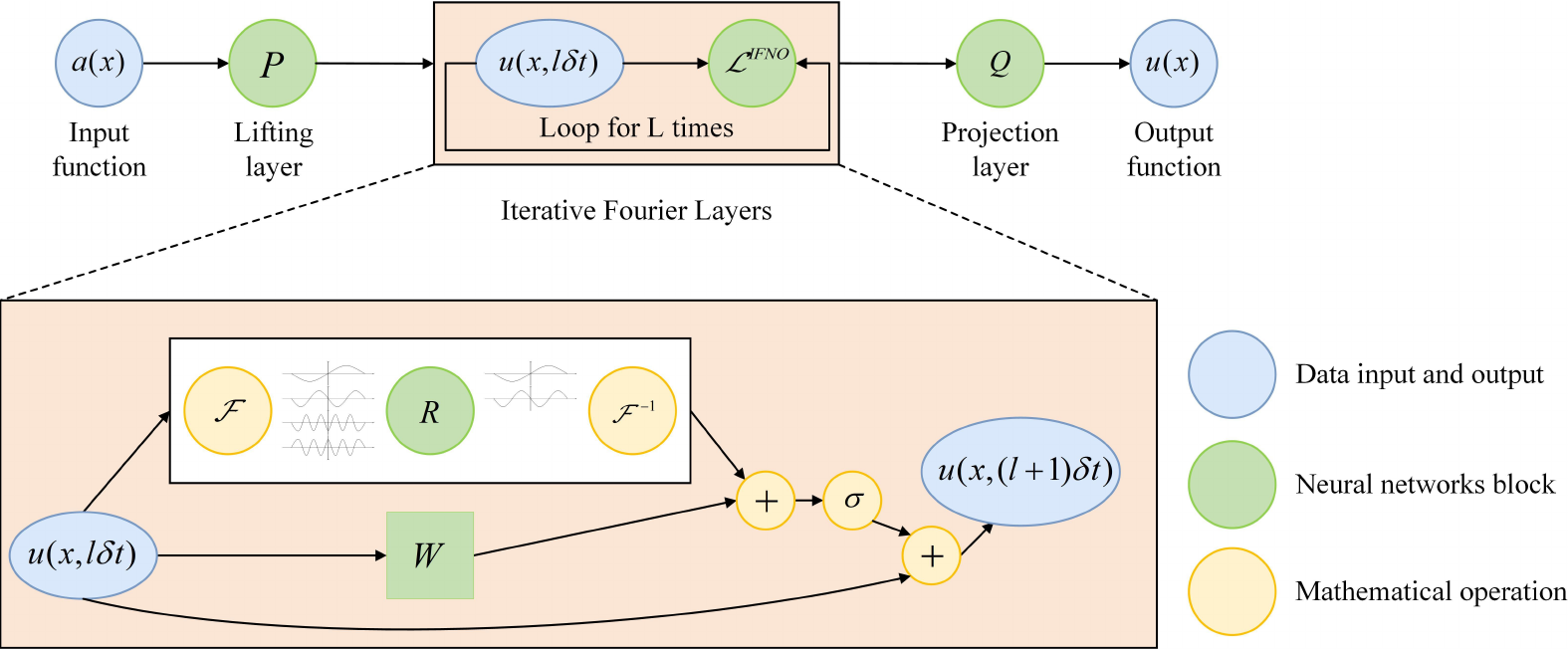}
\caption{The architecture of the implicit Fourier neural operator (IFNO).}
\label{fig2}
\end{figure}

\begin{equation}
\label{eq 9}
    \mathcal{L}_{\text{data}}(u,\mathcal{G}_\theta(a))=\|u-\mathcal{G}_\theta(a)\|_{\mathcal{U}}^2=\int_D|u(x)-\mathcal{G}_\theta(a)(x)|^2\mathrm{d}x.
\end{equation}

{In practice, the neural operator can be formulated as an iterative architecture in the following form \cite{FNO}}
 \begin{equation}
 \label{eq 10}
    v_{t+1}(x):=\sigma\Big(Wv_t(x)+\Big(\mathcal{K}(a;\phi)v_t\Big)(x)\Big),\quad\forall x\in D,
\end{equation}
where $\mathcal{K}:\mathcal{A}\times\Theta_{\mathcal{K}}\to\mathcal{L}(\mathcal{U}(D;\mathbb{R}^{d_{v}})$ maps to bounded linear operators on $\mathcal{U}(D;\mathbb{R}^{d_{v}})$ and is parameterized by $\phi\in\Theta_{K}$, $W:\mathbb{R}^{d_{v}}\to\mathbb{R}^{d_{v}}$ is a linear transformation. $\mathcal{K}$ is parameterized as a kernel integral transformation by a neural network $\kappa_{\phi}:\mathbb{R}^{2(d+d_{a})}\to\mathbb{R}^{d_{v}\times d_{v}}$ parameterized by $\phi\in\Theta_{K}$. Generally $\mathcal{K}$ is defined as follows:
\begin{equation}
    \big(\mathcal{K}(a;\phi)v_t\big)(x):=\int_D\kappa\big(x,y,a(x),a(y);\phi\big)v_t(y)\mathrm{d}y,\quad\forall x\in D.
\label{eq 11}
\end{equation}

In this work, we employ the Fourier neural operator (FNO) \cite{FNO} and implicit Fourier neural operator (IFNO) \cite{IFNO} for large-eddy simulation of turbulence. The Fourier neural operator establishes a system with mixed kernel integral and Fourier convolution layers. {The FNO architecture is shown in Fig. \ref{fig1}.} By mapping the input $\mathcal{A}$ into Fourier space for efficient computation and parameter learning, the Fourier neural operator achieves a very high computational efficiency \cite{FNO}.

Let $\mathcal{F}$ denotes the Fourier transform of a function $f:D\to\mathbb{R}^{d_{v}}$ and $\mathcal{F}^{-1}$ its inverse, then

\begin{equation*}
    (\mathcal{F}f)_{j}(k)=\int_{D}f_{j}(x)e^{-2i\pi\langle x,k\rangle}\mathrm{d}x,\quad(\mathcal{F}^{-1}f)_{j}(x)=\int_{D}f_{j}(k)e^{2i\pi\langle x,k\rangle}\mathrm{d}k.
\end{equation*}
The  Fourier integral operator $\mathcal{K}$ is given by

\begin{equation}
    \big(\mathcal{K}(\phi)v_t\big)(x)=\mathcal{F}^{-1}\Big(R_\phi\cdot(\mathcal{F}v_t)\Big)(x),\quad\forall x\in D.
\label{eq 12}
\end{equation}
By truncating the Fourier series at a specified maximum number of modes $k_{\mathrm{max}} = |Z_{k_{\mathrm{max}}}| = |\{k \in \mathbb{Z}^{d} : |k_{j}| \leq k_{\mathrm{max},j}, \mathrm{for} j = 1,\ldots,d\}|$, a finite-dimensional parameterization is achieved. $\mathcal{F}(v_{t}) \in \mathbb{C}^{n\times d_{v}}$ can be obtained by discretizing domain $D$ with $n\in\mathbb{N}$ points \cite{FNO}. $R_{\phi}$ is a Fourier transform of a periodic function, which is parameterized as complex-valued-tensor ($k_{\max}\times d_{v}\times d_{v}$) containing a collection of truncated Fourier modes $R_\phi\in\mathbb{C}^{k_{\max}\times d_v\times d_v}$. Therefore, the following equation can be derived by multiplying $R_{\phi}$ and $\mathcal{F}(v_{t})$:

\begin{equation}
\label{eq 13}
\left(R_\phi\cdot(\mathcal{F}v_t)\right)_{k,l}=\sum_{j=1}^{d_v}R_{\phi k,l,j}(\mathcal{F}v_t)_{k,j},\quad k=1,...,k_{\max},\quad j=1,...,d_v.
\end{equation}

{It has been demonstrated that with a sufficiently large depth ${L}$, FNO can serve as an universal approximator capable of accurately representing any continuous operator \cite{kovachki2021universal}. However, as the number of Fourier layers $L$ increases, the corresponding growth in the number of model parameters significantly compromises the computational efficiency of the FNO architecture. Subsequently, the implicit Fourier neural operator (IFNO) \cite{IFNO} is proposed as shown in Fig. \ref{fig2}. The IFNO reduces the number of parameters and computational memory by sharing parameters across Fourier layers, and its iteration process can be written as}
\begin{equation}
\label{eq 14}
    v_{t+1}(x):=v_t(x) + \delta{t} \sigma\Big(Wv_t(x)+\Big(\mathcal{K}(a;\phi)v_t\Big)(x)\Big),\quad\forall x\in D,
\end{equation}
where $\delta t$ is the time interval between input and output.

{As a supervised learning framework, data-driven operator learning requires a substantial amount of high-resolution data.} However, in the field of computational fluid dynamics, acquiring such high-resolution fluid data is highly costly and time-consuming, thereby compromising its inherent speed.

\subsection{Physics-informed neural operator}
\label{section2-4}
To combine  the respective advantages of PINNs and operator learning, physics-informed neural operator (PINO) \cite{PINO} has been proposed. {Using data loss $\mathcal{L}_{\text{data}}$ Eq. \eqref{eq 9} and PDE loss $\mathcal{L}_{\mathrm{pde}}$ Eq. \eqref{eq 7}, the neural operator $\mathcal{G_{\theta}}$ is able to approximate the target solution operator $\mathcal{G}^{\dagger}$. Using the PDE constraints of equations makes the neural operator adhere to physical laws, while the data loss also facilitates the optimization process.}

To fully leverage the advantages of operator learning and save computational time, PINO has developed an efficient and exact derivative computation method similar to automatic differentiation, which calculates full gradient field through the architecture of neural operators. Additionally, by applying the chain rule, derivatives in Fourier space can be directly calculated \cite{PINO}. The time derivative is calculated using a central difference method as

\begin{equation}
\label{eq new3}
    \{\frac{\partial u}{\partial t}\}^i=\frac{(u^{i+1}-u^{i-1})}{2dt},\quad i=2,...,T-1,
\end{equation}
where $i$ is the time node, $dt$ is the time interval, and $T$ is the total number of time instants. For non-periodic problems, the periodic boundary conditions can be recovered by padding the boundaries \cite{PINO}.

\begin{figure}[ht]
\centering
\includegraphics [width=1.0\textwidth]{ 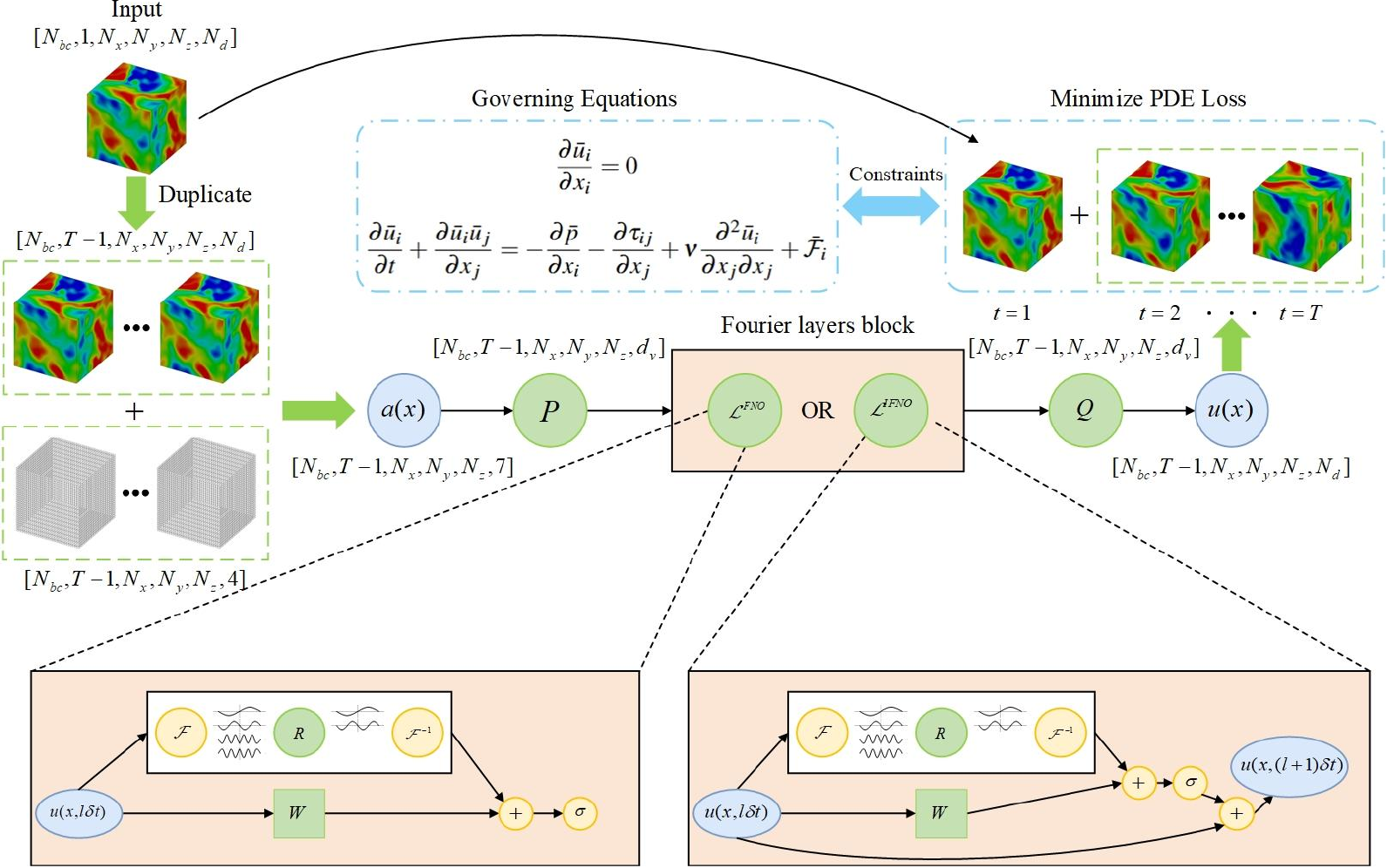}
\caption{The architecture of the Large-Eddy Simulation nets (LESnets) models, using the ideas of physics-informed neural operator to combined the FNO or IFNO model with LES equations constraints.}
\label{fig3}
\end{figure}

The original PINO encounters several significant challenges:

\begin{itemize} 
\item Based on the standard relative error $L_2={||u^*-u||_2}/{||u||_2}$ of two-dimensional turbulence, the model prediction at short-time ($T = 0.125$) gets $6\%$ error under the constraint of partial differential equations (PDEs), but almost $74\%$ error in the prediction of long time series ($T = 0.5$) \cite{PINO}. Here, $u$ is the true value, $u^*$ is the predicted value, and $||\cdot||_2$ is the $L_2$ norm.

\item {It optimizes the equation loss, initial condition loss and data loss concurrently, leading to the challenge for the selection of hyper-parameters for a multitude of losses.} Despite attempts to mitigate this issue through adaptive parameters \cite{wang2021understanding}, the coexistence of multiple losses complicates the optimization landscape.

\item {The training and prediction time periods for PINO overlap, implying the difficulty in generalizing to extended time periods.}
\end{itemize}

{The performance of original PINO model is discussed in \ref{Appendix A}}

\subsection{LESnets (Large-Eddy Simulation nets)}
\label{subsec5}

{In the present study, we develop LESnets models to learn a neural operator  $\mathcal{G}_\mathcal{\theta}$ by using the PDE loss of Eq. \eqref{eq 3} and Eq. \eqref{eq 4}. The architecture of LESnets models is shown in Fig. \ref{fig3}. The key components of the proposed method include:}

\begin{itemize} 
\item Employ the neural operator $\mathcal{G}_\mathcal{\theta}$ via Fourier neural operator (FNO) and implicit Fourier neural operator (IFNO) to verify the effectiveness of our method.

\item {Utilize the large-eddy simulation equations with a classical SGS model as physical constraints to construct the PDE loss for optimizing the operator without any labeled data.}

\item Use the initial conditions ${u_t}$ as known quantities rather than quantities to be learned, and directly combine it with the subsequently output fields ${u_{t+1}}\cdots{u_{t+T}}$ to calculate the PDE loss.

\item Train the model with the input-output pair from just one time node: ${u_t}\mapsto{u_{t+1}}\cdots{u_{t+T}}$. In the new prediction, the predicted ${u_{t+T}}$ is once again input into the model as the initial field to obtain ${u_{t+T}}\mapsto{u_{t+T+1}}\cdots{u_{t+2T}}$,$\cdots$. Therefore, we only need one initial field ${u_t}$ for both training and predicting. {This feature closely mimics the traditional CFD approach.}

\begin{algorithm}
    \caption{Training process of LESnets}
    \label{alg1}
    \begin{algorithmic}[1]
        \STATE \textbf{Input:}
        \STATE \qquad   ${\{{\bm{u}(t_n)}\}_{n=1}^{N_1}}\leftarrow$ Training Data
        \STATE \qquad   ${\{{t_{n=1}}\}}\leftarrow$ Time stamp
        \STATE \qquad   $\mathcal{G_{\theta}}\leftarrow$ Neural operator
        \STATE $\left({T,N_x,N_y,N_z,N_d}\right)\leftarrow\left({11,32,32,32,3}\right)$
        \STATE $\left({N_{ep},{\lambda},N_1,N_{bc}}\right)\leftarrow\left({30000,1e^{-3},5000,1}\right)$
        \STATE $\bm{\Theta}\leftarrow$ Empty List
        \FOR{$j=1$ to $N_{ep}$}
            \STATE ${\{{\bm{G}(t_n)}\}_{n=1}^{N_1}}\sim\mathbb{U}\left(N_{bc},T,N_x,N_y,N_z,4\right)\leftarrow$ Grid information
            \STATE ${\{{\bm{u}(t_n)}\}_{n=1}^{N_1}}={\{{\bm{u}(t_n)}\}_{n=1}^{N}}+{\{{\bm{G}(t_n)}\}_{n=1}^{N_1}}\leftarrow$ Add at dimension $N_d$
            \STATE ${\{{\bm{u}(t_n)}\}_{n=2,3,...T}^{N_1}}=\mbox{LESnets}\left({\mathcal{G_{\theta}},{\{{\bm{u}(t_n)}\}_{n=1}^{N_1}}}\right)$
            \STATE ${\{{\bm{u}(t_n)}\}_{n=1,2,...T}^{N_1}}={\{{\bm{u}(t_n)}\}_{n=1}^{N_1}}+{\{{\bm{u}(t_n)}\}_{n=2,3,...T}^{N_1}}\leftarrow$ Add at dimension $T$
            \STATE $\{{{\partial}_{t}}\bm{u}(t_n)\}_{n=2,...T-1}=\left({\bm{u}_{t+1}-\bm{u}_{t-1}}\right)/\left({2\Delta{t}}\right)$
            \STATE ${\{{\hat{\bm{u}}}\}^{k}} = \mathcal{F}\left({\bm{u}}\right)$
            \STATE $\mathcal{L}_1=\nabla\cdot\bm{u}=\mathcal{F}^{-1}\left({ik\hat{\bm{u}}}\right)$
            \STATE ${\{{\bm{\omega}(t_n)}\}_{n=1,...T}^{N_1}}=\mathcal{F}^{-1}\left({ik\times{\hat{\bm{u}}}}\right)$ 
            \STATE ${\{{\hat{\bm{R}}}\}^{k}}=\mathcal{F}\left({\bm{u}\times{\omega}}\right)-{ik\hat{\bm{\tau}}}$
            \STATE ${\{{\hat{\bm{p}}}\}^{k}}=-\frac{1}{k^2}ik\cdot\hat{R}$
            \STATE $\mathcal{L}_2={\partial}_{t}\bm{u}+\mathcal{F}^{-1}\left(\nu{k^2}\hat{\bm{u}}\right)-\mathcal{F}^{-1}\left({\hat{\bm{R}}}\right)+\mathcal{F}^{-1}\left({\hat{\bm{p}}}\right)$
            \STATE $\mathcal{L}_{PDE} = \sum_{n=2}^{T-1}||\mathcal{L}_1+\mathcal{L}_2||$
            \STATE $\theta\leftarrow\theta-\lambda\cdot\nabla_{\boldsymbol{\theta}}\mathcal{L}_{PDE}$
            \STATE Append $\theta$ to $\Theta$
        \ENDFOR
        \STATE Select proper $\theta^\ast$ from $\Theta$ based on relative error during training process.
    \end{algorithmic}  
\end{algorithm}

\end{itemize}

During the training of LESnets models, the loss function ${{\mathcal{L}}_{PDE}}$ is constructed as follows:

\begin{align}
\label{eq 17}
     \mathcal{L}_{1}&=\left\|\nabla \cdot \bar{u}\right\|_{L^{2}(T;D)}^{2} \nonumber,\\ 
  \mathcal{L}_{2}&=\left\|{{\partial }_{t}}\bar{u}+\bar{u}\cdot \nabla \bar{u}+\nabla \bar{p}-\nu {{\nabla }^{2}}\bar{u}-\nabla{\tau}\right\|_{L^{2}(T;D)}^{2} \nonumber,\\ 
  {{\mathcal{L}}_{PDE}}&=\mathcal{L}_{1}+\mathcal{L}_{2}. 
\end{align}
Here, ${\bar{u}}$ represents the coarse-grid filtered DNS (fDNS)
velocity and the forcing term $\mathcal{\bar{F}}_i = 0$ in the present study. Algorithm \ref{alg1} summarizes the training process of LESnets. The training process is briefly summarized as:

\begin{enumerate}[i]
\item {Input the prepared initial field ${\{{
\bm{u}(t_n)}\}_{n=1}^{N_1}}$ and add the grid information ${\{{\bm{G}(t_n)}\}_{n=1}^{N_1}}$.}
\item {Output the subsequent field ${\{{\bm{u}(t_n)}\}_{n=2,3,...T}^{N_1}}$ through neural operator $\mathcal{G_{\theta}}$.}
\item Calculate the PDE loss $\mathcal{L}_{PDE}$ as the target to optimize the model parameters and select the appropriate model parameters in the prediction stage.
\end{enumerate}

{We define $N_1$ as the number of training datasets, $T$ as the number of time steps of LESnets, $N_{ep}$ as the number of training epochs, $\lambda$ as the learning rate, and $N_{bc}$ as the batch-size of the LESnets. The PDE loss $\mathcal{L}_{PDE}$, as defined in Eq. \eqref{eq 17}, serves as the optimization objective for updating the model parameters. The merits of LESnets models in learning a neural operator includes:}

\begin{itemize}
\item The efficiency  of using only one initial field for training and using spectral method for computing PDE loss.
\item The flexibility of output time period ${T}$.
\item {The generalization ability to unseen regime of flow fields.}
\end{itemize}


\begin{figure}[htbp]
\centering
\includegraphics[width=0.8\textwidth]{ 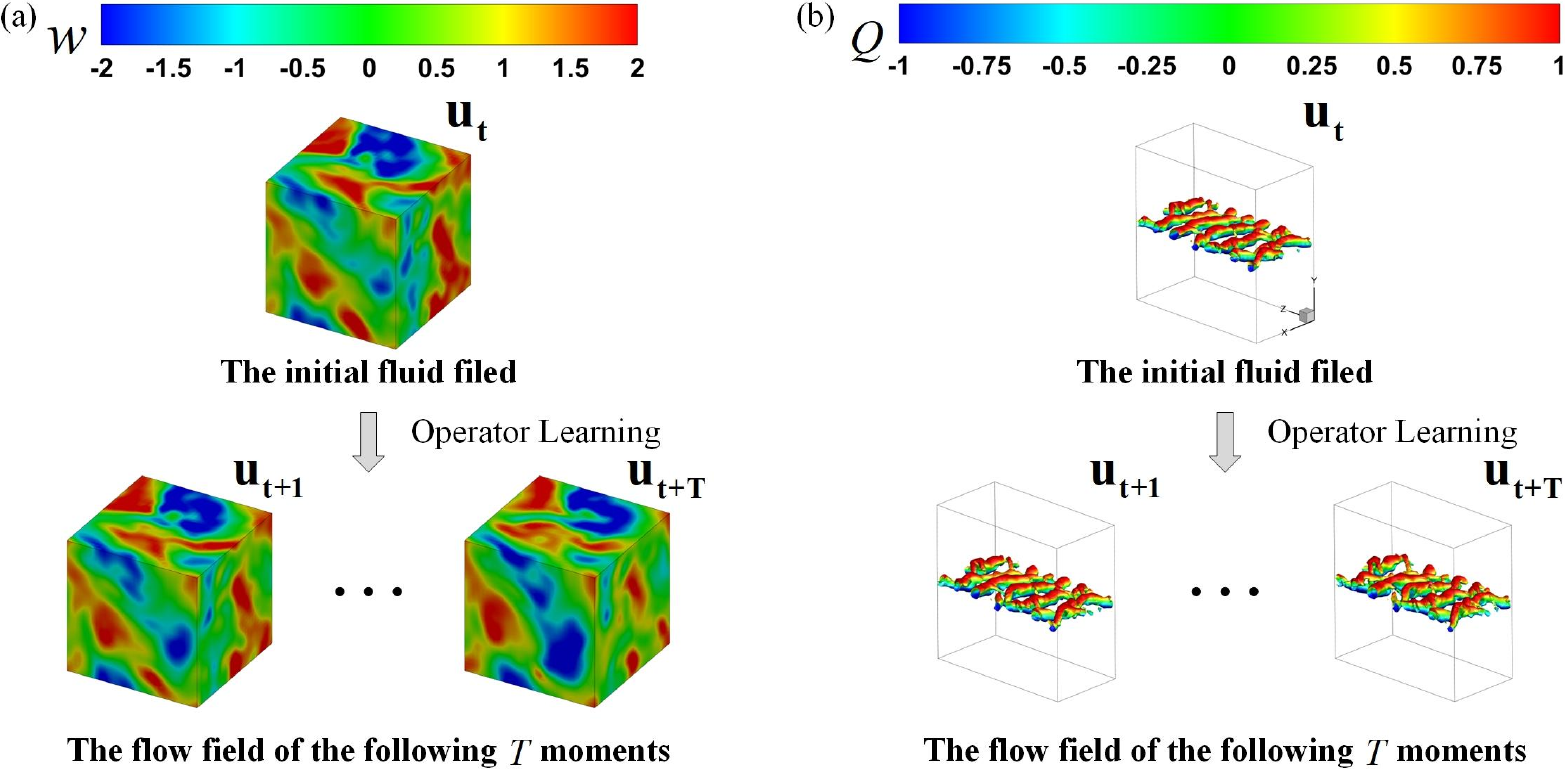}

\caption{The inputs and outputs of the neural operators for the two incompressible flow simulation cases of this study. (a) the contours of velocity w in decaying homogeneous isotropic turbulence; (b) the $Q$ criterion in temporally evolving turbulent mixing layer. }
\label{fig 4}
\end{figure}

{In the following sections, we will consider two cases of LES of incompressible turbulence: a 3D decaying homogeneous isotropic turbulence, and a 3D temporally evolving turbulent mixing layer. In Fig. \ref{fig 4}, the flow fields are visualized by using contours of velocity w in the Z direction and the $Q$ criterion respectively. The definition of $Q$ is defined as \cite{hunt1988eddies,dubief2000coherent}}

\begin{equation}
    Q=\frac{1}{2}\left(\bar{\Omega}_{ij}\bar{\Omega}_{ij}-\bar{S}_{ij}\bar{S}_{ij}\right),
\label{eq 18}
\end{equation}
{where $\bar{\Omega}_{ij}=(\partial\bar{u}_{i}/\partial x_{j}-\partial\bar{u}_{j}/\partial x_{i})/2$ is the filtered rotation-rate tensor. In addition, we only learn the mapping once in the training, and give the subsequent flow fields with longer time steps through iteration during the prediction.}

\section{Numerical experiments}
\label{sec3}

{In this section, a series of numerical experiments are carried out to assess the performance of the LESnets models in predictions of 3D decaying homogeneous isotropic turbulence and 3D temporally evolving turbulent mixing layer. The training and testing loss curves of four machine learning models are presented in the training stage. In the $a$ $posteriori$ test, five independent simulations with different initial fields are performed to evaluate the performance of the LESnets models. By comparing LESnets models with traditional large-eddy simulation using the Smagorinsky model, as well as with data-driven methods including FNO and IFNO, the accuracy and generalization capabilities of LESnets models are comprehensively examined.}

\begin{figure}[htbp]
\centering
\begin{minipage}{0.45\linewidth}
\centerline{\includegraphics[width=\textwidth]{ 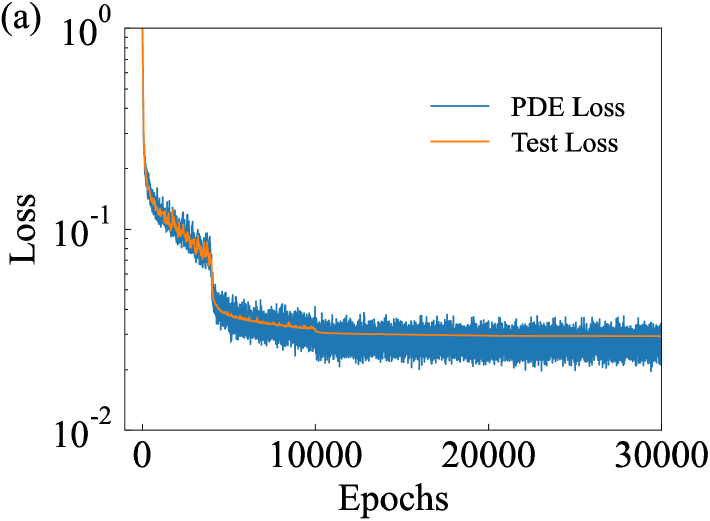}}
\end{minipage}
\hspace{1pt}
\begin{minipage}{0.45\linewidth}
\centerline{\includegraphics[width=\textwidth]{ 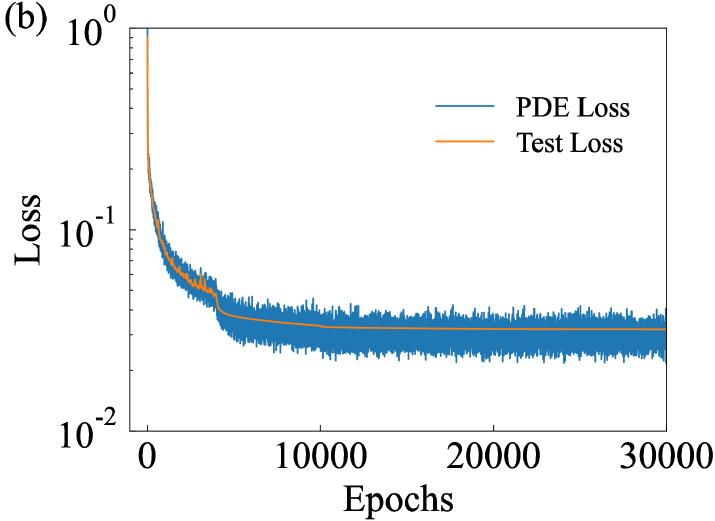}}
\end{minipage}
\vspace{-1pt}
\begin{minipage}{0.45\linewidth}
\centerline{\includegraphics[width=\textwidth]{ 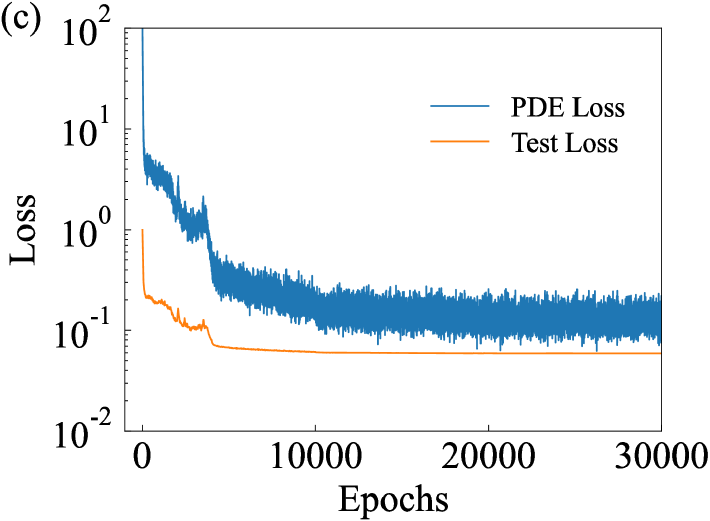}}
\end{minipage}
\hspace{1pt}
\begin{minipage}{0.45\linewidth}
\centerline{\includegraphics[width=\textwidth]{ 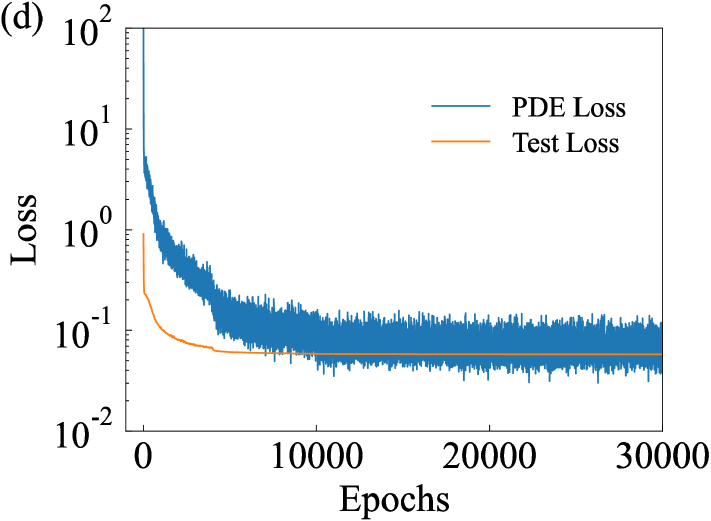}}
\end{minipage}
\caption{The evolutions of the loss curves in decaying HIT: (a) FNO; (b) IFNO; (c) LESnets and (d) LESnets-I.}
\label{fig5}
\end{figure}

\subsection{Decaying homogeneous isotropic turbulence}
\label{subsec3_1}

The direct numerical simulation of forced homogeneous isotropic turbulence (HIT) is performed with the uniform grid resolutions of ${128^3}$ in a cubic box of $(2\pi)^3$ with periodic boundary conditions \cite{xie2020modeling,yuan2020deconvolutional}. The initial velocity
field is randomly generated with the Gaussian distribution in
spectral space. The initial velocity spectrum of the random velocity
field is given by \cite{yuan2020deconvolutional}

\begin{equation}
    E(k)=A_0\left(\frac{k}{k_0}\right)^4\exp\left[-2\left(\frac{k}{k_0}\right)^2\right],
\label{eq 19}
\end{equation}
where ${E(k)}$ is the spectrum of of kinetic energy per unit mass, $k$ is the wavenumber magnitude in the spectral space. Here, $A_0 = 2.7882$ and $k_0=4.5786$.

The governing equations Eq. \eqref{eq 1} and Eq. \eqref{eq 2} are spatially discretized using the pseudo-spectral
method, and a second-order two-step Adams–Bashforth explicit
scheme is utilized for time integration \cite{spectralmethod,xie2020modeling,yuan2020deconvolutional}. {The aliasing error is eliminated by truncating the high wavenumbers of Fourier modes using the two-thirds rule.}

{The large-scale force is constructed by amplifying the velocity field in the wavenumber space to maintain the kinetic energy spectrum in the first two wavenumber shells at the prescribed values ${E_0{(1)}=1.242477}$ and ${E_0{(2)}=0.391356}$, respectively \cite{yuan2020deconvolutional}. The force $\hat{\mathcal{F}}^f_i(k)$ is calculated as} 

\begin{equation}
    \hat{\mathcal{F}}^f_i(k)=\alpha\hat{\mathcal{F}}_i(k),\quad\text{where  }\alpha=\begin{cases} \sqrt{E_0(1)/E(1)},&0.5\le k\le1.5,\\ \sqrt{E_0(2)/E(2)},&1.5\le k\le2.5,\\ 1&\text{otherwise.}\end{cases}
    \label{eq 20}
\end{equation}

\begin{figure}[htbp]
\centering
\begin{minipage}{0.45\linewidth}
\centerline{\includegraphics[width=\textwidth]{ 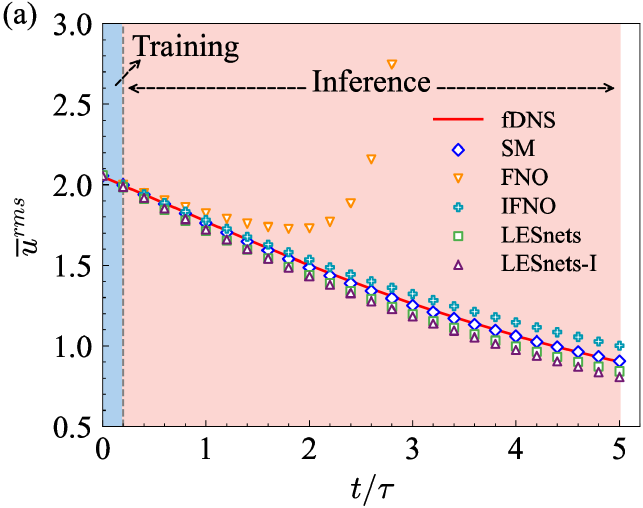}}
\end{minipage}
\hspace{1pt}
\begin{minipage}{0.45\linewidth}

\centerline{\includegraphics[width=\textwidth]{ 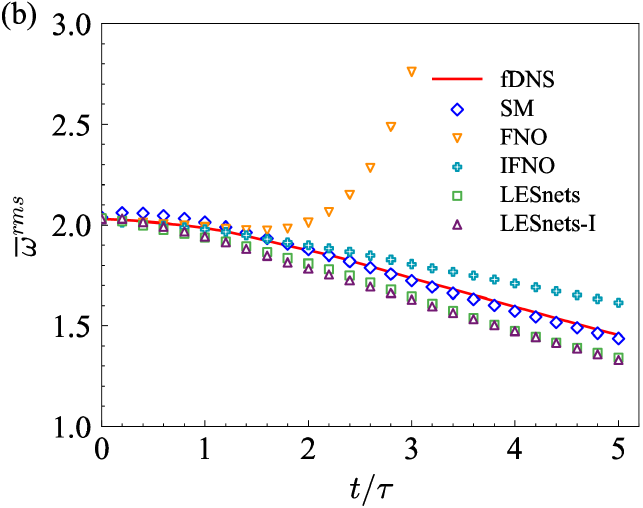}}
\end{minipage}
\caption{Temporal evolutions of (a) the rms velocity and (b) the rms vorticity for different models in decaying HIT.}
\label{fig6}
\end{figure}

{The kinematic viscosity $\nu = 0.015625$, and the corresponding Taylor Reynolds number
${Re_{\lambda} \approx 60}$. The DNS time step ${\Delta{t}}$ is set to 0.001. The force term is maintained for 10,000 DNS time steps until the flow becomes statistically steady. Subsequently, the forcing term is removed, allowing the turbulence to decay over 5,000 DNS time steps, leading to the development of decaying homogeneous isotropic turbulence. In this numerical experiment, we generate 5,000 distinct cases with varying initial fields for training, complemented by additional 20 cases for validation. Velocity fields of the numerical simulations are saved at every 20 time DNS steps.}

\begin{figure}[htbp]
\centering
\begin{minipage}{0.45\linewidth}
\centerline{\includegraphics[width=\textwidth]{ 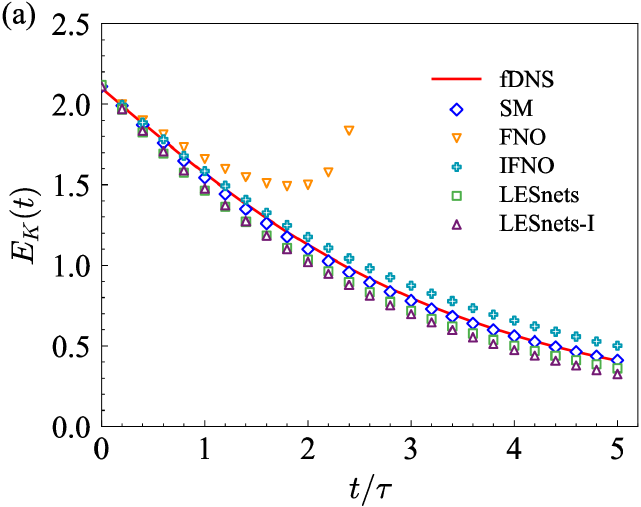}}
\end{minipage}
\hspace{1pt}
\begin{minipage}{0.45\linewidth}

\centerline{\includegraphics[width=\textwidth]{ 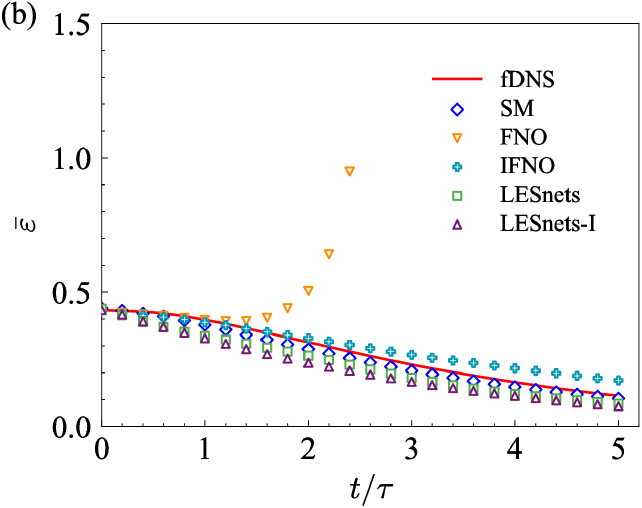}}
\end{minipage}
\caption{Temporal evolutions of (a) the turbulent kinetic energy $E_K(t)$ and (b) the average dissipation rate $\mathit{\bar{\varepsilon}}$ for different models in decaying HIT.}
\label{fig7}
\end{figure}

\begin{table}[H]
\captionsetup{font=small,labelfont=bf, width=.69\textwidth}

\setlength{\abovecaptionskip}{0pt}
\setlength{\belowcaptionskip}{1pt}
\caption{Parameters and statistics for DNS and fDNS of decaying HIT.}
\label{table1}
\centering

\begin{tabular}{cccccccc}
\toprule
Reso.(DNS) & Reso.(fDNS) & Domain & $Re_{\lambda}$ & $\nu$ & ${\Delta{t}}$ &$k_c$ & $\tau$\\
\midrule
$128^3$ & $32^3$ & $(2\pi)^{3}$ & $60$ & $0.015625$ & $0.001$ &$10$ & $1.00$\\

\bottomrule
\end{tabular}
\end{table}

In this work, we consider LES equations Eq. \eqref{eq 3} and Eq. \eqref{eq 4} with the Smagorinsky model as the physics constraints of the flow. We use the sharp spectral filter: $\hat{G}(k) = H(k_c-|k|)$, where $k_c=10$ is the cutoff wavenumber. {The SGS stress in Eq. \eqref{eq 4} is modeled by the SM model according to Eq. \eqref{eq 5}, with the Smagorinsky coefficient $C_{\mathrm{Smag}}=0.1$. The fDNS dataset comprises high-fidelity flow fields organized in tensors of dimensions [$T$ $\times$ $N_x$ $\times$ $N_y$ $\times$ $N_z$ $\times$ $N_d$], where $T$ is number of time steps of the input to the models, $N_x=N_y=N_z=32$ are the grid resolutions and $N_d=3$ is the number of the velocity components. For training the LESnets and LESnets-I, we employ 5,000 cases using only one ($T=1$) velocity field at $t=10,000\Delta t$. In contrast, for training FNO and IFNO, we employ 5,000 cases, with $T=11$ velocity fields taken from $t=10,000\Delta t$ to $10,200\Delta t$, collected every 20 DNS steps. Additionally, 20 cases with $T=11$ velocity fields are employed as testing data for the training of the four models. The detailed simulation parameters are given in Table \ref{table1}.}

{The four neural operator models employ a given number of Fourier modes, specifically $k_{max}=12$. For LESnets and FNO, we utilize $L=6$ Fourier layers with a channel space `width' $d_v$ of 80 to enhance memory usage and improve learning efficiency. For LESnets-I and IFNO, we enhance the depth of the models by increasing the number of Fourier layers to $L=20$ and adjusting the channel space `width' $d_v$ to 150.} In order to ensure a fair comparison, we take the same optimization strategy: the initial learning rate for Adam \cite{Adam} decays from $10^{-3}$ (4,000 training epochs) to $10^{-4}$ (6,000 training epochs), $10^{-5}$ (10,000 training epochs), $10^{-6}$ (10,000 training epochs), with a total of 30,000 epochs. {The GELU function \cite{GELU} is chosen as the activation function. LESnets models minimizes the PDE loss according to Eq.\eqref{eq 17}. For the FNO and IFNO models, the training and testing losses are both defined as}

\begin{equation}
    Loss=\frac{||u^*-u||_2}{||u||_2},\quad\text{where } ||\mathbf{A}||_2=\frac{1}{n}\sqrt{\sum_{k=1}^n|\mathbf{A_k}|^2}.
\label{eq 21}
\end{equation}

Here, $u^*$ denotes the prediction of velocity fields and $u$ is the ground truth. {A comparison of the training and testing loss cruve is given in Fig. \ref{fig5}. It is worth noting that the magnitude of PDE loss will reach the order of $10^2$ in the early training stage, while the data loss given by Eq. \eqref{eq 21} is always less than 1.0. 
It is shown that the test loss of LESnets models are similar to that of the FNO and IFNO models. Furthermore, the implicit-based models (i.e., IFNO and LESnets-I) exhibit a faster convergence.}


\begin{figure}[htbp]
\center

\begin{minipage}{0.45\linewidth}
\centerline{\includegraphics[width=\textwidth]{ 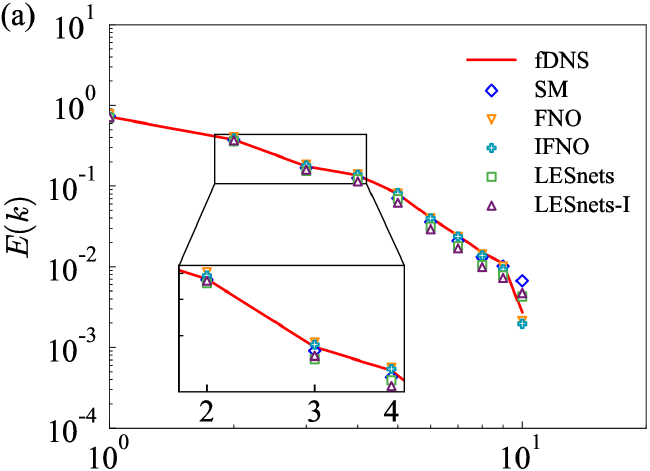}}
\end{minipage}
\hspace{1pt}
\begin{minipage}{0.45\linewidth}
\centerline{\includegraphics[width=\textwidth]{ 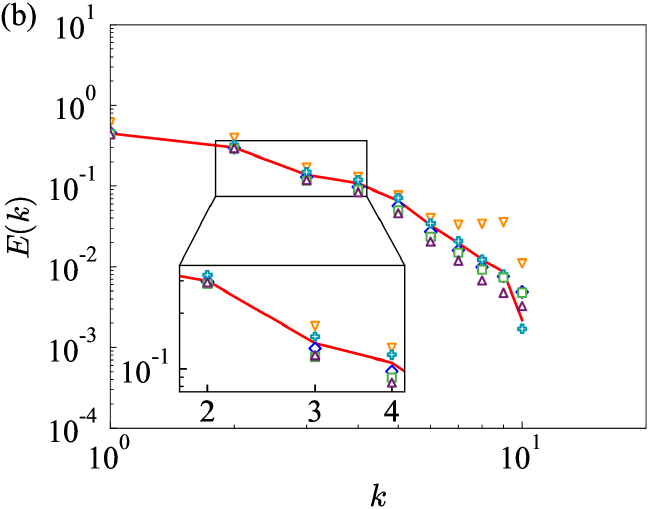}}
\end{minipage}

\vspace{-1pt}

\begin{minipage}{0.45\linewidth}
\centerline{\includegraphics[width=\textwidth]{ 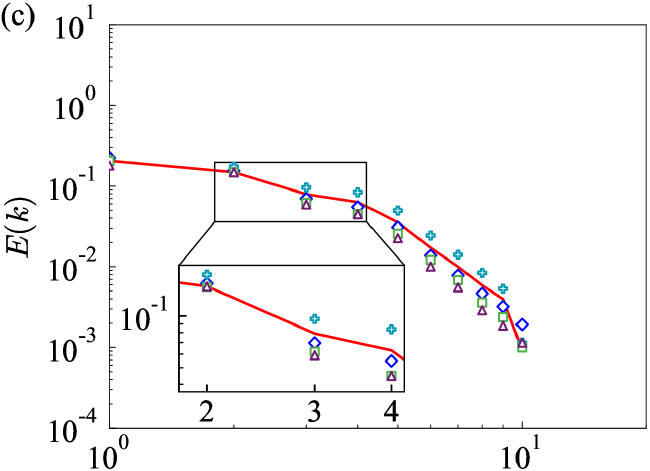}}
\end{minipage}
\hspace{1pt}
\begin{minipage}{0.45\linewidth}
\centerline{\includegraphics[width=\textwidth]{ 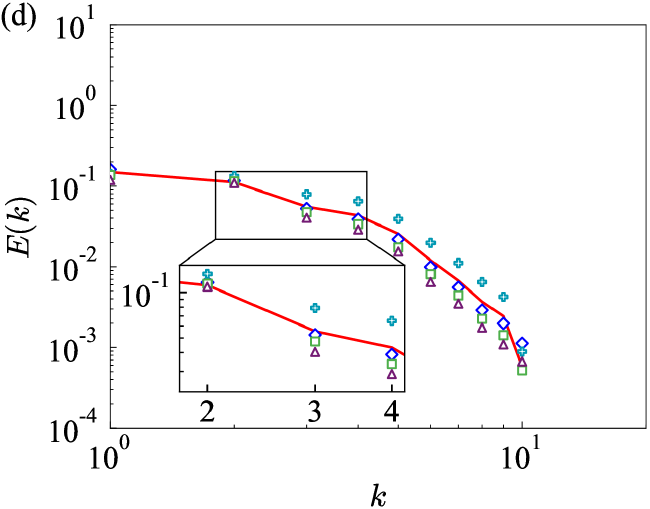}}
\end{minipage}

\caption{Spectra of turbulent kinetic energy of fDNS, SM, FNO, IFNO, LESnets and LESnets-I at (a) $t\approx\tau$; (b) $t\approx2\tau$ ; (c) $t\approx4\tau$ and (d) $t\approx5\tau$. }
\label{fig8}
\end{figure}

\begin{table}[H]
\captionsetup{font=small,labelfont=bf, width=.47\textwidth}

\setlength{\abovecaptionskip}{0pt}
\setlength{\belowcaptionskip}{1pt}
\caption{{Comparison of the minimum training and testing loss of different models in decaying homogeneous isotropic turbulence.}}
\label{table2}
\centering

\begin{tabular}{llll}
\toprule
Model & Data Loss & PDE Loss & Testing Loss \\
\midrule
FNO & $\pmb{0.019680}$ & N/A & $\pmb{0.029363}$ \\
IFNO & $0.021713$ & N/A & $0.031973$ \\
LESnets & N/A & $0.062677$ & {$0.059160$} \\
LESnets-I & N/A & $0.030308$ & {$0.057906$} \\
\bottomrule%
\end{tabular}
\end{table}

{Table \ref{table2} shows the minimum data loss, PDE loss and test loss in the training process. Although FNO achieved the best results, it will be shown in the subsequent verification that FNO can not be generalized to the subsequent flow field evolution in the decaying turbulence, and  its stability is poor in the long-term prediction. Additionally, we conduct numerical experiments with the LESnets models using both data loss and PDE loss in Section \ref{sec5}.}

{In the $a$ $posteriori$ test, five distinct turbulent flow fields at statistically steady time instant $t=10,000\Delta t$, are extracted from newly generated fDNS datasets that are independent of the training set. These fields are used as initial conditions for DNS generated from decaying homogeneous isotropic turbulence over 5,000 DNS time steps. The Numerical solutions of each initial condition are sampled every 20 DNS time steps, resulting in $N_t=250$ velocity fields ${\{{\bm{u}(t_n)}\}_{n=1,2,...N_t}}$ for analysis. The interval of learning time steps is $\Delta t_{n}=t_{n+1}-t_n=20\Delta t$. Specifically, the models are trained on time series data $N_t= 11$ velocity fields ${\{{\bm{u}(t_n)}\}_{n=1,2,...N_t}}$, while inference is performed across the extended temporal domain of $N_t=250$ learning time steps. The ensemble average is performed across all five cases to analyze flow statistics. Additionally, the temporal evolutions of flow fields are shown at intervals of $0.2\tau$.}

We show the predicted root mean square (rms) values of velocity and vorticity at different time instants in Fig. \ref{fig6}. They represent the statistical characteristics of the fluctuation of fluid velocity and the vortex motion, respectively. {In the flow regime consistent with the training process ($t{\leq}0.2\tau$), all models demonstrate an excellent predictive performance. However, the FNO exhibits a significant deviation from the true values of rms velocity and rms vorticity at the 10th iteration (i.e., $t\approx2\tau$). While IFNO shows an improved performance over conventional FNO architectures, the implicit methodology yields a negligible enhancement for LESnets. Both LESnets models exhibit remarkable stability, with predictions closely consistent with fDNS. Interestingly, the data-driven model overestimates values of rms velocity and rms vorticity, whereas LESnets models underestimate these values.}


\begin{figure}[htbp]
\center
\begin{minipage}{0.3\linewidth}
\centerline{\includegraphics[width=\textwidth]{ 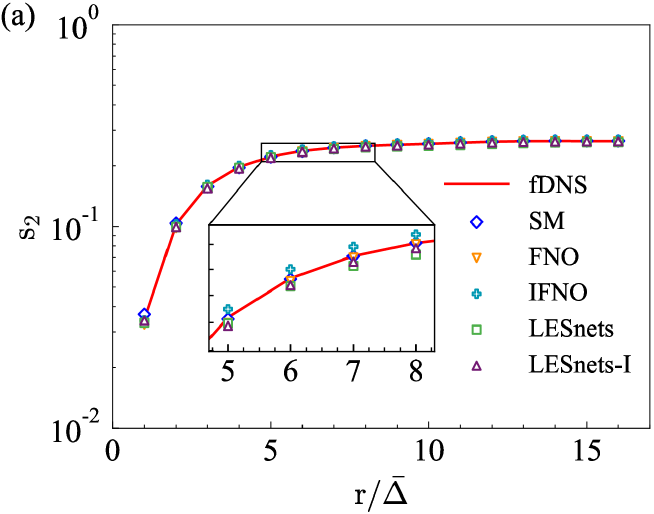}}
\end{minipage}
\hspace{-1pt}
\begin{minipage}{0.3\linewidth}
\centerline{\includegraphics[width=\textwidth]{ 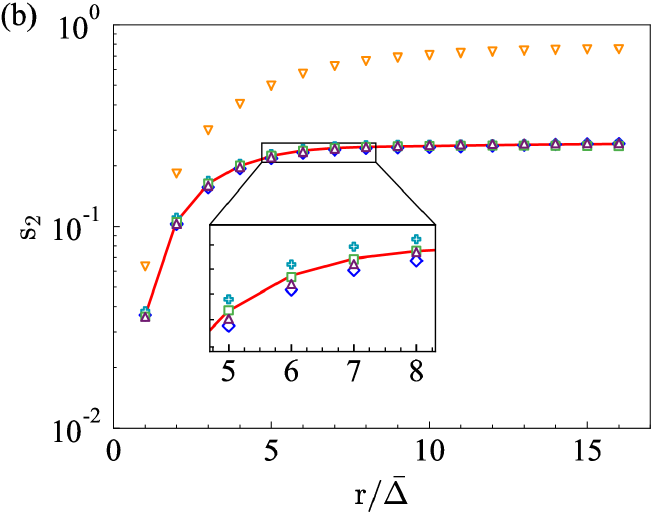}}
\end{minipage}
\hspace{-1pt}
\begin{minipage}{0.3\linewidth}
\centerline{\includegraphics[width=\textwidth]{ 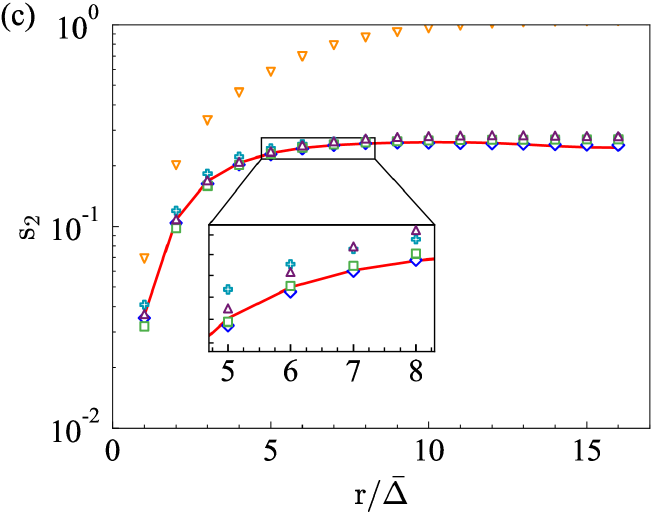}}
\end{minipage}

\vspace{-1pt}

\begin{minipage}{0.3\linewidth}
\centerline{\includegraphics[width=\textwidth]{ 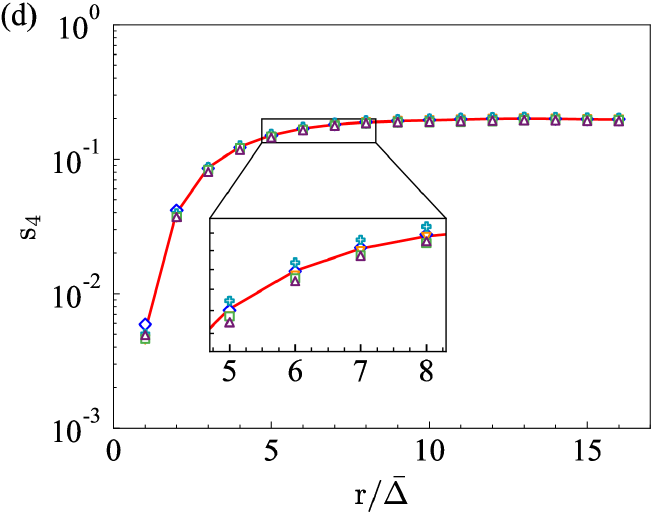}}
\end{minipage}
\hspace{-1pt}
\begin{minipage}{0.3\linewidth}
\centerline{\includegraphics[width=\textwidth]{ 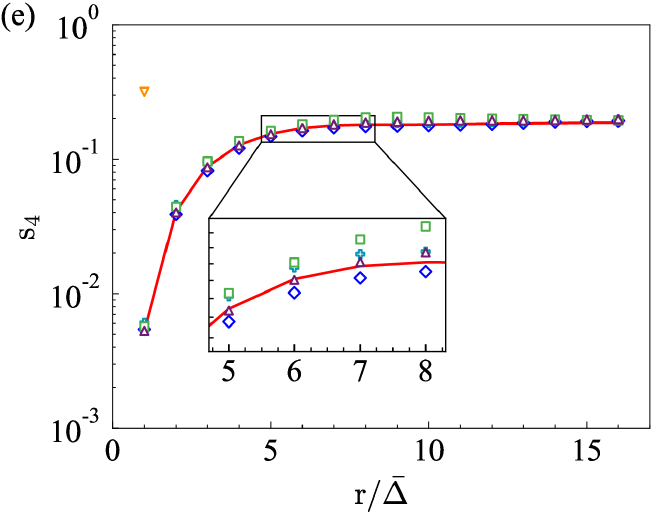}}
\end{minipage}
\hspace{-1pt}
\begin{minipage}{0.3\linewidth}
\centerline{\includegraphics[width=\textwidth]{ 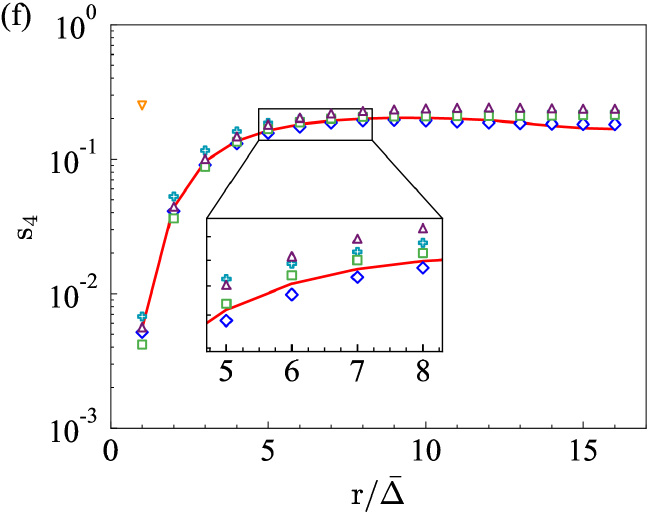}}
\end{minipage}

\caption{Second-order and fourth-order structure functions of the LES using different models in decaying HIT at different time instants (a) second-order, $t\approx\tau$; (b) second-order, $t\approx3\tau$; (c) second-order, $t\approx5\tau$; (d) fourth-order, $t\approx\tau$; (e) fourth-order, $t\approx3\tau$ and (f) fourth-order, $t\approx5\tau$.}
\label{fig9}
\end{figure}
{In Fig. \ref{fig7}, we present the temporal evolutions of the turbulent kinetic energy $E_K(t)=\int_0^\infty E(k)dk=\frac{1}{2}(u^{\mathrm{rms}})^2$ and the predicted dissipation rate $\mathit{\bar{\varepsilon}}$ by the SM, FNO, IFNO, LESnets models along with the fDNS data. It can be seen that the kinetic energy gradually decays from the initial state over time, and the FNO model fails to predicting the turbulent kinetic energy at $t\approx\tau$, suggesting an early divergence. Here, LESnets models show a similar accuracy to that of SM. The turbulent kinetic energy and energy dissipation rate predicted by the IFNO model tend to be significantly larger than fDNS, whereas those predicted by the LESnets models are smaller than fDNS.}

{Fig. \ref{fig8} shows the spectra of kinetic energy $E(k)$ at different instants. It can be seen that the turbulent kinetic energy of turbulence decreases with the time at each wavenumber. The energy spectra predicted by the traditional LES with SM are slightly lower than fDNS. The FNO model is no longer consistent the fDNS at $t\approx4\tau$. IFNO model exhibits higher values than the fDNS results at all wavenumbers, and conversely, the LESnets-I exhibits slightly lower than the fDNS results. The energy spectra predicted by LESnets are very close to the SM results at $t\approx4\tau$ and $t\approx5\tau$, and slightly deviate from the SM results at $t\approx\tau$ and $t\approx2\tau$.}


\begin{figure}[htbp]
\center

\begin{minipage}{0.45\linewidth}
\centerline{\includegraphics[width=\textwidth]{ 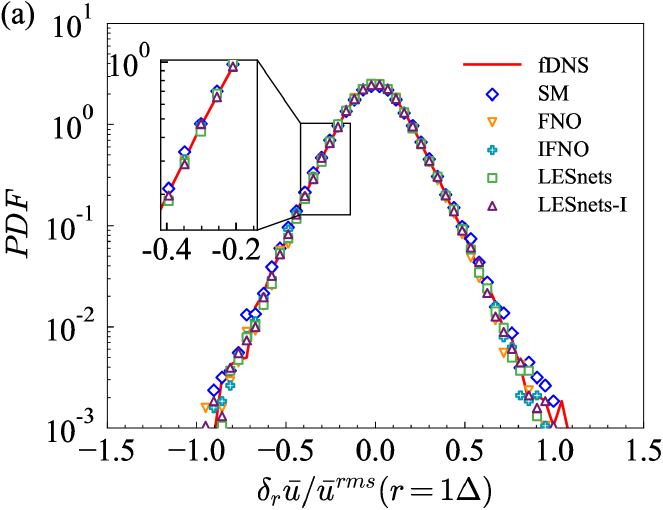}}
\end{minipage}
\hspace{1pt}
\begin{minipage}{0.45\linewidth}
\centerline{\includegraphics[width=\textwidth]{ 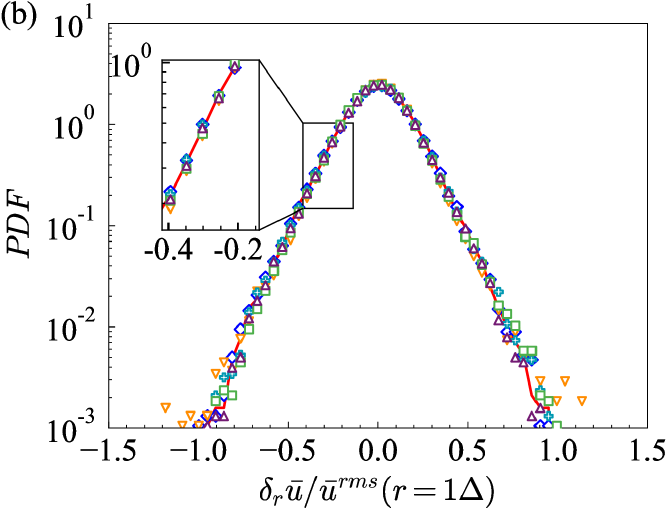}}
\end{minipage}

\vspace{-1pt}

\begin{minipage}{0.45\linewidth}
\centerline{\includegraphics[width=\textwidth]{ 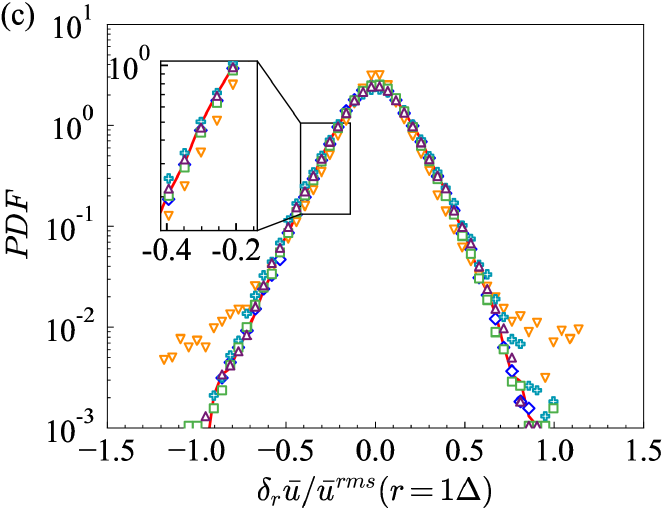}}
\end{minipage}
\hspace{1pt}
\begin{minipage}{0.45\linewidth}
\centerline{\includegraphics[width=\textwidth]{ 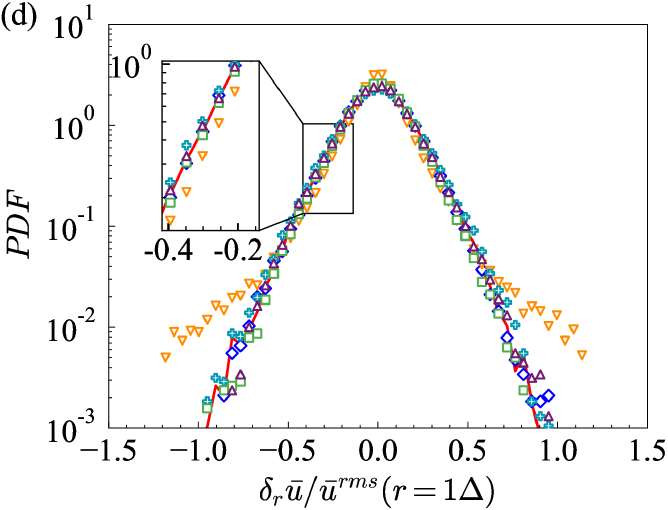}}
\end{minipage}
\caption{PDFs of the normalized velocity increments $\delta_{r}\bar{u}/\bar{u}^{\mathrm{rms}}$ of fDNS, SM, FNO, IFNO, LESnets and LESnets-I at (a) $t\approx\tau$; (b) $t\approx2\tau$ ; (c) $t\approx4\tau$ and (d) $t\approx5\tau$. }
\label{fig10}
\end{figure}

{To further examine the LESnets models in predicting the multi-scale properties of turbulence, we compute the longitudinal structure functions of the filtered velocity field as}

\begin{equation}
    \bar{S}_n(r)=\left\langle\left|\frac{\delta_r\bar{u}}{\bar{u}^\mathrm{rms}}\right|^n\right\rangle, 
\label{eq22}
\end{equation}
where $n$ denotes the order of structure function and $\delta_r\bar{u}=[\bar{\mathbf{u}}(\mathbf{x}+\mathbf{r})-\bar{\mathbf{u}}(\mathbf{x})]\cdot\widehat{\mathbf{r}}$ represents the longitudinal increment of the velocity at the separation ${\mathbf{r}}$. Here, $\widehat{\mathbf{r}} = \mathbf{r}/|\mathbf{r}|$ is the unit vector.

{Fig. \ref{fig9} compares the second-order and fourth-order structure functions of the filtered velocity field for different models at $t\approx\tau$, $t\approx3\tau$ and $t\approx5\tau$. It can be seen that the SM model gives consistent results with fDNS, and the LESnets and LESnets-I have the good predictions similar to SM. Moreover, FNO still gets a divergent result and the IFNO overestimates the structure functions at all distances compared to those of the fDNS data at the last time instants (i.e., $t\approx5\tau$). In contrast, at all time instants, LESnets gives predictions similar to those by the SM model at distances from $r/\bar{\Delta}=2$ to $r/\bar{\Delta}=8$.}

Furthermore, we compare PDFs of the normalized velocity increments $\delta_{r}\bar{u}/\bar{u}^{\mathrm{rms}}$ at distance $r=\delta$ at different time instants in Fig. \ref{fig10}. {It can be seen that the tails of PDFs of the normalized velocity increments predicted by FNO become longer than the fDNS result as the time increases ($t\approx4\tau$). The tails of PDFs of the normalized velocity increments predicted by the IFNO model are also slightly higher than the fDNS results. The LESnets and LESnets-I demonstrate a comparable accuracy to the Smagorinsky model (SM) on the prediction of PDFs of the normalized velocity increments, and are nearly identical to those of the fDNS results. }

\begin{figure}[ht]
\centering
\includegraphics [width=0.9\textwidth]{ 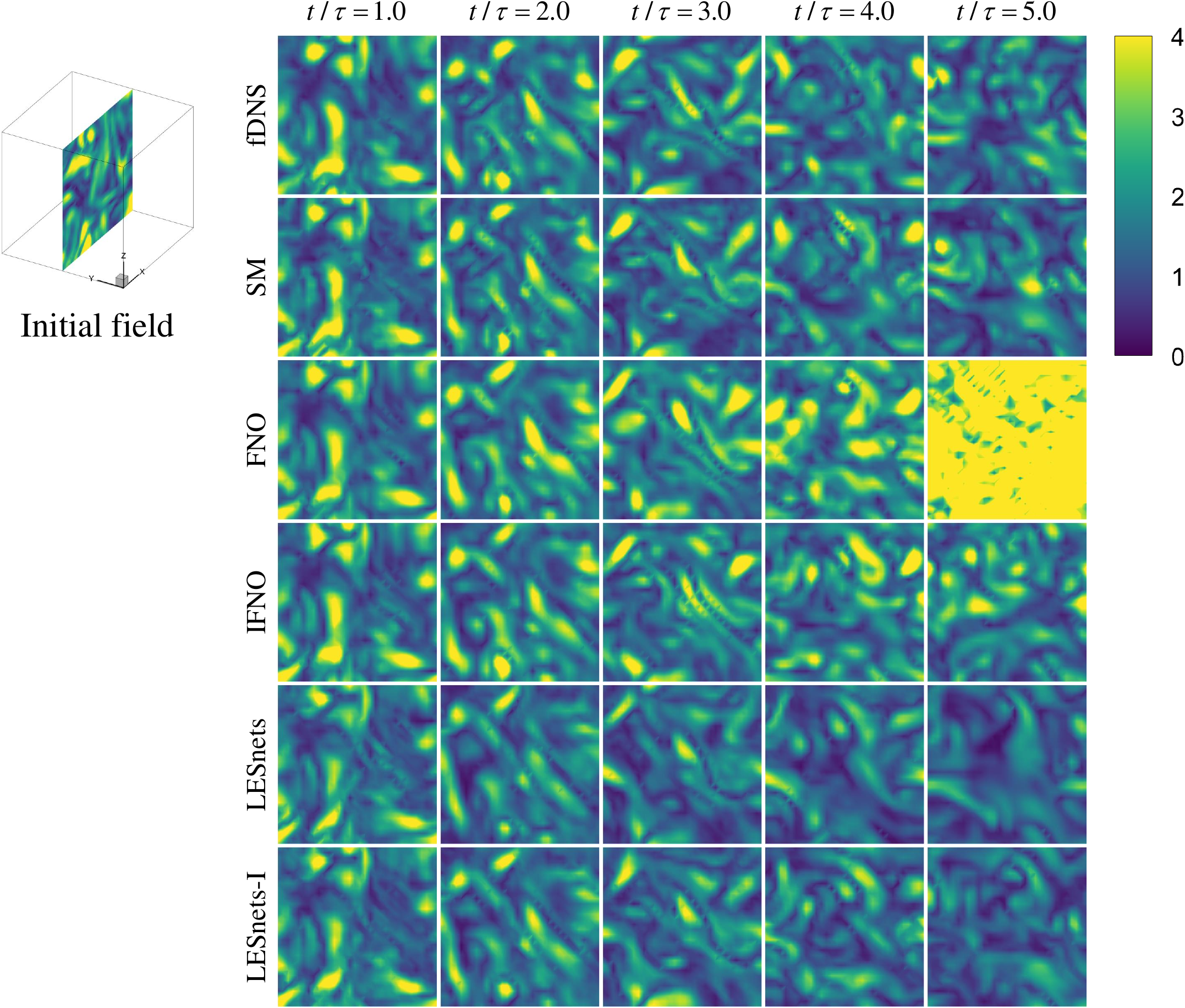}
\caption{{The normalized vorticity magnitude fields ${|\omega|}/{\bar{\omega}^{rms}_{fDNS}}$ (at the center x–z plane) from $t\approx\tau$ to $t\approx5\tau$.}}
\label{fig11}
\end{figure}

{The forgoing comparisons on the prediction accuracy of turbulent statistics show that the neural operators with physical constraints perform better than their purely data driven counterparts.}

{Fig. \ref{fig11} illustrates the contours of the normalized vorticity magnitude fields ${|\omega|}/{\bar{\omega}^{rms}_{fDNS}}$ predicted by different models. The instantaneous snapshots are selected at the center x-z plane at five different time instants. The first row presents the true values of ${|\omega|}/{\bar{\omega}^{rms}_{fDNS}}$ and the other five rows correspond to the predicted results by the SM, FNO, IFNO, LESnets and LESnets-I models. It can be seen that the traditional SM model gives flow structures slightly different from the results of fDNS in the later stage. It is shown that the FNO model completely deviates from the true value in the later stage while IFNO overestimates the evolution of vorticity, which may lead to more divergent results in the future. LESnets models can achieve a similar result as traditional large-eddy simulation and data-driven models in the initial stage, and only slightly lower than fDNS in the later stage. Since the large-eddy simulation equations are used as the physics constraints, it is observed that the results of LESnets models are closer to SM.}

\begin{figure}[ht]
\centering
\includegraphics [width=0.8\textwidth]{ 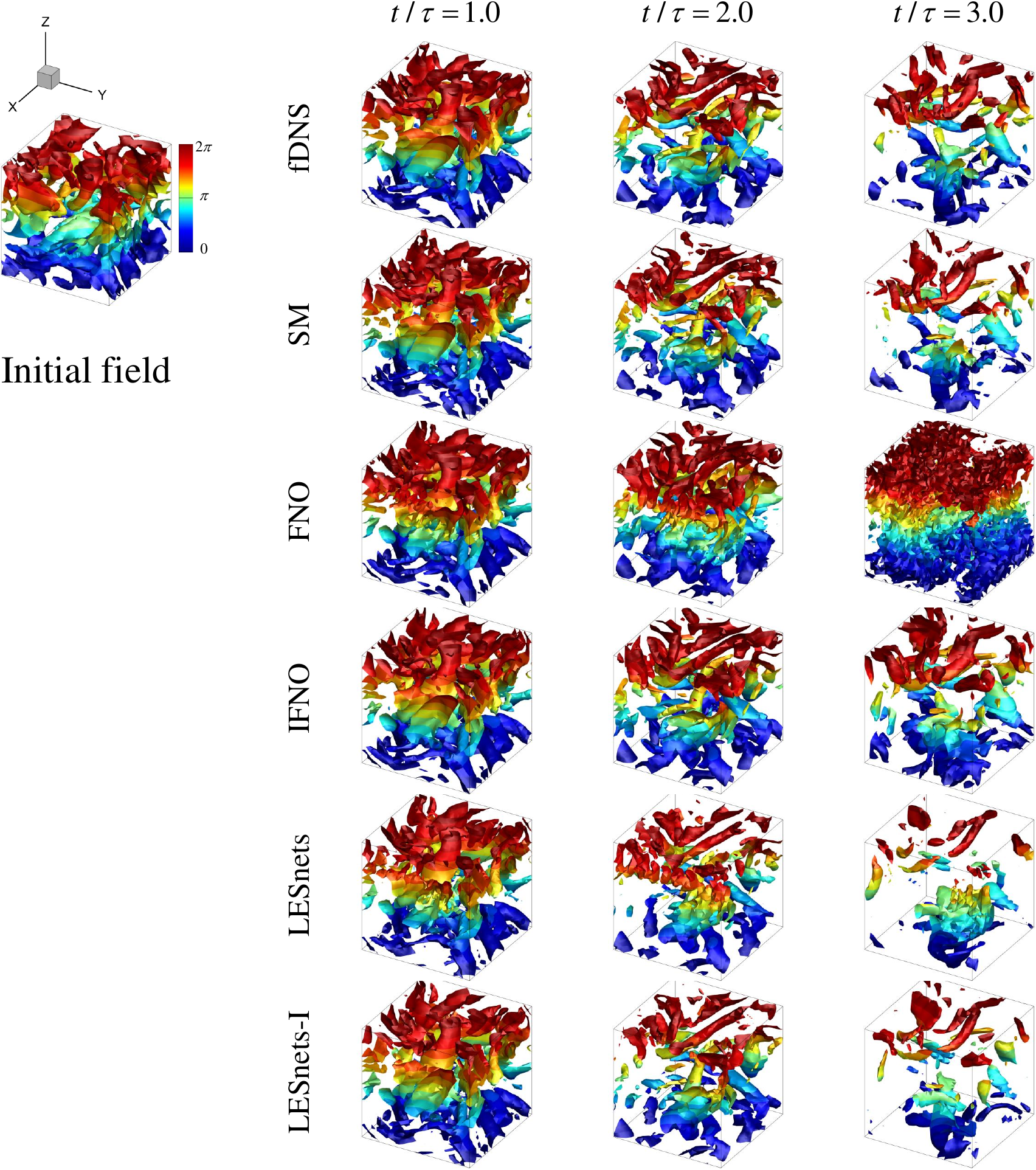}
\caption{{Iso-surface of the normalized vorticity $\bar{\omega}/\bar{\omega}_{\mathrm{fDNS}}^{\mathrm{rms}}=1.2$ (colored by the z-coordinate values) predicted by different models from $t\approx\tau$ to $t\approx3\tau$.}}
\label{fig12}
\end{figure}

{In addition, we also examine the predicted iso-surfaces by different models at a specific normalized vorticity value $\bar{\omega}/\bar{\omega}_{\mathrm{fDNS}}^{\mathrm{rms}}=1.2$ , as shown in Fig. \ref{fig12}. It can be seen that LESnets models exhibit a similar accuracy compared with traditional large-eddy simulation methods and data-driven models for iso-surfaces of normalized vorticity.}


\subsection{Temporally evolving turbulent mixing layer}
\label{subsec3-2}

{We study the performance of neural operators on the large-eddy simulation of a 3D temporally evolving turbulent mixing layer (TML).} The turbulent mixing layer provides a suitable example for studying the effects of non-uniform turbulent shear and mixing on the accuracy of operator learning. The free-shear turbulent mixing layer is governed by Navier–Stokes equations Eq. \eqref{eq 1} and Eq. \eqref{eq 2} without the forcing term. The mixing layer is numerically simulated in a cuboid domain with lengths $L_1\times L_2\times L_3 = 8\pi\times8\pi\times4\pi$ using a uniform grid resolution of $N_1\times N_2\times N_3 =256\times256\times128$. {Here, $x_1\in[-L_1/2,L_1/2],x_2\in[-L_2/2,L_2/2],x_3\in[-L_3/2,L_3/2]$ denote the streamwise, normal, and spanwise coordinates, respectively.}

The initial streamwise velocity is given by \cite{wang2022constant,wang2022compressibility}.

\begin{equation}
    u_1=\frac{\Delta U}{2}\left[\tanh\left(\frac{x_2}{2\delta_\theta^0}\right)-\tanh\left(\frac{x_2+L_2/2}{2\delta_\theta^0}\right)-\tanh\left(\frac{x_2-L_2/2}{2\delta_\theta^0}\right)\right]+\lambda_{1}.
\label{eq 22}
\end{equation}

Here, $-L_2/2\leqslant x_2 \leqslant L_2/2$, ${\delta}_{\theta}^0=0.08$ is the initial momentum thickness and $\Delta U = U_2-U_1=2$ is the velocity difference between two equal and opposite free streams across the shear layer \cite{wang2022constant}. The momentum thickness quantifies the size of the turbulence region in the mixing layer, which is given by \cite{rogers1994direct}

\begin{equation}
    \delta_\theta=\int_{-L_2/4}^{L_2/4}\left[\frac{1}{4}-\left(\frac{\langle\bar{u}_1\rangle}{\Delta U}\right)^2\right]dx_2.
\label{eq 23}    
\end{equation}

\begin{table}[H]
\captionsetup{font=small,labelfont=bf, width=.98\textwidth}

\setlength{\abovecaptionskip}{0pt}
\setlength{\belowcaptionskip}{1pt}
\caption{{Parameters and statistics for DNS and fDNS of TML.}}
\label{table3}
\centering

\begin{tabular}{cccccccc}
\toprule
Reso.(DNS:$N_x\times N_y \times N_z$) & Reso.(fDNS:$N_x\times N_y \times N_z$) & Domain & $Re_{\theta}^0$ & {${\nu}_{\infty}$} & ${\Delta{t}}$ & ${\delta}_{\theta}^0$ & $\Delta U$ \\
\midrule 
$256\times256\times128$ & $64\times64\times32$ & $8\pi \times 8\pi \times 4\pi$ & {$20$} & $0.008$ & $0.001$ & $0.08$ & $2$\\

\bottomrule
\end{tabular}
\end{table}

\begin{figure}[htbp]
\centering

\begin{minipage}{0.45\linewidth}
\centerline{\includegraphics[width=\textwidth]{ 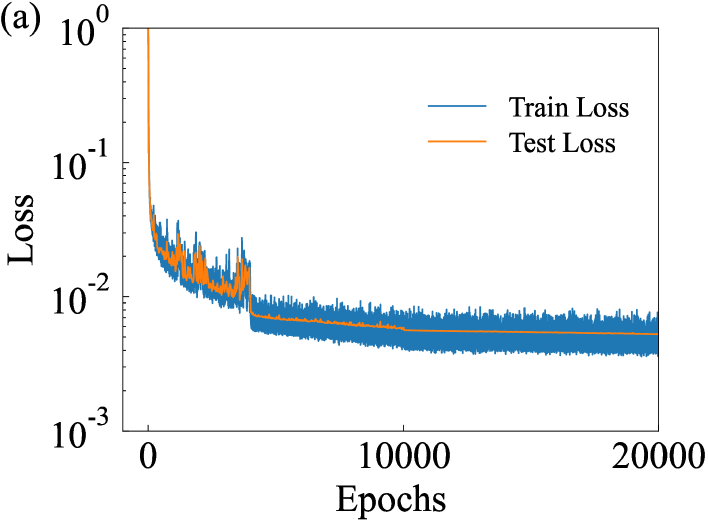}}
\end{minipage}
\hspace{1pt}
\begin{minipage}{0.45\linewidth}
\centerline{\includegraphics[width=\textwidth]{ 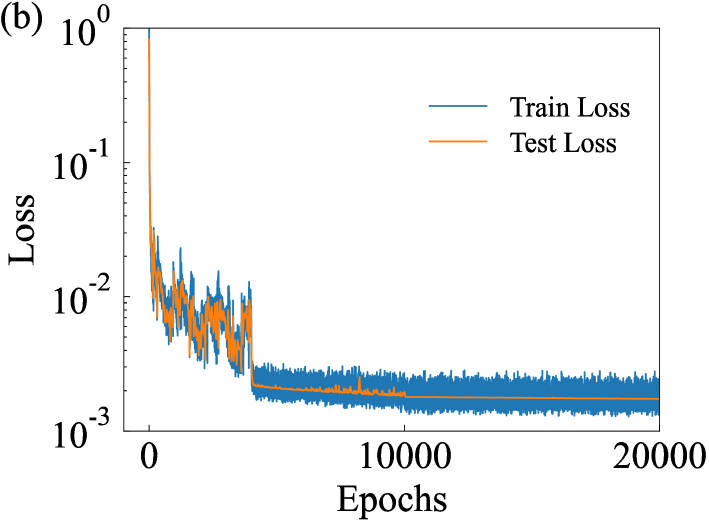}}
\end{minipage}

\vspace{-1pt}

\begin{minipage}{0.45\linewidth}
\centerline{\includegraphics[width=\textwidth]{ 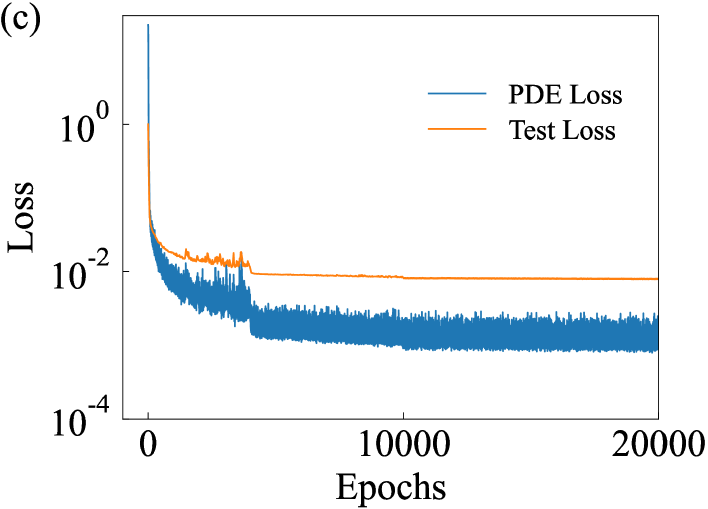}}
\end{minipage}
\hspace{1pt}
\begin{minipage}{0.45\linewidth}
\centerline{\includegraphics[width=\textwidth]{ 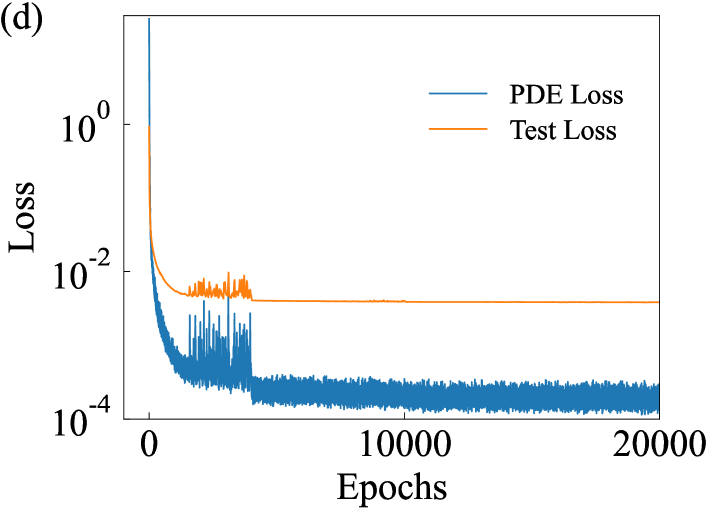}}
\end{minipage}
\caption{The evolutions of the loss curves in the TML: (a) FNO; (b) IFNO; (c) LESnets and (d) LESnets-I.}
\label{fig13}
\end{figure}

{The initial normal and spanwise velocity perturbations are given as $u_2={\lambda}_2$ and $u_3={\lambda}_3$, respectively.} Here, ${\lambda}_1,{\lambda}_2,{\lambda}_3 \sim \mathcal{N}\left({\mu,{\sigma}_2}\right)$, i.e., ${\lambda}_1,{\lambda}_2,{\lambda}_3$ satisfy the Gaussian random distribution. The expectation of the distribution is $\mu=0$, and the variance of the distribution is ${\sigma}_2=0.01$. The Reynolds number based on the momentum thickness $Re_{\theta}$ is defined as $Re_{\theta}={\Delta}U{\delta}_{\theta}/{\nu}_\infty$. {Here, the kinematic viscosity of shear layer is set to ${\nu}_{\infty}=0.008$, so the initial momentum thickness Reynolds number is $Re_{\theta}^0=20$. To mitigate the impact of the top and bottom boundaries on the central mixing layer, two numerical diffusion buffer zones are configured near the vertical edges of the computational domain \cite{yuan2023adjoint}.} The periodic boundary conditions in all three directions are utilized, and the pseudo-spectral method with the two-thirds dealiasing rule is employed. An explicit two-step Adam–Bashforth scheme is chosen as the time advancing scheme.

\begin{figure}[htbp]
\centering

\begin{minipage}{0.47\linewidth}
\centerline{\includegraphics[width=\textwidth]{ 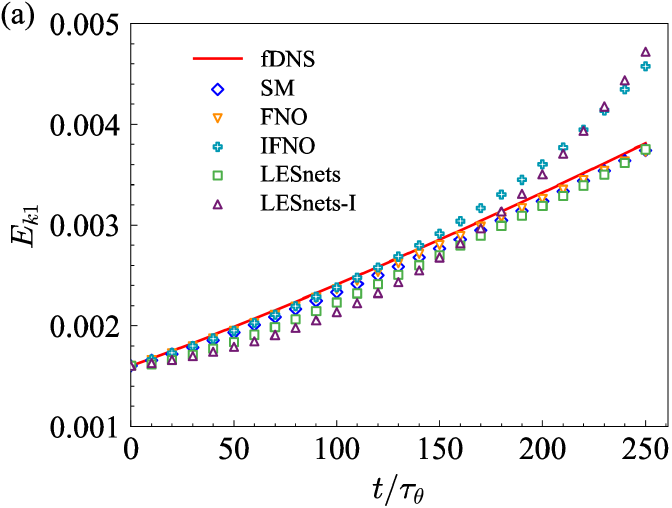}}
\end{minipage}
\hspace{1pt}
\begin{minipage}{0.45\linewidth}

\centerline{\includegraphics[width=\textwidth]{ 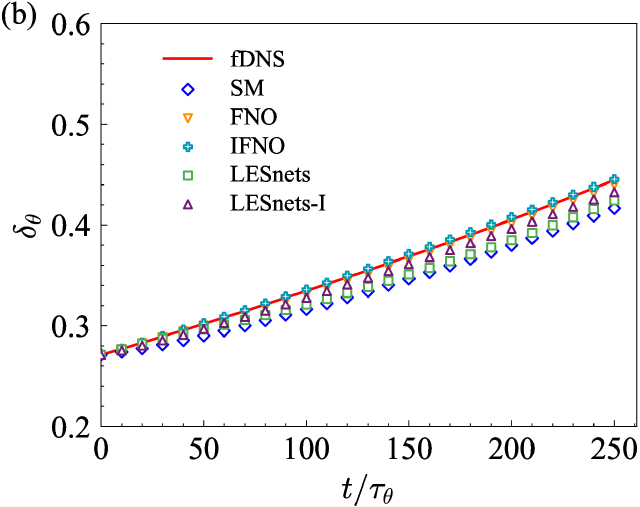}}
\end{minipage}
\caption{{Temporal evolutions of (a) the streamwise turbulent kinetic energy $E_{k1}$ and (b) the momentum thickness $\delta_{\theta}$ for the LES using different models in TML.}}
\label{fig14}
\end{figure}


\begin{figure}[ht]
\center

\begin{minipage}{0.45\linewidth}
\centerline{\includegraphics[width=\textwidth]{ 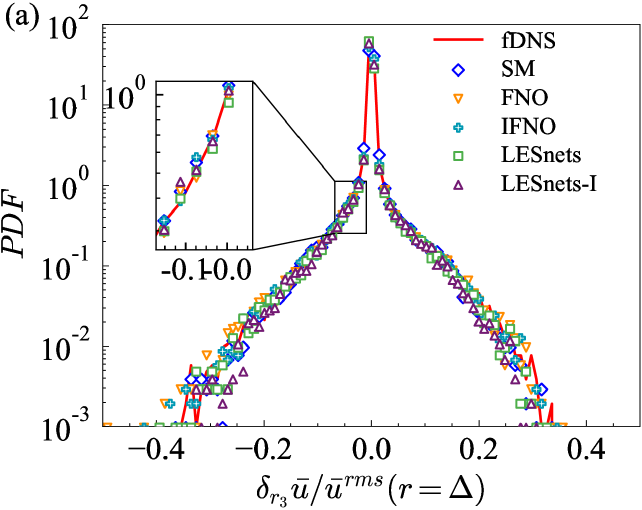}}
\end{minipage}
\hspace{1pt}
\begin{minipage}{0.45\linewidth}
\centerline{\includegraphics[width=\textwidth]{ 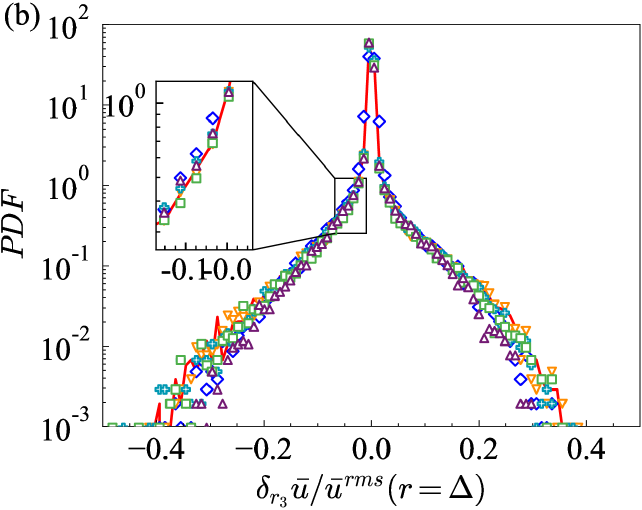}}
\end{minipage}

\vspace{-1pt}

\begin{minipage}{0.45\linewidth}
\centerline{\includegraphics[width=\textwidth]{ 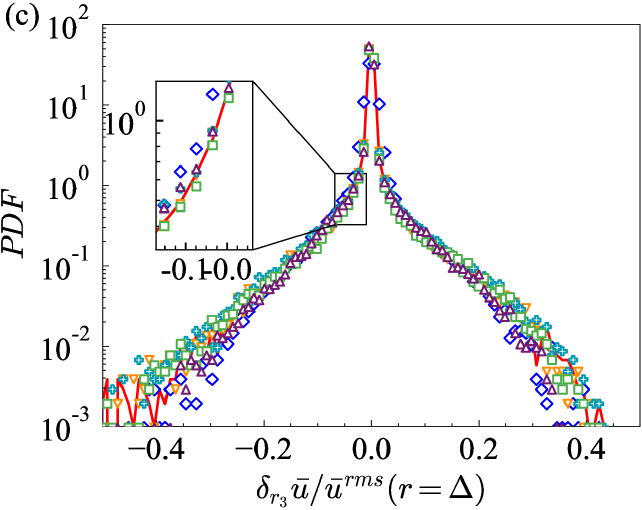}}
\end{minipage}
\hspace{1pt}
\begin{minipage}{0.45\linewidth}
\centerline{\includegraphics[width=\textwidth]{ 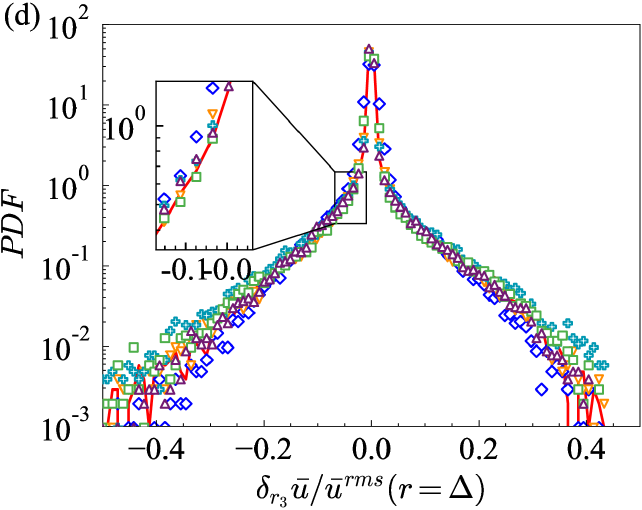}}
\end{minipage}
\caption{The PDFs of the spanwise velocity increment of fDNS, SM, FNO, IFNO, LESnets and LESnets-I at (a) $t\approx50\tau_\theta$; (b) $t\approx100\tau_\theta$ ; (c) $t\approx200\tau_\theta$ and (d) $t\approx250\tau_\theta$. }
\label{fig15}
\end{figure}

{Similarly, we use the same sharp spectral filter: $\hat{G}(k) = H(k_c-|k|)$ to filter the DNS data as for the decaying HIT. Here, the filtering scale is $\bar{\Delta}=8h_{DNS}$, where $h_{DNS}$ is the grid spacing of DNS. The filter-to-grid ratio $FGR=\bar{\Delta}/h_{LES}=2$ is utilized, and the corresponding grid resolution of fDNS is $64\times 64 \times 32$. The SGS model SM is used with the Smagorinsky coefficient $C_{\mathrm{Smag}}=\sqrt{0.001}$.}

\begin{figure}[ht]
\centering
\includegraphics [width=0.82\textwidth]{ 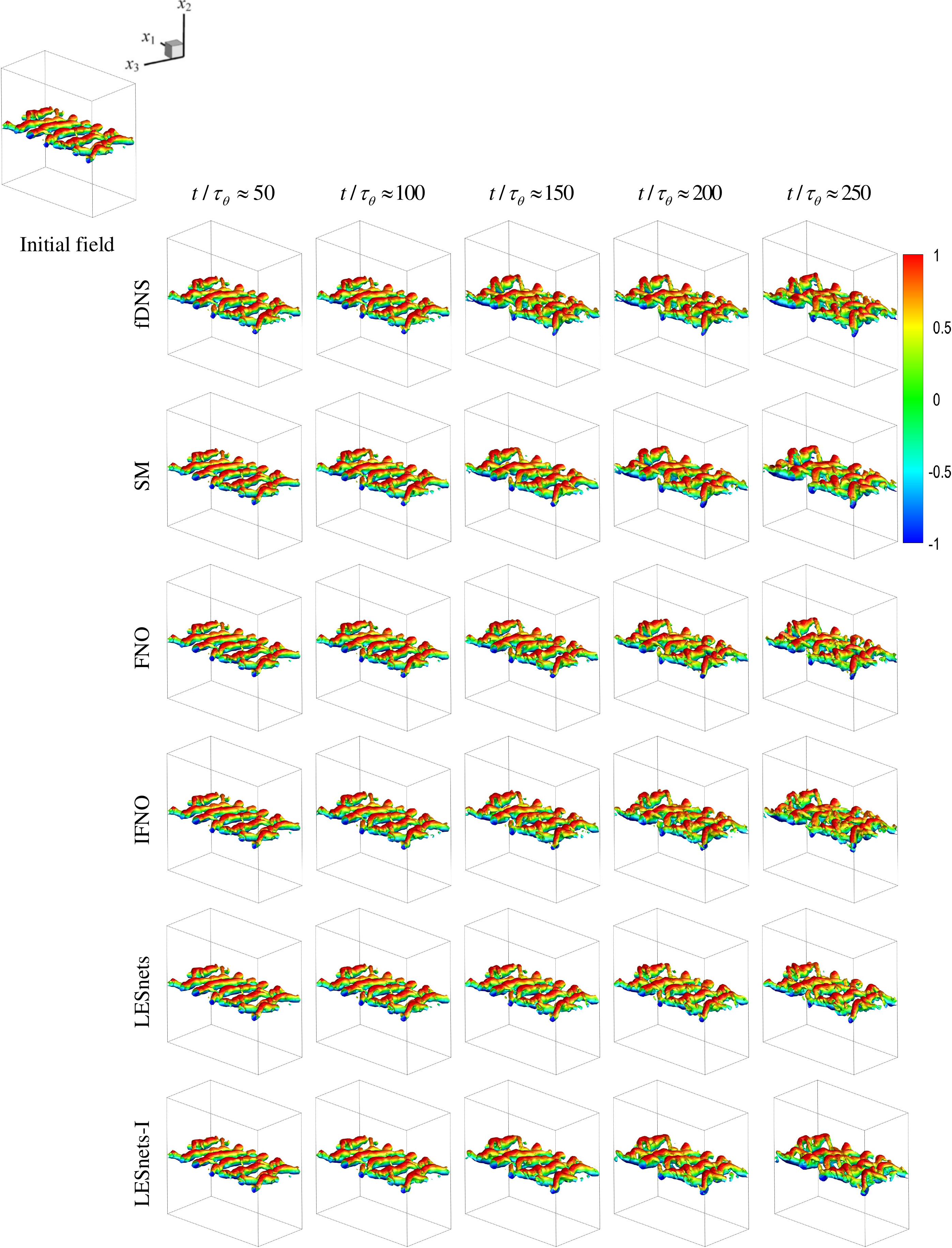}
\caption{{The iso-surface of the $Q$ criterion at $Q=0.2$ colored by the streamwise velocity $u$ at $t/{\tau_{\theta}}\approx50, 100, 150, 200, 250$ in TML.}}
\label{fig16}
\end{figure}

{In this numerical experiment, we simulate the 3D temporally evolving turbulent mixing layer (TML) with 2,000 DNS time steps. The velocity fields of the numerical simulations are saved every $20\Delta t$, and then $N_t=101$ velocity fields ${\{{\bm{u}(t_n)}\}_{n=1,2,...N_t}}$ are obtained. We generate 200 distinct flow fields with varying initial fields for training, and 5 additional flow fields for validation. These cases are divided into $N_t=11$ velocity fields ${\{{\bm{u}(t_n)}\}_{n=1,2,...N_t}}$, resulting in 2000 cases for training and 50 cases for validation. The dataset comprises high-fidelity flow fields organized in tensors of dimensions [$T$ $\times$ $N_x$ $\times$ $N_y$ $\times$ $N_z$ $\times$ $N_d$], where $T$ is number of time steps of the input to the models, $N_x=N_y=64,N_z=32$ are the grid resolutions, and $N_d=3$ is the number of velocity components. In the training of LESnets models, we employ 2,000 cases, each containing only one ($T=1$) velocity field as initial field of the model. In contrast, for training FNO and IFNO model, we employ 2,000 cases, each with $T=11$  velocity fields. Moreover, 50 cases with $T=11$ velocity fields are employed as testing data for the four models.}

{In this problem, we maintain the same parameter settings, activation function and optimizer as in the decaying HIT problem.} {The initial learning rate for Adam decays from $10^{-3}$ (4,000 training epochs) to $10^{-4}$ (6,000 training epochs), $10^{-5}$ (10,000 training epochs), with a total of 20,000 training epochs. The LESnets architecture employs the loss function in Eq. \eqref{eq 17}, whereas the FNO and IFNO frameworks utilize the data loss in Eq. \eqref{eq 21}.}

\begin{figure}[ht]
\centering
\includegraphics [width=1.0\textwidth]{ 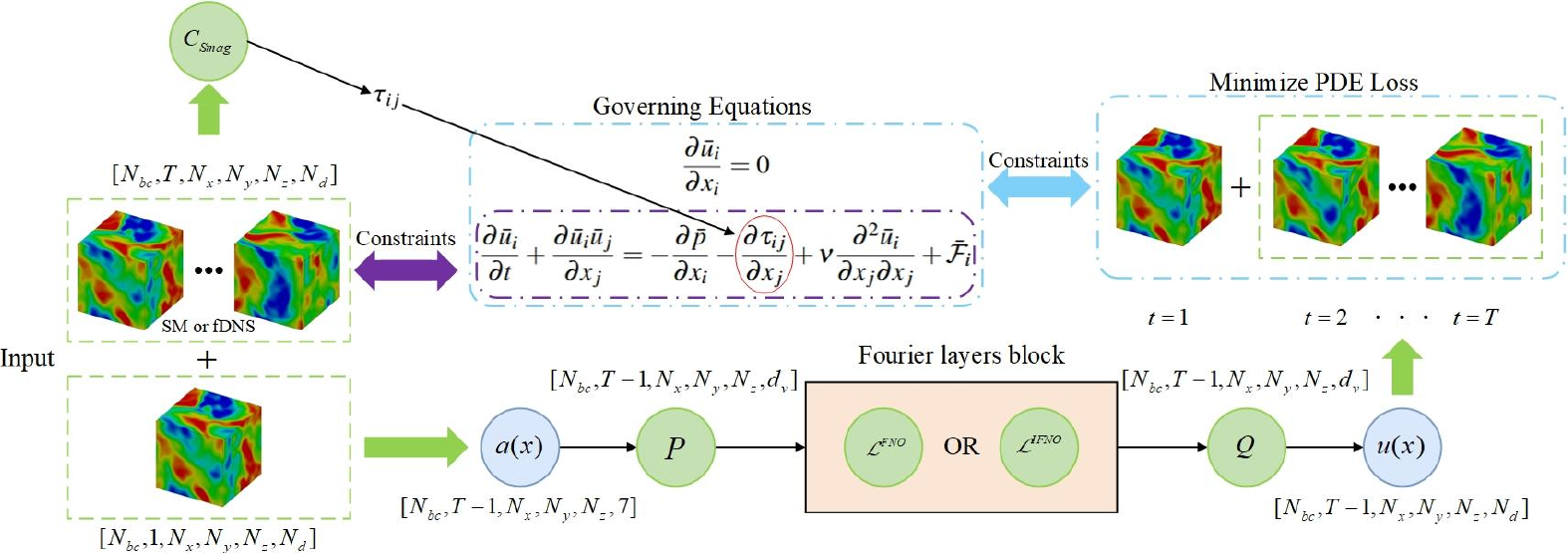}
\caption{The architecture of the Large-Eddy Simulation nets (LESnets) models for automatic learning of $C_{\mathrm{Smag}}$.}
\label{fig17}
\end{figure}

{The comparisons of the training and testing loss curves are given in Fig. \ref{fig13}. As can be seen, the test error of LESnets models are larger than data-driven models, similar to the phenomenon previously observed in decaying HIT.}

{After training, five groups of flow data with different initial fields are generated to perform the $a$ $posteriori$ test, resulting in $N_t=250$ velocity fields ${\{{\bm{u}(t_n)}\}_{n=1,2,...N_t}}$ for analysis. Fig. \ref{fig14} compares the temporal evolutions of the streamwise turbulent kinetic energy $E_{k1}=\frac12(\sqrt{\langle u_1u_1\rangle})^2$ and the momentum thickness $\delta_{\theta}$ of the SM, FNO, IFNO, LESnets and LESnets-I models.} It is interesting to note that FNO does not diverge in the TML problem. {In addition, the implicit iteration method is similar to the traditional FNO model for the streamwise turbulent kinetic energy. In terms of the momentum layer thickness, the data-driven model has a better prediction accuracy than the large-eddy simulation method. LESnets achieve comparable results to those of the SM method.}

\begin{algorithm}
    \caption{Training process of LESnets for automatically learning of $C_{\mathrm{Smag}}$ (a brief version)}
    \label{alg2}
    \begin{algorithmic}[1]
        \STATE \textbf{Input:}
        \STATE \qquad   ${\{{\bm{u}(t_n)_{fDNS}}\}_{n=1}^{N_1}, \{{\bm{u}(t_n)_{SM/fDNS}}\}_{n=1,2,...T}^{N_2}}\leftarrow$ Training Data
         \STATE \qquad  $\left({N_1,N_2,T}\right)\leftarrow\left({5000,1,11}\right)$
        \STATE \qquad   $\mathcal{G_{\theta}}\leftarrow$ Neural operator (Adam optimizer \uppercase\expandafter{\romannumeral1})
        \STATE \qquad   $C_{\mathrm{Smag}}=10^{-8}\leftarrow$ An external model parameter (Adam optimizer \uppercase\expandafter{\romannumeral2})
        \STATE \textbf{Output:}
        \STATE \qquad   $\{{\bm{u}(t_n)_{predict}}\}_{n=1,2,...T}^{N_1}$, $C_{\mathrm{Smag}}$
            \STATE \qquad For ${{\mathcal{L}}_{PDE}}$, using $\bm{u} = \{{\bm{u}(t_n)_{predict}}\}_{n=1,2,...T}^{N_1}, C_{\mathrm{Smag}}$
            
            \STATE \qquad \qquad $\mathcal{L}_1=\nabla\cdot\bm{u}=\mathcal{F}^{-1}\left({ik\hat{\bm{u}}}\right)$
            \STATE \qquad \qquad 
            $\mathcal{L}_2={\partial}_{t}\bm{u}+\mathcal{F}^{-1}\left(\nu{k^2}\hat{\bm{u}}\right)-\mathcal{F}^{-1}\left({\hat{\bm{R}}}\right)+\mathcal{F}^{-1}\left({\hat{\bm{p}}}\right)$
            \STATE \qquad \qquad 
            ${{\mathcal{L}}_{PDE}}=\mathcal{L}_{1}+\mathcal{L}_{2}$
            \STATE \qquad For ${{\mathcal{L}}_{cs}}$, using $\bm{u} = \{{\bm{u}(t_n)_{SM/fDNS}}\}_{n=1,2,...T}^{N_2}, C_{\mathrm{Smag}}$
            
            \STATE \qquad \qquad $\mathcal{L}_{cs}={\partial}_{t}\bm{u}+\mathcal{F}^{-1}\left(\nu{k^2}\hat{\bm{u}}\right)-\mathcal{F}^{-1}\left({\hat{\bm{R}}}\right)+\mathcal{F}^{-1}\left({\hat{\bm{p}}}\right)$
            \STATE \qquad $\mathcal{L} = \sum_{n=2}^{T-1}||\mathcal{L}_{PDE}+\gamma'\mathcal{L}_{cs}||$
    
    \end{algorithmic}  
\end{algorithm}

Furthermore, we compare PDFs of the spanwise velocity increment $\delta_{r_3}\bar{u}/\bar{u}^{\mathrm{rms}}$ with distance $r=\Delta$ at different time instants in Fig. \ref{fig15}. {It can be seen that the PDFs of the spanwise velocity increment predicted by the five models are closely consistent with the fDNS results.}

{Finally, we compare the spatial structures of vortex predicted by different models with fDNS data. The $Q$ criterion has been widely used for visualizing vortex structures in turbulent flows \cite{hunt1988eddies,dubief2000coherent}, where $Q$ is defined by Eq. \eqref{eq 19}. Fig. \ref{fig16} displays the instantaneous Iso-surfaces at $Q=0.2$ at five learning time steps colored by the streamwise velocity. In the prediction of vortex structure, LESnets models exhibit a similar accuracy compared to the large-eddy simulation method and data-driven models.}

\section{{Automatically learning the Smagorinsky coefficient}}
\label{sec4}

{
The coefficient $C_{\mathrm{Smag}}$ for Smagorinsky model in large eddy simulation is known $a$ $priori$ in previous LESnets models. When the value of the coefficient is unknown $a$ $priori$, data assimilation is an effective approach to improve subgrid scale (SGS) models by automatically optimizing model coefficients \cite{wang2023ensemble,KATO2015559}. Notably, PINN can also learn unknown model coefficients using the neural networks \cite{PINN,chen2021physics}. By automatically learning the coefficients of PDE through the neural networks based on sparse datasets, PINN can avoid the intricate processes typically involved in traditional data assimilation. }

\begin{table}[H]
\captionsetup{font=small,labelfont=bf, width=.65\textwidth}
\belowrulesep=0pt
\aboverulesep=0pt
\setlength{\abovecaptionskip}{0pt}
\setlength{\belowcaptionskip}{1pt}
\caption{The $C_{\mathrm{Smag}}$ values obtained by using five weights $\gamma'$ and four different learning rates with SM or fDNS dataset in decaying HIT.}
\label{table4}
\centering
\begin{tabular}{c|l|l|l|l|l|l}
\toprule
\multicolumn{2}{c|}{$\gamma'$} & 0.5 & 5 & 25 & 50 & 100 \\ 
\midrule
\multirow{2}{*}{$C_{\mathrm{Smag}}$} & SM data& 0.1797 & 0.1123 & \textbf{0.0998} & \textbf{0.0998} & 0.0982 \\ %
\cmidrule{2-7} %
 & fDNS data& 0.1983 & 0.1892 & 0.1866 & 0.1859 & 0.1856 \\ %
\cmidrule{1-7} %
\multicolumn{2}{c|}{Learning rate} & $10^{-3}$ & $10^{-4}$ & $10^{-5}$ & $10^{-6}$ \\
\cmidrule{1-6} %
\multirow{2}{*}{$C_{\mathrm{Smag}}$} & SM data& 0.0987 & \textbf{0.0989} & 0.0988 & 0.0661 \\
\cmidrule{2-6} %
 & fDNS data & 0.1868 & 0.1868 & 0.1859 & 0.0680   \\

\bottomrule
\end{tabular}
\end{table}

\begin{figure}[ht]
\center

\begin{minipage}{0.45\linewidth}
\centerline{\includegraphics[width=\textwidth]{ 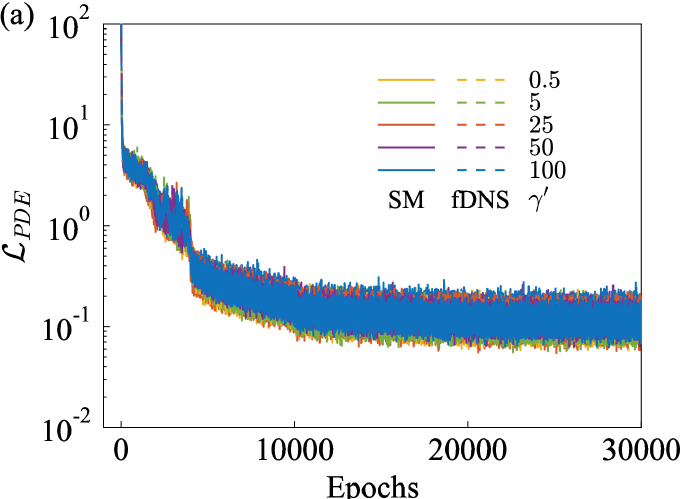}}
\end{minipage}
\hspace{1pt}
\begin{minipage}{0.45\linewidth}
\centerline{\includegraphics[width=\textwidth]{ 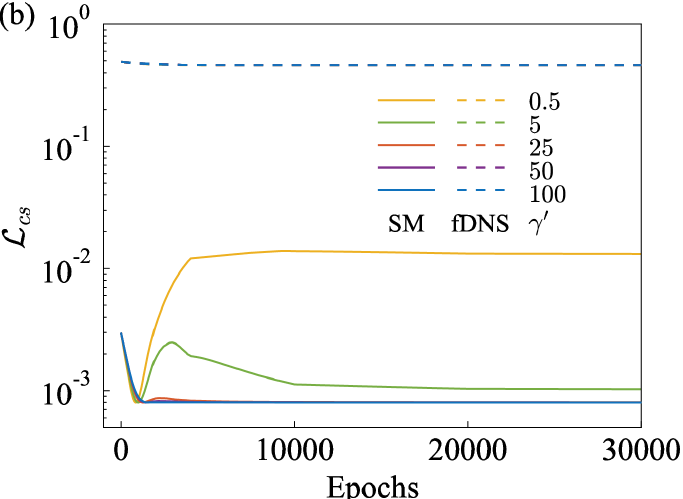}}
\end{minipage}

\vspace{-1pt}

\begin{minipage}{0.45\linewidth}
\centerline{\includegraphics[width=\textwidth]{ 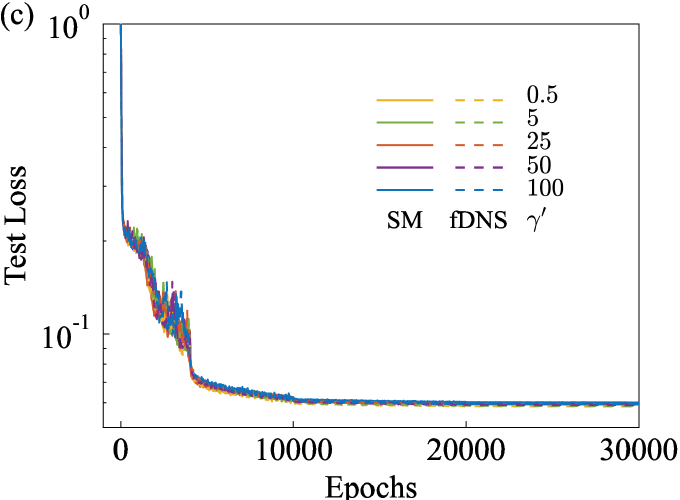}}
\end{minipage}
\hspace{1pt}
\begin{minipage}{0.45\linewidth}
\centerline{\includegraphics[width=\textwidth]{ 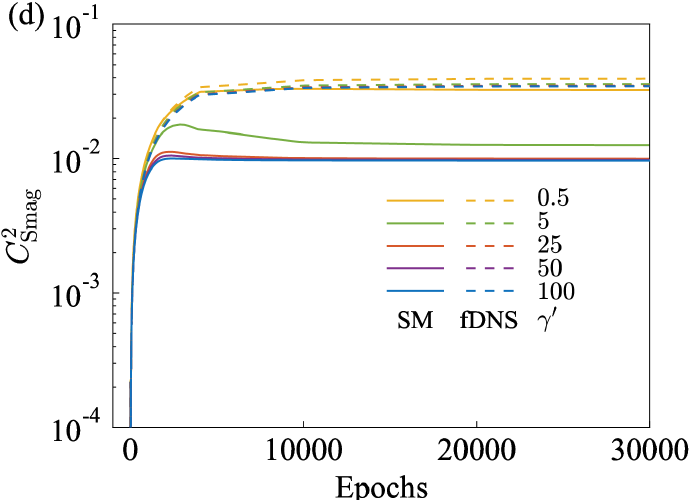}}
\end{minipage}
\caption{The evolutions of (a) ${{\mathcal{L}}_{PDE}}$; (b) ${{\mathcal{L}}_{cs}}$; (c) test loss and (d) $C^{2}_{\mathrm{Smag}}$ for different weights $\gamma'$ with SM or fDNS datasets. }
\label{fig18}
\end{figure}

{
Inspired by previous data assimilation and PINN methods, we propose an approach to automatically learn the coefficient $C_{\mathrm{Smag}}$ of Smagorinsky model during the training process of LESnets model. Considering the decaying HIT, we incorporate a single set of SM or fDNS datasets, which have dimensions of [$T$ $\times$ $N_x$ $\times$ $N_y$ $\times$ $N_z$ $\times$ $N_d$]. These datasets are obtained from LES simulations based on the Smagorinsky model with $C^2_{\mathrm{Smag}}=0.01$ and DNS respectively. The SM and fDNS datasets are simulated from $t=10,000\Delta t$ to $10,200\Delta t$, collected every 20$\Delta t$, with $T=11$, $N_x=N_y=N_z=32$ and $N_d=3$. The architecture of LESnets for the automatic learning of $C_{\mathrm{Smag}}$ is shown in Fig \ref{fig17}. During the training of LESnets, the ${{\mathcal{L}}_{cs}}$ is constructed as follows:}

\begin{equation}
\label{eq 25}
     {{\mathcal{L}}_{cs}}=\left\|{{\partial }_{t}}\bar{u}+\bar{u}\cdot \nabla \bar{u}+\nabla \bar{p}-\nu {{\nabla }^{2}}\bar{u}-\nabla{\tau}\right\|_{L^{2}(T;D)}^{2}. 
\end{equation}

{The training process of LESnets for the automatic learning of $C_{\mathrm{Smag}}$ is briefly described in Algorithm \ref{alg2}, and briefly summarized as follows: }

\begin{enumerate}[i]
\item {Input the original fDNS dataset $\{{\bm{u}(t_n)_{fDNS}}\}_{n=1}^{N_1}$ and additional SM or fDNS datasets $\{{\bm{u}(t_n)_{SM/fDNS}}\}_{n=1,2,...T}^{N_2}$, where $N_1=5000$, $N_2=1$ and $T=11$.}
\item {Two separate Adam optimizers are employed to handle distinct learning schemes: one for the neural operator $\mathcal{G_{\theta}}$ and another for the coefficient $C_{\mathrm{Smag}}$.}
\item {The output velocity fields $\{{\bm{u}(t_n)_{predict}}\}_{n=1,2,...T}^{N}$ and coefficient $C_{\mathrm{Smag}}$ are used to compute ${{\mathcal{L}}_{PDE}}$. The ${{\mathcal{L}}_{cs}}$ is computed using the SM or fDNS datasets $\{{\bm{u}(t_n)_{SM/fDNS}}\}_{n=1,2,...T}^{N_2}$ and the coefficient $C_{\mathrm{Smag}}$.}

\end{enumerate}

{The coefficient $C_{\mathrm{Smag}}$ is defined as a model parameter and initialized with the value of $10^{-8}$. It is important to emphasize that the additional SM or fDNS datasets are only used for computing the ${{\mathcal{L}}_{cs}}$. Meanwhile, the ${{\mathcal{L}}_{cs}}$ loss does not directly affect the other parameters of the LESnets model, except for $C_{\mathrm{Smag}}$. Consequently, the LESnets model employs the following composite loss function for optimization:}

\begin{equation}
\label{eq 26}
   \mathcal{L} = {{\mathcal{L}}_{PDE}} + \gamma'{{\mathcal{L}}_{cs}}. 
\end{equation}

\begin{figure}[ht]
\center
\begin{minipage}{0.45\linewidth}
\centerline{\includegraphics[width=\textwidth]{ 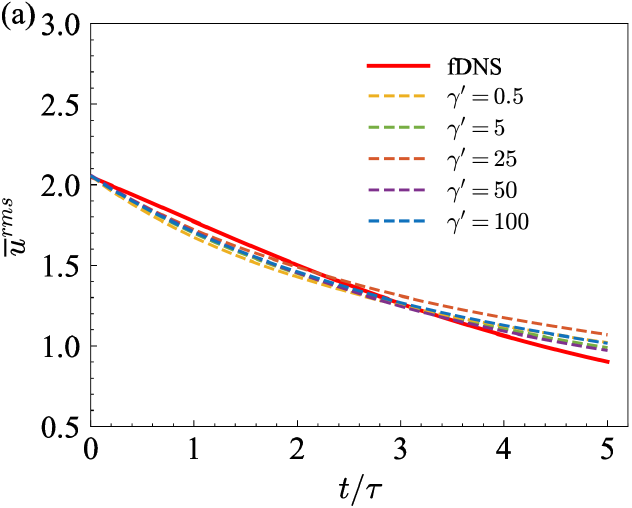}}
\end{minipage}
\hspace{1pt}
\begin{minipage}{0.46\linewidth}
\centerline{\includegraphics[width=\textwidth]{ 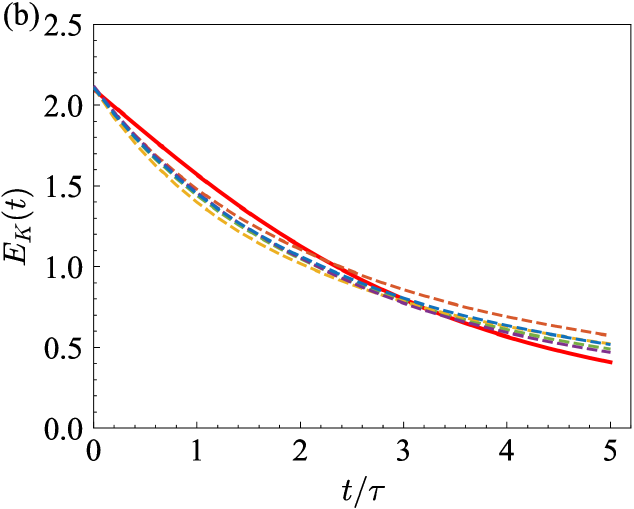}}
\end{minipage}

\vspace{-1pt}

\begin{minipage}{0.46\linewidth}
\centerline{\includegraphics[width=\textwidth]{ 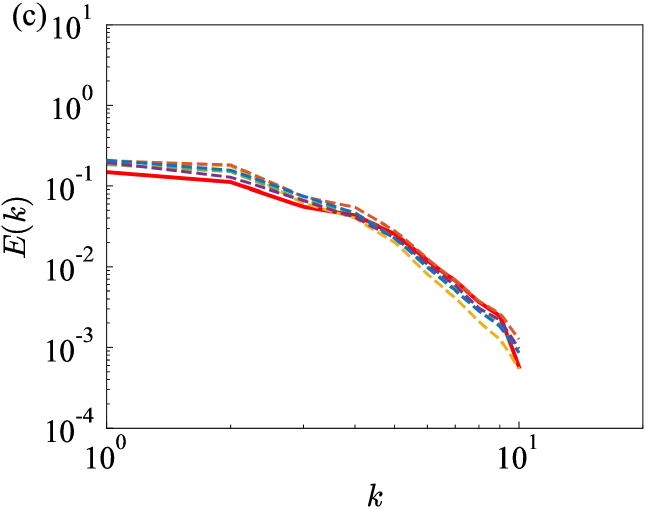}}
\end{minipage}
\hspace{-1pt}
\begin{minipage}{0.48\linewidth}
\centerline{\includegraphics[width=\textwidth]{ 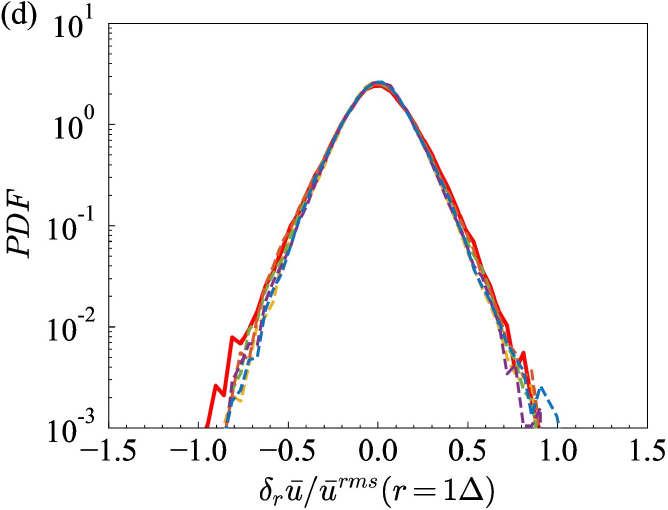}}
\end{minipage}
\caption{Influence of weight $\gamma'$ for ${{\mathcal{L}}_{cs}}$ with SM dataset: The temporal evolutions of (a) the rms velocity; (b) the turbulent kinetic energy $E_K(t)$; (c) the spectra of turbulent kinetic energy at $t\approx5\tau$ and (d) PDFs of the normalized velocity increments $\delta_{r}\bar{u}/\bar{u}^{\mathrm{rms}}$ at $t\approx5\tau$.}
\label{fig19}
\end{figure}

\begin{figure}[ht]
\center

\begin{minipage}{0.45\linewidth}
\centerline{\includegraphics[width=\textwidth]{ 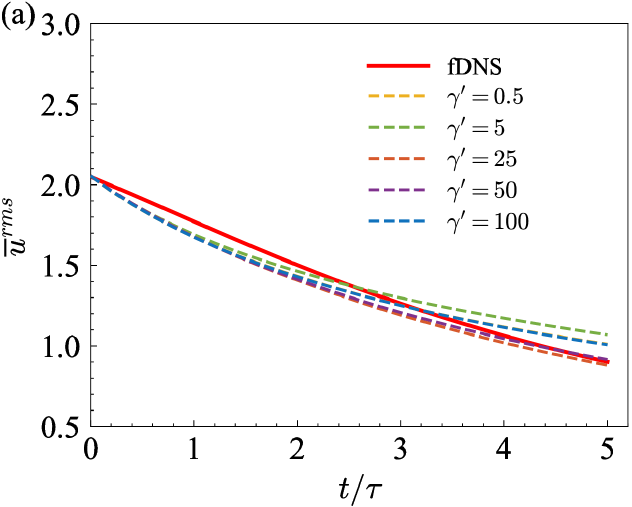}}
\end{minipage}
\hspace{1pt}
\begin{minipage}{0.46\linewidth}
\centerline{\includegraphics[width=\textwidth]{ 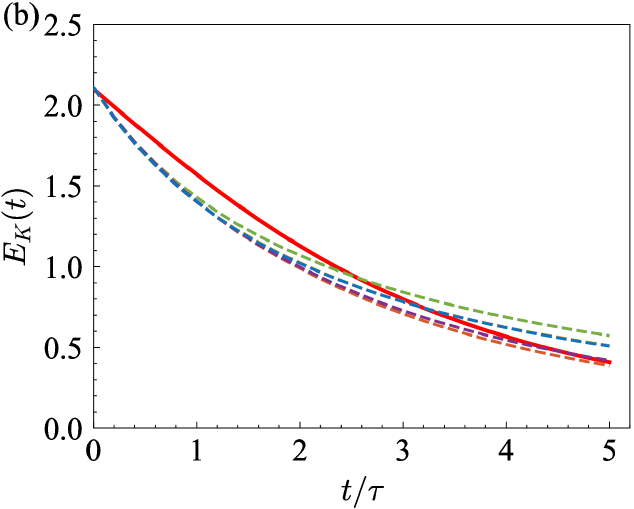}}
\end{minipage}

\vspace{-1pt}

\begin{minipage}{0.46\linewidth}
\centerline{\includegraphics[width=\textwidth]{ 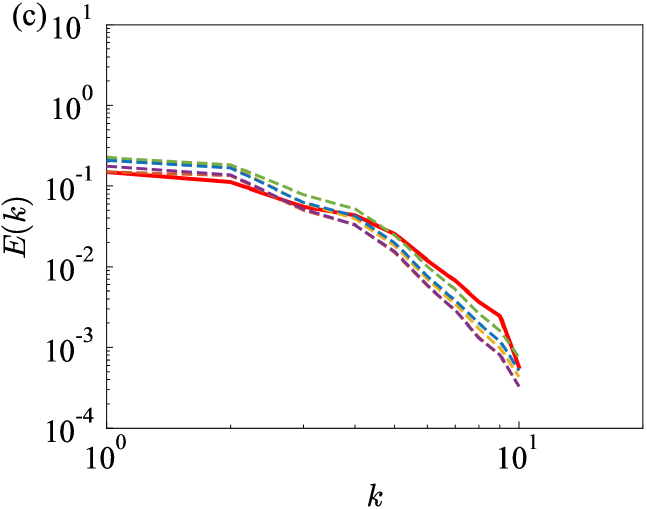}}
\end{minipage}
\hspace{1pt}
\begin{minipage}{0.48\linewidth}
\centerline{\includegraphics[width=\textwidth]{ 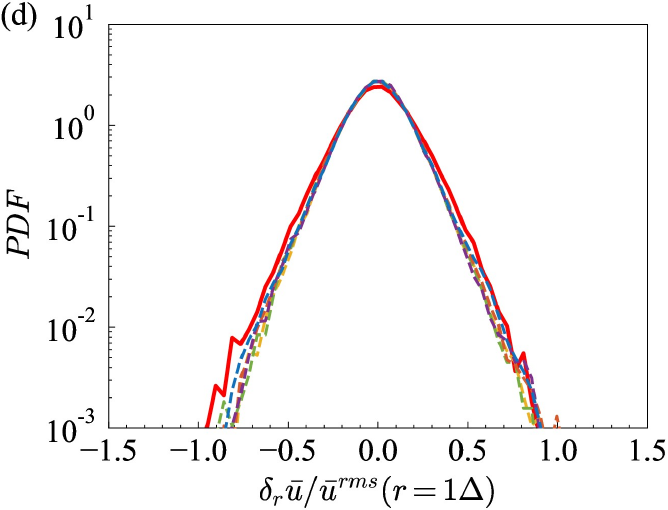}}
\end{minipage}
\caption{Influence of weight $\gamma'$ for ${{\mathcal{L}}_{cs}}$ with fDNS dataset: The temporal evolutions of (a) the rms velocity; (b) the turbulent kinetic energy $E_K(t)$; (c) the spectra of turbulent kinetic energy at $t\approx5\tau$ and (d) PDFs of the normalized velocity increments $\delta_{r}\bar{u}/\bar{u}^{\mathrm{rms}}$ at $t\approx5\tau$.}
\label{fig20}
\end{figure}

\begin{figure}[ht]
\center

\begin{minipage}{0.45\linewidth}
\centerline{\includegraphics[width=\textwidth]{ 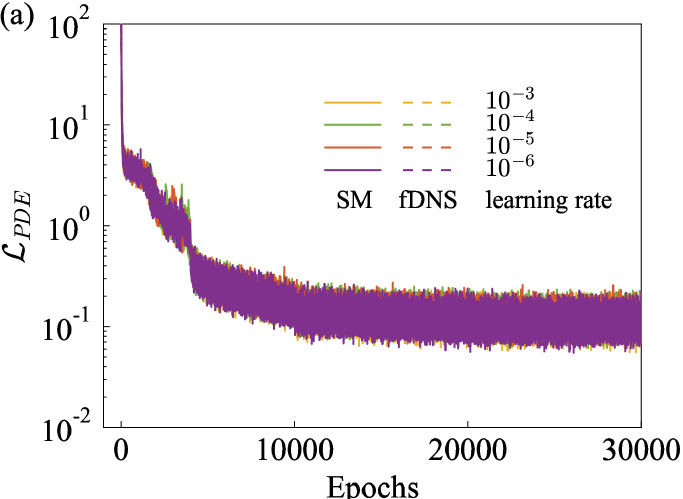}}
\end{minipage}
\hspace{1pt}
\begin{minipage}{0.45\linewidth}
\centerline{\includegraphics[width=\textwidth]{ 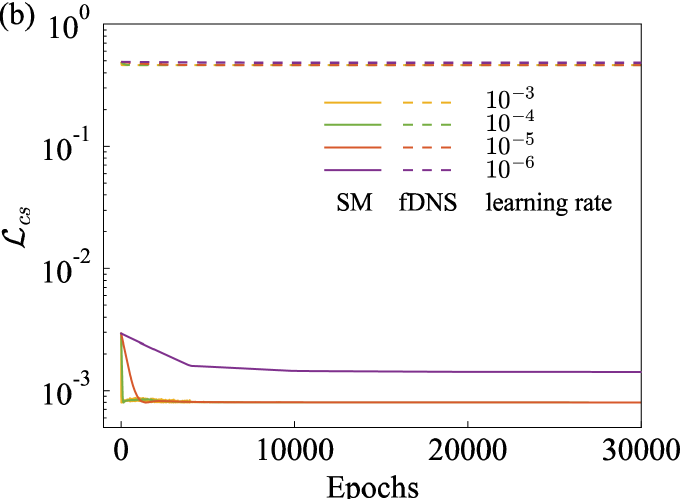}}
\end{minipage}

\vspace{-1pt}

\begin{minipage}{0.45\linewidth}
\centerline{\includegraphics[width=\textwidth]{ 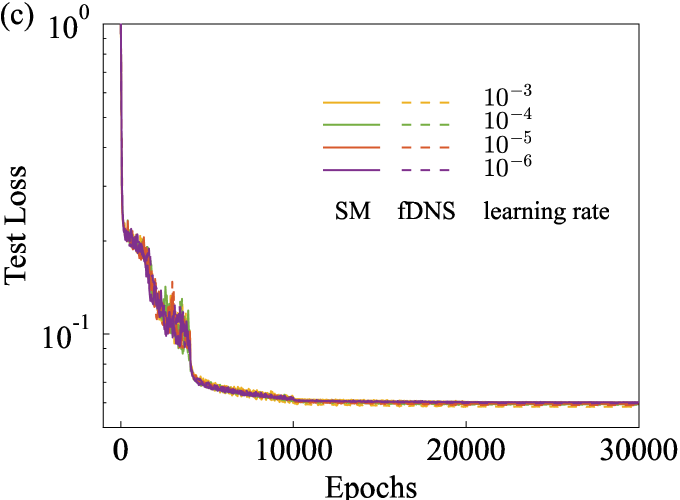}}
\end{minipage}
\hspace{1pt}
\begin{minipage}{0.45\linewidth}
\centerline{\includegraphics[width=\textwidth]{ 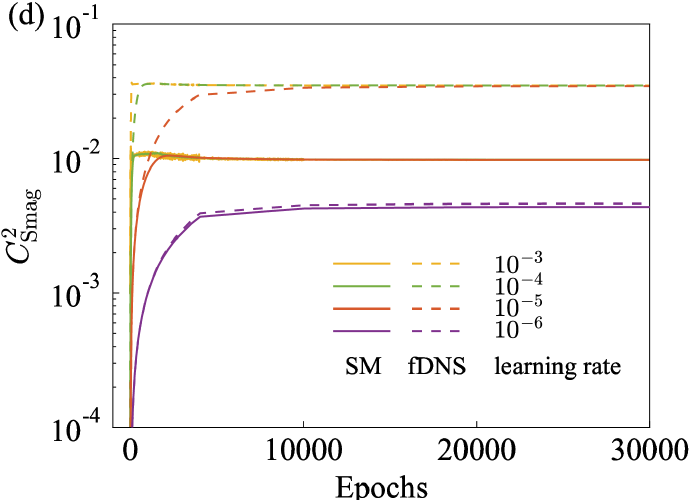}}
\end{minipage}
\caption{The evolutions of (a) ${{\mathcal{L}}_{PDE}}$; (b) ${{\mathcal{L}}_{cs}}$; (c) test loss and (d) $C^{2}_{\mathrm{Smag}}$ for different learning rates with SM or fDNS datasets. }
\label{fig21}
\end{figure}

\subsection{{Tests at different loss weights}}
\label{subsec4-1}
{We set five loss weights of $\gamma'$ from 0.5 to 100 at fixed learning rate $10^{-5}$ for the training process of $C_{\mathrm{Smag}}$. The loss curves and learning process of $C^{2}_{\mathrm{Smag}}$ for LESnets during the training process are shown in Fig. \ref{fig18}. The $C_{\mathrm{Smag}}$ values obtained by five schemes are shown in Table \ref{table4}. The $C_{\mathrm{Smag}}$ values obtained by the four loss weights $\gamma'=5, 25,50$ and $100$ with SM dataset are close to the value of 0.1. The $C_{\mathrm{Smag}}$ values obtained by the five loss weights $\gamma'$ with fDNS dataset range from 0.1856 to 0.1983. Although these values remain around 0.19, deviating from the value of $C_{\mathrm{Smag}}=0.1$, the results are comparable to those with SM dataset in the $a$ $posterior$ test. It is noted from Fig. \ref{fig18} that the loss ${{\mathcal{L}}_{cs}}$ with weight $\gamma'=0.5$ decreases around the 1,000th epoch when adding SM dataset, but gradually increases afterward. This observation suggests that ${{\mathcal{L}}_{cs}}$ decreases rapidly during the initial stage of training. However, a smaller value of $\gamma'$ leads to ${{\mathcal{L}}_{cs}}$ deviating from the optimization direction of the coefficient $C_{\mathrm{Smag}}$ in the later stages of learning. When adding fDNS datasets, the initial value of ${{\mathcal{L}}_{cs}}$ is larger than that of SM datasets, and it only decreases around the 3,000th epoch, corresponding to the trend of the $C^2_{\mathrm{Smag}}$ curves. We provide similar initial fields for LESnets model as in Section \ref{subsec3_1} in the $a$ $posterior$ test. The temporal evolutions of the rms velocity and the turbulent kinetic energy $E_K(t)$ values for LESnets with SM dataset of weight $\gamma'$ from $0.5$ to $100$ are shown in Fig. \ref{fig19} (a) and (b). The spectra of turbulent kinetic energy and PDFs of the normalized velocity increments $\delta_{r}\bar{u}/\bar{u}^{\mathrm{rms}}$ at $t\approx5\tau$ for LESnets with SM dataset of weight $\gamma'$ from $0.5$ to $100$ are shown in Fig. \ref{fig19} (c) and (d). Similarly, for LESnets with fDNS dataset (weight $\gamma'$ from 0.5 to 100), these quantities are shown in Fig. \ref{fig20}. Although changing loss weights $\gamma'$ results in different values for ${{\mathcal{L}}_{cs}}$ and $C_{\mathrm{Smag}}$ during the $a$ $priori$ test, the influence of different weight $\gamma'$ on the model is relatively minor in the $a$ $posterior$ test. Overall, the best result for LESnets model is obtained when applying a weight of $\gamma'=50$, both for SM or fDNS datasets. This setting effectively balances the learning of the coefficient $C_{\mathrm{Smag}}$ and the optimization of the PDE loss ${{\mathcal{L}}_{PDE}}$.}

\begin{figure}[ht]
\center

\begin{minipage}{0.45\linewidth}
\centerline{\includegraphics[width=\textwidth]{ 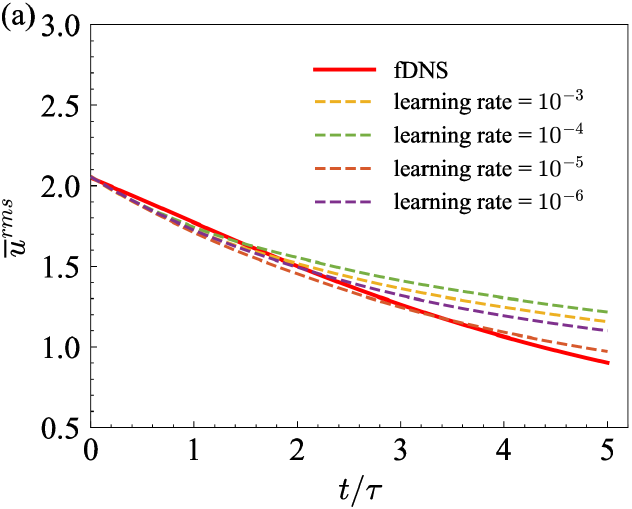}}
\end{minipage}
\hspace{1pt}
\begin{minipage}{0.46\linewidth}
\centerline{\includegraphics[width=\textwidth]{ 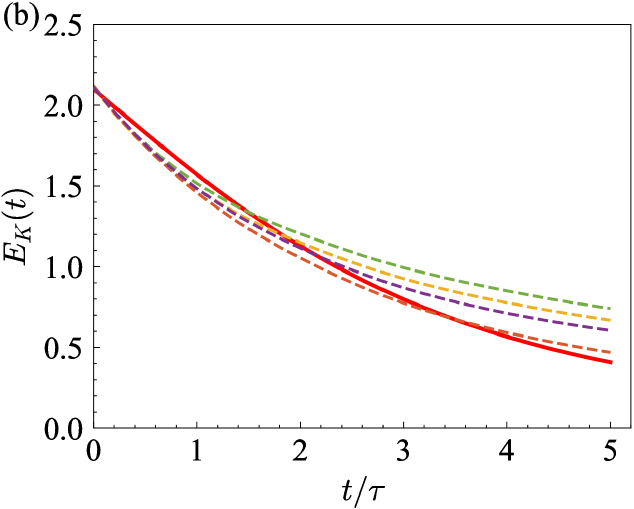}}
\end{minipage}

\vspace{-1pt}

\begin{minipage}{0.46\linewidth}
\centerline{\includegraphics[width=\textwidth]{ 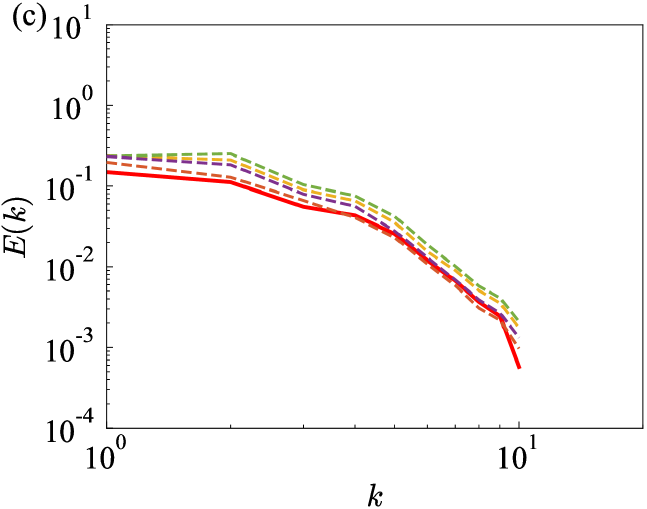}}
\end{minipage}
\hspace{1pt}
\begin{minipage}{0.48\linewidth}
\centerline{\includegraphics[width=\textwidth]{ 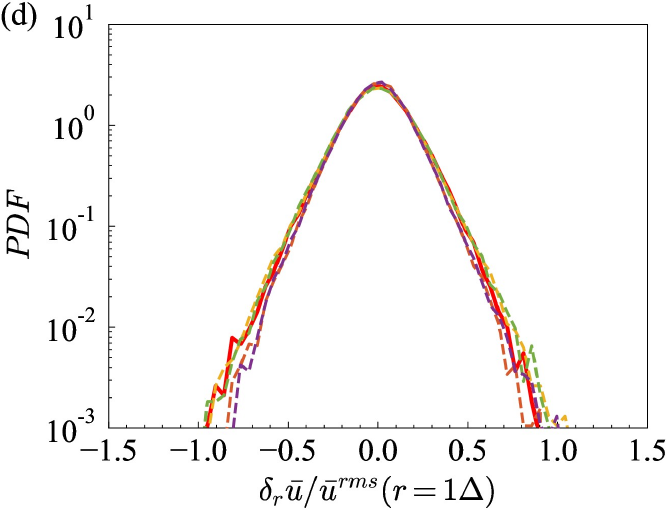}}
\end{minipage}
\caption{Influence of learning rate for ${{\mathcal{L}}_{cs}}$ with SM dataset: The temporal evolutions of (a) the rms velocity; (b) the turbulent kinetic energy $E_K(t)$; (c) the spectra of turbulent kinetic energy at $t\approx5\tau$ and (d) PDFs of the normalized velocity increments $\delta_{r}\bar{u}/\bar{u}^{\mathrm{rms}}$ at $t\approx5\tau$.}
\label{fig22}
\end{figure}

\begin{figure}[ht]
\center

\begin{minipage}{0.45\linewidth}
\centerline{\includegraphics[width=\textwidth]{ 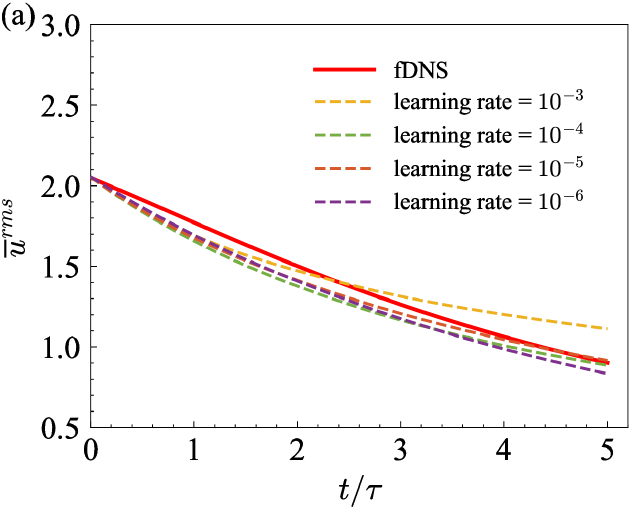}}
\end{minipage}
\hspace{1pt}
\begin{minipage}{0.46\linewidth}
\centerline{\includegraphics[width=\textwidth]{ 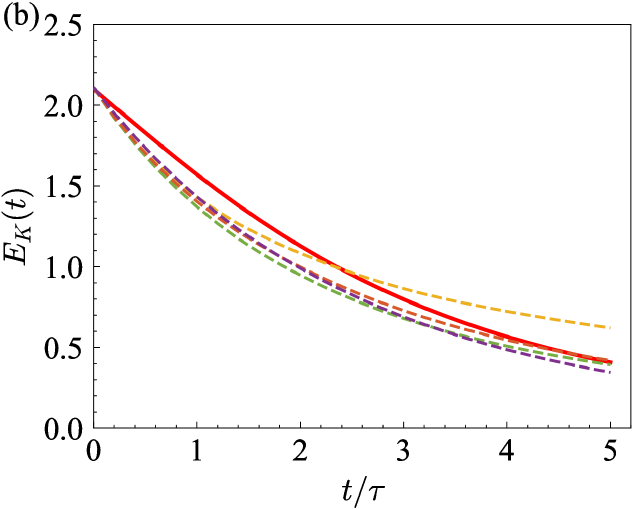}}
\end{minipage}

\vspace{-1pt}

\begin{minipage}{0.46\linewidth}
\centerline{\includegraphics[width=\textwidth]{ 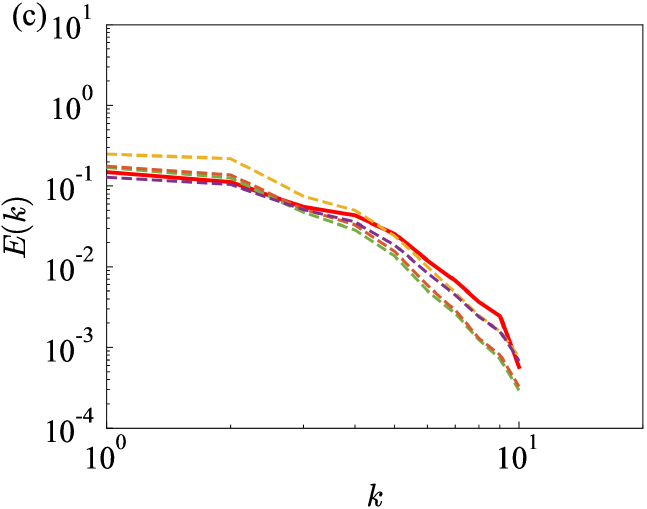}}
\end{minipage}
\hspace{1pt}
\begin{minipage}{0.48\linewidth}
\centerline{\includegraphics[width=\textwidth]{ 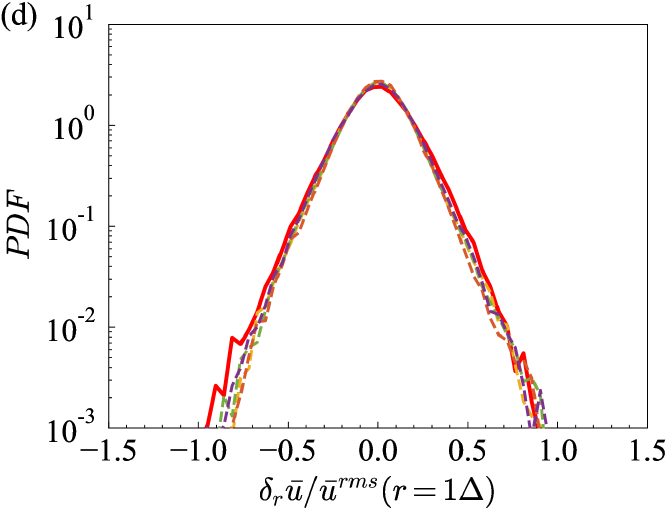}}
\end{minipage}
\caption{Influence of learning rate for ${{\mathcal{L}}_{cs}}$ with fDNS dataset: The temporal evolutions of (a) the rms velocity; (b) the turbulent kinetic energy $E_K(t)$; (c) the spectra of turbulent kinetic energy at $t\approx5\tau$ and (d) PDFs of the normalized velocity increments $\delta_{r}\bar{u}/\bar{u}^{\mathrm{rms}}$ at $t\approx5\tau$.}
\label{fig23}
\end{figure}

\subsection{{Tests at different learning rates}}
\label{subsec4-2}

{
We employ four different learning rates from $10^{-3}$ to $10^{-6}$ for the training process of $C_{\mathrm{Smag}}$ at fixed loss weight of $\gamma'=50$. The $C_{\mathrm{Smag}}$ values obtained by four schemes are shown in Table \ref{table4}. The loss curves and learning process of $C^{2}_{\mathrm{Smag}}$ for LESnets during the training process are shown in Fig. \ref{fig21}. The $C_{\mathrm{Smag}}$ values obtained by three initial learning rates $10^{-3}$, $10^{-4}$ and $10^{-5}$ are close to the value of 0.1 when adding SM dataset. The $C_{\mathrm{Smag}}$ values obtained by the four learning rates with fDNS dataset range from 0.068 to 0.1868. We provide similar initial fields for LESnets model as in Section \ref{subsec3_1} in the $a$ $posterior$ test. The temporal evolutions of the rms velocity and the turbulent kinetic energy values $E_K(t)$ for LESnets using different learning rates for $C_{\mathrm{Smag}}$ are shown in Fig. \ref{fig22} (a) and (b). The spectra of turbulent kinetic energy and PDFs of the normalized velocity increments $\delta_{r}\bar{u}/\bar{u}^{\mathrm{rms}}$ at $t\approx5\tau$ for LESnets using different learning rates for $C_{\mathrm{Smag}}$ are shown in Fig. \ref{fig22} (c) and (d). Similarly, for LESnets with fDNS dataset, these quantities are shown in Fig. \ref{fig23}. Overall, the LESnets model achieves its best performance with a learning rate of $10^{-5}$ for the SM dataset and $10^{-6}$ for the fDNS dataset. These settings are optimal to balance the convergence speed and stability of the model under the respective data conditions.}
 
{Since only one instance comprising $N_t=11$ velocity fields ${\{{\bm{u}(t_n)}\}_{n=1,2,...N_t}}$ of SM or fDNS datasets are added, the learned value of $C_{\mathrm{Smag}}$ are not always unique. However, the learned coefficients lead to similar results, indicating that the LESnets can optimize the model coefficients even with very limited data.}


\section{{Discussion and some further experiments}}
\label{sec5}

{In this section, we compare the computational efficiency of LESnets models with data-driven models and traditional LES. We also examine the impact of the LESnets parameters and the size of the dataset. Finally, we integrate data and PDE loss to optimize the LESnets model. Here, the coefficient $C_{\mathrm{Smag}}$ for all LESnets model is known as $a$ $priori$.}

\subsection{Computational efficiency}
\label{subsec5-1}
{We present the training and inference times for the four types of models across two turbulence prediction tasks, and compare them with traditional LES method to evaluate the efficiency of each model. Table \ref{table6} compares the training cost of one epoch, the inference cost of 250 prediction steps (i.e., 5,000 DNS time steps) by averaging across 100 test cases, number of parameters of the model, and GPU memory-usage for different models on the predictions of decaying HIT. The neural network models are trained and tested on an Nvidia A100 40G PCIe GPU, while the CPU used for loading model parameters and data is an Intel(R) Xeon(R) Gold 6248R CPU @3.00 GHz. The LES simulations are implemented on a computing cluster, where the type of CPU is Intel Xeon Gold 6148 @2.40 GHz. Table \ref{table6} illustrates that LESnets spend only 1\% to 5\% more time on training than their corresponding data-driven models, and the number of model parameters, computational memory, and inference time are similar to those of the data-driven models. In terms of inference efficiency, LESnets is 30 times faster than LES simulations. Although data-driven models have a similar inference efficiency, the time to prepare the data will also be enormous. In contrast, LESnets only needs initial fields for training. It is important to emphasize that, the present LESnets models are less efficient than traditional LES methods when the training times are considered. However, that could be highly efficient when performing many new inference tasks using the well-trained model.
}
\begin{table}[H]
\captionsetup{font=small,labelfont=bf, width=.98\textwidth}

\setlength{\abovecaptionskip}{0pt}
\setlength{\belowcaptionskip}{1pt}
\caption{Computational efficiency of different approaches in decaying HIT.}
\label{table6}
\centering

\begin{tabular}{lllll}
\toprule
Method & Training (s/epoch) & Number of parameters ($\times 10^6$) & GPU memory-usage (GB) & Inference (s) \\
\midrule
SM & N/A & N/A & N/A & 45.62\\

FNO & 0.39 & 1061.7 & 38.2 & {1.51}\\

IFNO & 2.17 & 622.1 & 37.7 & {11.80}\\

LESnets & 0.41 & 1061.7 & 38.0 & {1.51}\\

LESnets-I & 2.20 & 622.1 & 37.6 & {11.80}\\
\bottomrule
\end{tabular}
\end{table}

{Since the size of the flow field data in temporally evolving turbulent mixing layer is larger, the neural network models are trained and tested on an Nvidia A100 80G PCIe GPU, where the CPU type is AMD EPYC 7763 @2.45 GHz. The LES simulations are implemented on the same computing cluster as in the case of decaying HIT. Table \ref{table7} illustrates that LESnets models exhibit the same efficiency as data-driven models in the training time, and the prediction time of LESnets model is merely 3.18 seconds by averaging across 100 test cases, which is 1/40 of the cost of the LES.}

\begin{table}[H]
\captionsetup{font=small,labelfont=bf, width=.98\textwidth}

\setlength{\abovecaptionskip}{0pt}
\setlength{\belowcaptionskip}{1pt}
\caption{Computational efficiency of different approaches in temporally evolving turbulent mixing layer.}
\label{table7}
\centering

\begin{tabular}{lllll}
\toprule
Method & Training (s/epoch) & Number of parameters ($\times 10^6$) & GPU memory-usage (GB) & Inference (s) \\
\midrule 
SM & N/A & N/A & N/A & 126.79\\

FNO & 0.69 & 1061.7 & 53.5 & {3.18}\\

IFNO & 3.95 & 622.1 & 77.4 & {18.32}\\

LESnets & 0.70 & 1061.7 & 53.5 & {3.18}\\

LESnets-I & 3.96 & 622.1 & 77.4 & {18.32}\\
\bottomrule
\end{tabular}
\end{table}


\subsection{{Sensitivity study of LESnets}}
\label{subsec5-2}

{In this subsection, we examine the impact of the number of Fourier layers $L$, channel space `width' $d_v$ and the number of Fourier modes $k_{max}$, on the accuracy of LESnets. The test cases are summarized in Table \ref{table8}:}

\begin{table}[H]
\captionsetup{font=small,labelfont=bf, width=.6\textwidth}
\setlength{\abovecaptionskip}{0pt}
\setlength{\belowcaptionskip}{1pt}
\caption{LESnets with varying parameters in the case of decaying HIT.}
\label{table8}
\centering
\begin{tabular}{lllll}
\toprule
Case & Fourier layers & channel space `width' & modes \\
\midrule
Case 1 & 2 & 80 & 12\\
Case 2 & 4 & 80 & 12\\
Case 3 & 6 & 60 & 12\\
Case 4 & 6 & 40 & 12\\
Case 5 & 6 & 80 & 4\\
Case 6 & 6 & 80 & 8\\
Case 7 (selected) & 6 & 80 & 12\\
\bottomrule
\end{tabular}
\end{table}

\begin{figure}[htbp]
\centering
\begin{minipage}{0.48\linewidth}
\centerline{\includegraphics[width=\textwidth]{ 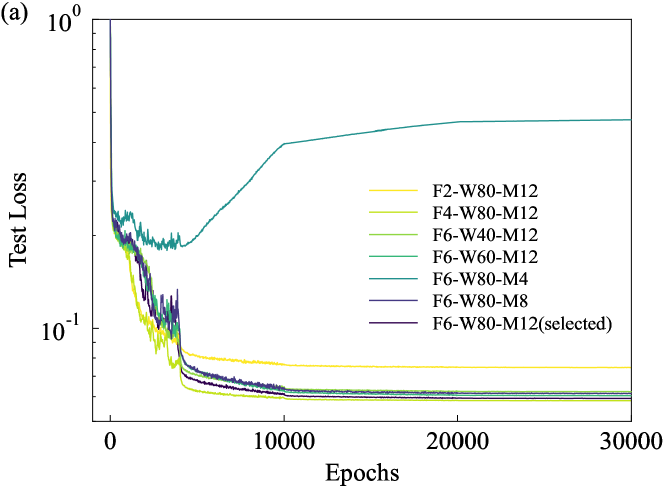}}
\end{minipage}
\hspace{1pt}
\begin{minipage}{0.45\linewidth}

\centerline{\includegraphics[width=\textwidth]{ 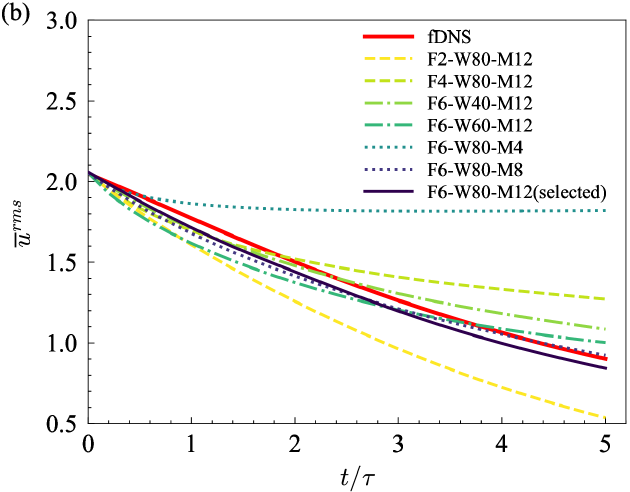}}
\end{minipage}

\hspace{1pt}
\begin{minipage}{0.48\linewidth}

\centerline{\includegraphics[width=\textwidth]{ 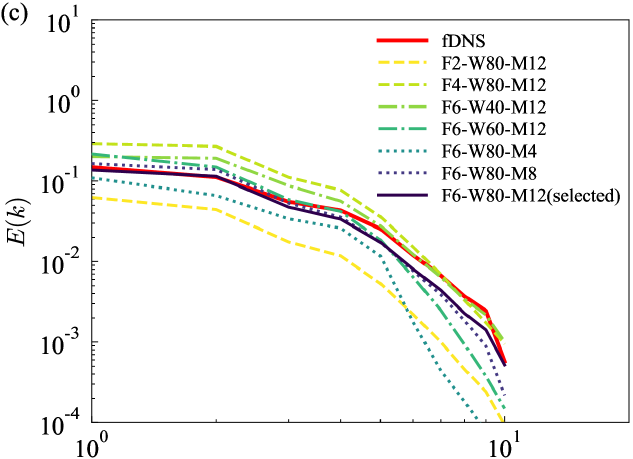}}
\end{minipage}

\caption{Influence of LESnets parameters: (a) temporal evolutions of $L_2$ test loss curves; (b) temporal evolutions of the rms velocity and (c) the spectra of turbulent kinetic energy at $t\approx5\tau$ for LESnets in decaying HIT.}
\label{fig24}
\end{figure}

{
We consider the same decaying HIT as in Section \ref{subsec3_1} and keep identical training parameter configurations as previously described, with the exception of the three model parameters. The $L_2$ test loss curves of LESnets at seven different configurations are shown in Fig. \ref{fig24} (a). In the $a$ $posteriori$ test, the temporal evolutions of the root mean square (rms) values of velocity are shown in Fig. \ref{fig24} (b). The spectra of turbulent kinetic energy at $t\approx5\tau$ for LESnets are shown in Fig. \ref{fig24} (c). The number of Fourier layers $L$ determines the depth of the LESnets. The prediction ability is significantly weakened for $L<6$. The channel space `width' $d_v$ represents the model's complexity in the feature space; small values $d_v<80$ could impair its representational capacity. Similarly, the number of Fourier modes $k_{max}$ determines the wavenumber range of multi-scale features the model can extract. A limited number of modes $k_{max}\le8$ may hinder the model's ability to capture sufficient features. When the values of parameters are adequate (i.e., $L=6$, $d_v=80$, $k_{max}=12$  ), LESnets can achieve convergent results.}

\begin{figure}[htbp]
\centering

\begin{minipage}{0.48\linewidth}
\centerline{\includegraphics[width=\textwidth]{ 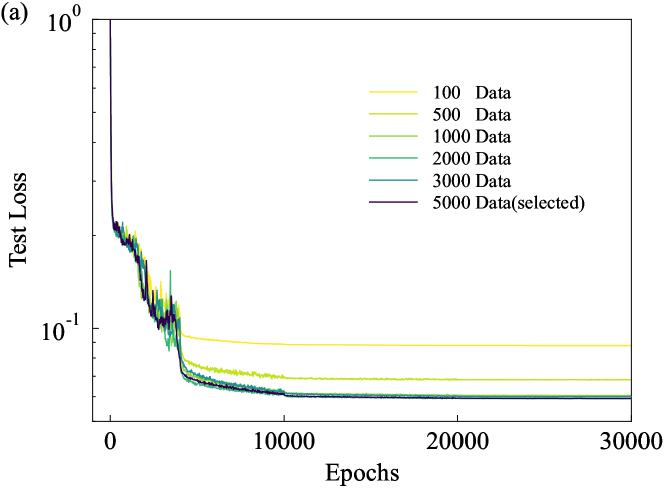}}
\end{minipage}
\hspace{1pt}
\begin{minipage}{0.45\linewidth}

\centerline{\includegraphics[width=\textwidth]{ 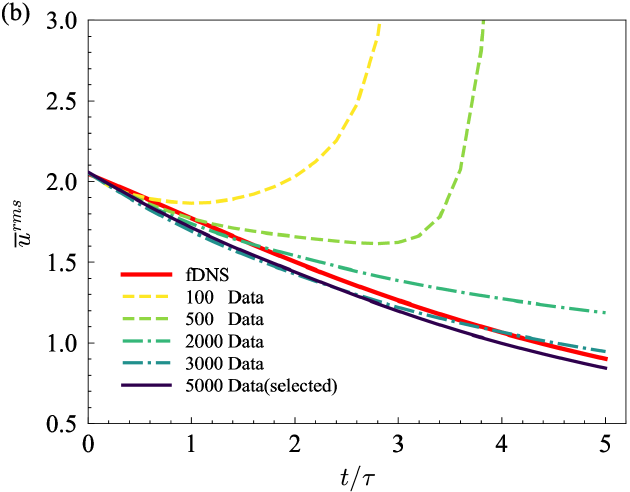}}
\end{minipage}

\hspace{1pt}
\begin{minipage}{0.48\linewidth}

\centerline{\includegraphics[width=\textwidth]{ 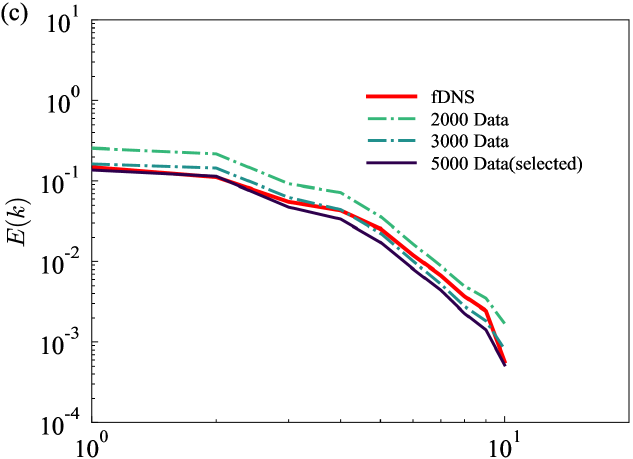}}
\end{minipage}
\caption{Influence of the number of initial fields on: (a) temporal evolutions of $L_2$ test loss curves; (b) temporal evolutions of the rms velocity and (c) the spectra of turbulent kinetic energy at $t\approx5\tau$ for LESnets in decaying HIT.}
\label{fig25}
\end{figure}

\subsection{{The effect of the number of initial fields on LESnets}}
\label{subsec5-3}

{We further test the present LESnets model using different number of initial fields, focusing on the same decaying HIT discussed in Section \ref{subsec3_1}. All datasets contain only the initial fields for training. Fig. \ref{fig25} (a) presents the $L_2$ test loss curves different number of initial conditions from 100 to 5000. In the $a$ $posteriori$ test, the temporal evolutions of the root mean square (rms) velocity profiles are shown in Fig. \ref{fig25} (b). The spectra of turbulent kinetic energy at $t\approx5\tau$ for LESnets are shown in Fig. \ref{fig25} (c). The $L_2$ loss decreases to a very small value around 0.059 when the dataset size reaches 2,000, indicating the well convergence of the model. However, this is insufficient for LESnets in the $a$ $posteriori$ test, since higher values of root mean square (rms) velocity and spectra of turbulent kinetic energy than the fDNS results are observed. Only when the number of dataset is increased to 3,000, the model gives the convergent results, that is comparable to the results using a dataset size of 5,000.}

\subsection{{LESnets with both data and PDE loss}}
\label{subsec5-4}

{We further test the present LESnets with both PDE loss and data loss. We employ 5,000 cases, with $T=11$ velocity fields taken from 10,000$\Delta t$ to 10,200$\Delta t$, collected every 20 DNS steps as label datasets for computation of data loss ${\mathcal{L}}_{data}$ followed by Eq. \eqref{eq 21}. The loss function of the modified LESnets model can be written as:}

\begin{equation}
\label{eq 27}
   \mathcal{L} = {{\mathcal{L}}_{PDE}} + \gamma{{\mathcal{L}}_{data}}. 
\end{equation}
{$\gamma$ denotes the the weight of the data loss.} 

\begin{figure}[htbp]
\centering

\begin{minipage}{0.48\linewidth}
\centerline{\includegraphics[width=\textwidth]{ 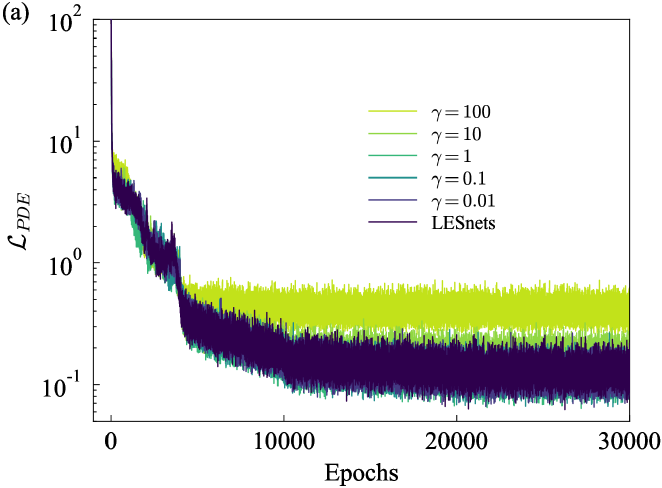}}
\end{minipage}
\hspace{1pt}
\begin{minipage}{0.48\linewidth}
\centerline{\includegraphics[width=\textwidth]{ 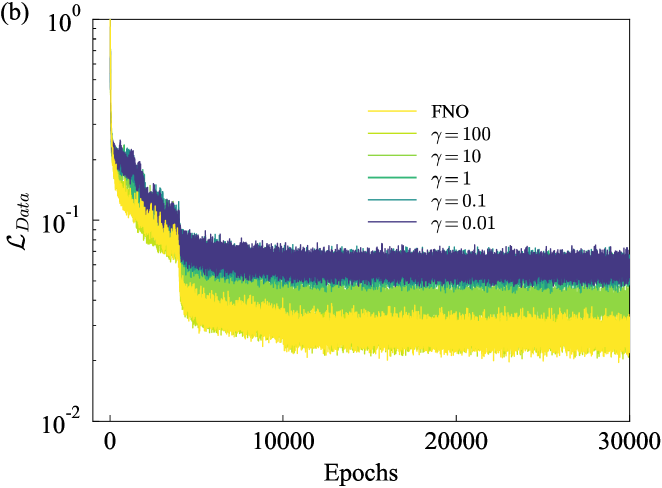}}
\end{minipage}
\hspace{1pt}
\begin{minipage}{0.48\linewidth}
\centerline{\includegraphics[width=\textwidth]{ 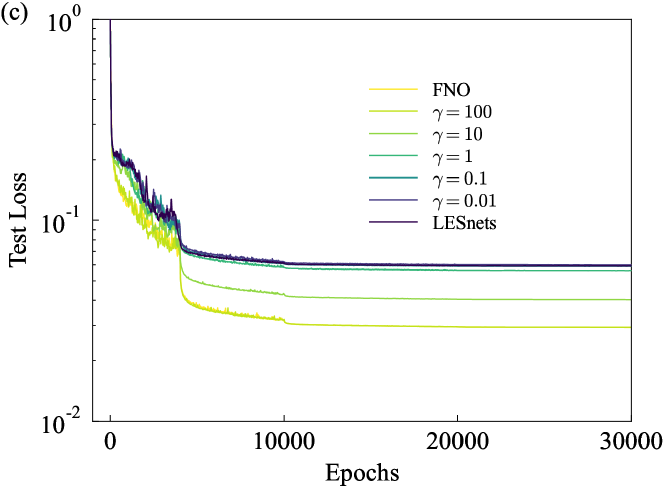}}
\end{minipage}

\caption{Influence of $\gamma$: the temporal evolutions of (a) ${{\mathcal{L}}_{PDE}}$; (b) ${{\mathcal{L}}_{Data}}$ and (c) $L_2$ test loss curves.}
\label{fig26}
\end{figure}

\begin{figure}[htbp]
\centering
\begin{minipage}{0.45\linewidth}
\centerline{\includegraphics[width=\textwidth]{ 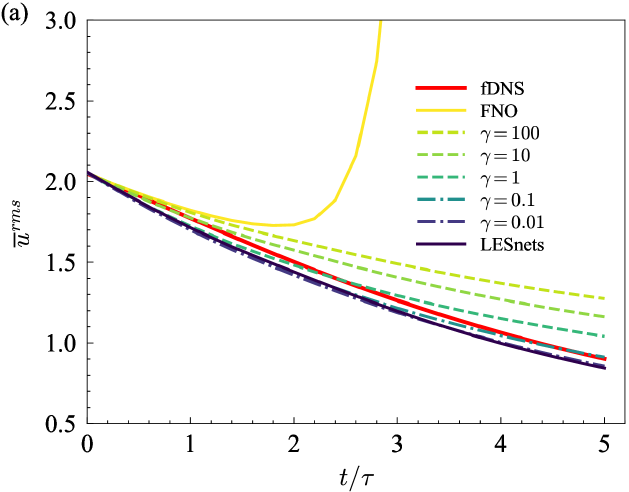}}
\end{minipage}
\hspace{1pt}
\begin{minipage}{0.46\linewidth}
\centerline{\includegraphics[width=\textwidth]{ 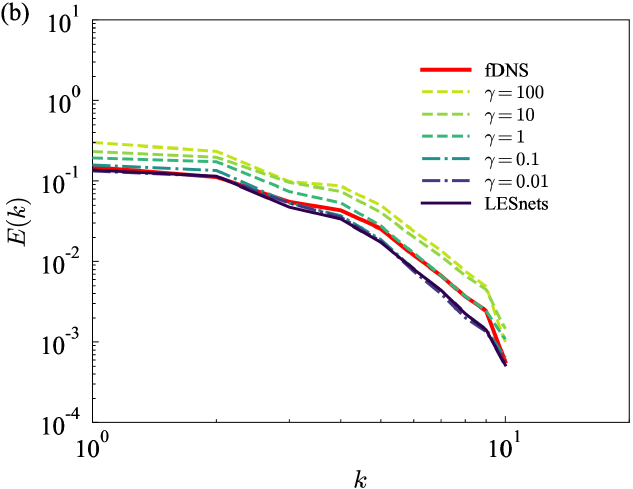}}
\end{minipage}

\caption{Influence of $\gamma$: (a) temporal evolutions of the rms velocity and (b) the spectra of turbulent kinetic energy at $t\approx5\tau$ for LESnets in decaying HIT.}
\label{fig27}
\end{figure}

{We evaluate the modified LESnets in the decaying HIT. Fig. \ref{fig26} (a)-(c) show the PDE loss ${{\mathcal{L}}_{PDE}}$, data loss ${{\mathcal{L}}_{Data}}$ and $L_2$ test loss curves for values of $\gamma$ ranging from 0.01 to 100. For comparison, loss curves of FNO and LESnets (only PDE loss) are also presented. As $\gamma$ decreases from $\gamma=100$ to $\gamma=0.01$, the loss value shifts from that of LESnets to that of FNO. It is important to emphasize that a lower test loss does not always indicate a better performance in long-term prediction. Fig. \ref{fig27} (a) presents the temporal evolutions of the root mean square (rms) velocity profile. The spectra of turbulent kinetic energy at $t\approx5\tau$ for LESnets are shown in Fig. \ref{fig27} (b). It is shown that the results of the modified LESnets using data loss and PDE loss, transition from FNO to LESnets when $\gamma$ decreases from $\gamma=100$ to $\gamma=0.01$. Among these results, the case with $\gamma=0.1$ demonstrates the closest results compared to the true value. }


\section{Conclusion}
\label{sec6}

Simulations of three-dimensional (3D) nonlinear partial differential equations (PDEs) are of great importance in engineering applications. Physics constraints have been widely used to enhance neural networks or operator learning. {To the best of our knowledge, no prior study has implemented the large-eddy simulation equations as PDE constraints to train a physics-informed neural operator for three-dimensional turbulence.}

In this study, we explore the effectiveness of physics-informed neural operator (PINO) to directly predict 3D incompressible turbulent flows, including decaying homogeneous isotropic turbulence and temporally evolving turbulent mixing layer. {We develop the LESnets models based on the constraint of large-eddy simulation equations and two neural operator models: FNO and IFNO. The input and output of operator learning are the filtered three-dimensional velocity fields at coarse grids. The PDE loss is then used as an optimization target to optimize network parameters, allowing the initial field as supervised data to train neural operator. Moreover, since there is no need for the label data, the length and interval of LESnets output time period are arbitrary.} Thus, there is no need to prepare a large amount of data due to the change of the output target. {This merit gives LESnets models a similar ability to generalize to the unseen flow regime compared with the data-driven model.}

{By using only physical constraints to train LESnets models with the two neural operators exhibit a high accurate prediction similar to the data-driven model and large-eddy simulation method in two turbulence prediction tasks. Additionally, we proposed a hybrid method that automatically adjusts the coefficients of the subgrid scale model during the training process. In terms of computational efficiency, LESnets models only take 1\% to 5\% more training time than data-driven models, and exhibit a similar inference efficiency as data-driven models, achieving a 40 times acceleration compared to traditional LES simulations. Moreover, the sensitivity analysis of LESnets parameters, the impact of the size of the dataset, and the combined effects of data and PDE losses are comprehensively examined. }

The current study for predicting turbulence using LESnets models is the first attempt to evaluate the PINO performance on LES of 3D turbulence. While the present results are encouraging, it is crucial to develop the PINO method for more challenging problems. The accuracy of LESnets models also rely on the traditional large-eddy simulation methods and the network architecture of data-driven models. Thus, using better SGS models and neural operator models can make LESnets models more accurate. In addition, it is necessary to use other discrete methods to calculate the loss function in the situations of non-uniform grids and complicated boundary conditions.

\section*{CRediT authorship contribution statement}

\textbf{Sunan Zhao:} Conceptualization, Methodology, Investigation, Coding, Validation, Writing - original draft preparation, Writing - reviewing and editing. \textbf{Zhijie Li:} Conceptualization, Methodology, Investigation, Coding, Writing - reviewing and editing. \textbf{Boyu Fan:} Conceptualization, Methodology, Investigation, Coding. \textbf{Yunpeng Wang:} Conceptualization, Investigation, Writing - reviewing and editing. \textbf{Huiyu Yang:} Conceptualization, Investigation, Writing - reviewing and editing. \textbf{Jianchun Wang:} Conceptualization, Methodology, Investigation, Supervision, Writing - reviewing and editing, Project administration, Funding acquisition.

\section*{Declaration of competing interest}
The authors declare that they have no known competing financial interests or personal relationships that could have appeared to influence the work reported in this paper.

\section*{{Code availability}}
{Code and dataset in this study are available on GitHub at} \href{https://github.com/Sunan-zhao/LESnets}{https://github.com/Sunan-zhao/LESnets}.

\section*{Acknowledgments}
This work was supported by the National Natural Science Foundation of China (NSFC Grant Nos. 12172161, 12302283, 92052301, and 12161141017), by the Shenzhen Science and Technology Program (Grant No. KQTD20180-
411143441009), and by Department of Science and Technology of Guangdong Province (Grant No. 2019B21203001, No. 2020B1212030001, and No. 2023B1212060001). This work was also supported by Center for Computational Science and Engineering of Southern University of Science and Technology, and by National Center for Applied Mathematics Shenzhen (NCAMS).

\appendix
\setcounter{table}{0}   
\setcounter{figure}{0}
\setcounter{equation}{0}

\renewcommand{\thetable}{A\arabic{table}}
\renewcommand{\thefigure}{A\arabic{figure}}
\renewcommand{\theequation}{A\arabic{equation}}

\section{{Performance of the original PINO model}}
\label{Appendix A}

\begin{figure}[htbp]
\centering

\begin{minipage}{0.48\linewidth}
\centerline{\includegraphics[width=\textwidth]{ 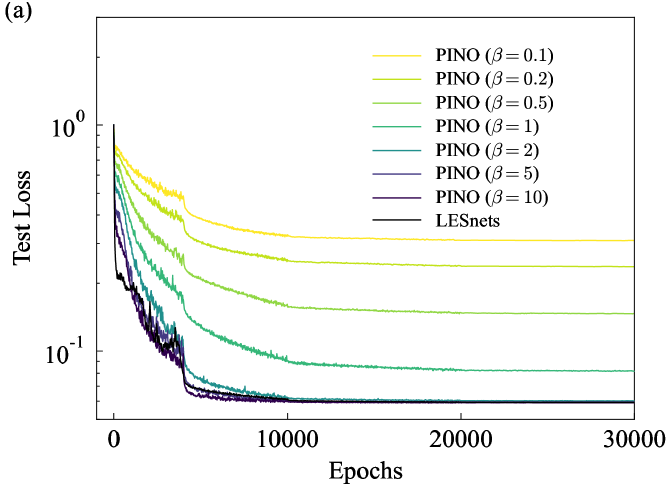}}
\end{minipage}
\hspace{1pt}
\begin{minipage}{0.45\linewidth}

\centerline{\includegraphics[width=\textwidth]{ 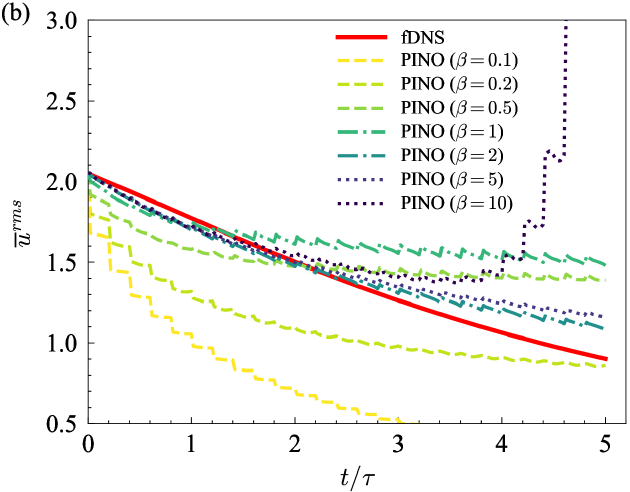}}
\end{minipage}

\caption{PINO models for different weights $\beta$: (a) temporal evolutions of $L_2$ test loss curves and (b) temporal evolutions of the rms velocity for PINO models with varying $\beta$ in decaying HIT. }
\label{figA1}
\end{figure}

{As discussed in Section \ref{section2-4}, the original PINO \cite{PINO} fails for long-term predictions of turbulence. One reason is that the original PINO predicts the entire time series at once from a given initial flow field. A straightforward solution is to train the model over shorter time periods and adopt an iterative prediction approach during the inference, similar to the FNO model \cite{FNO}. However, the loss of original PINO includes the initial condition loss, making the iterative prediction impractical. Unless the initial fields at later time steps must have a similar distribution or characteristics as those used during training, the iterative process will fail. The LESnets models avoids incorporating the initial condition as part of the loss function, and instead directly learns the flow fields at subsequent time instants, making iterative predictions feasible.}

{Following the exact same parameters and the same calculations of PDE loss as in Section \ref{subsec3_1}, we trained the PINO model in decaying HIT using the loss function including both the initial loss and the PDE loss. The loss function is as follows:}

\begin{equation}
\label{eq A1}
   \mathcal{L} = {{\mathcal{L}}_{PDE}} + \beta{{\mathcal{L}}_{ic}}. 
\end{equation}

{We use different weights of $\beta$ from 0.1 to 10. Fig. \ref{figA1} (a) shows the loss curves of PINO model. Although PINO models converge to solutions with $\beta=2,5$ and $10$, all the PINO models show divergent results in Fig. \ref{figA1} (b) in the $a$ $posterior$ test. PINO models with $\beta=0.1$, $0.2$, and $0.5$ can not accurately predict the turbulent flow fields during the first iteration. This observation suggests that the weight $\beta$ assigned to the initial condition loss is too small, and the model fails to give a correct initial field. Additionally, strong oscillations are observed during iterative predictions, though the amplitude of these oscillations is reduced when $\beta$ is increased to $2$ and $5$. These observations demonstrate that the original PINO model faces challenges in handling iterative prediction tasks.} 

\newgeometry{top=1.0cm}

\section{{The table of Nomenclature}}
\label{Appendix B}
{A summary of key mathematical notations is presented below.}

\begin{figure}[htbp]
\centering
\begin{minipage}{0.95\linewidth}
\centerline{\includegraphics[width=\textwidth]{ 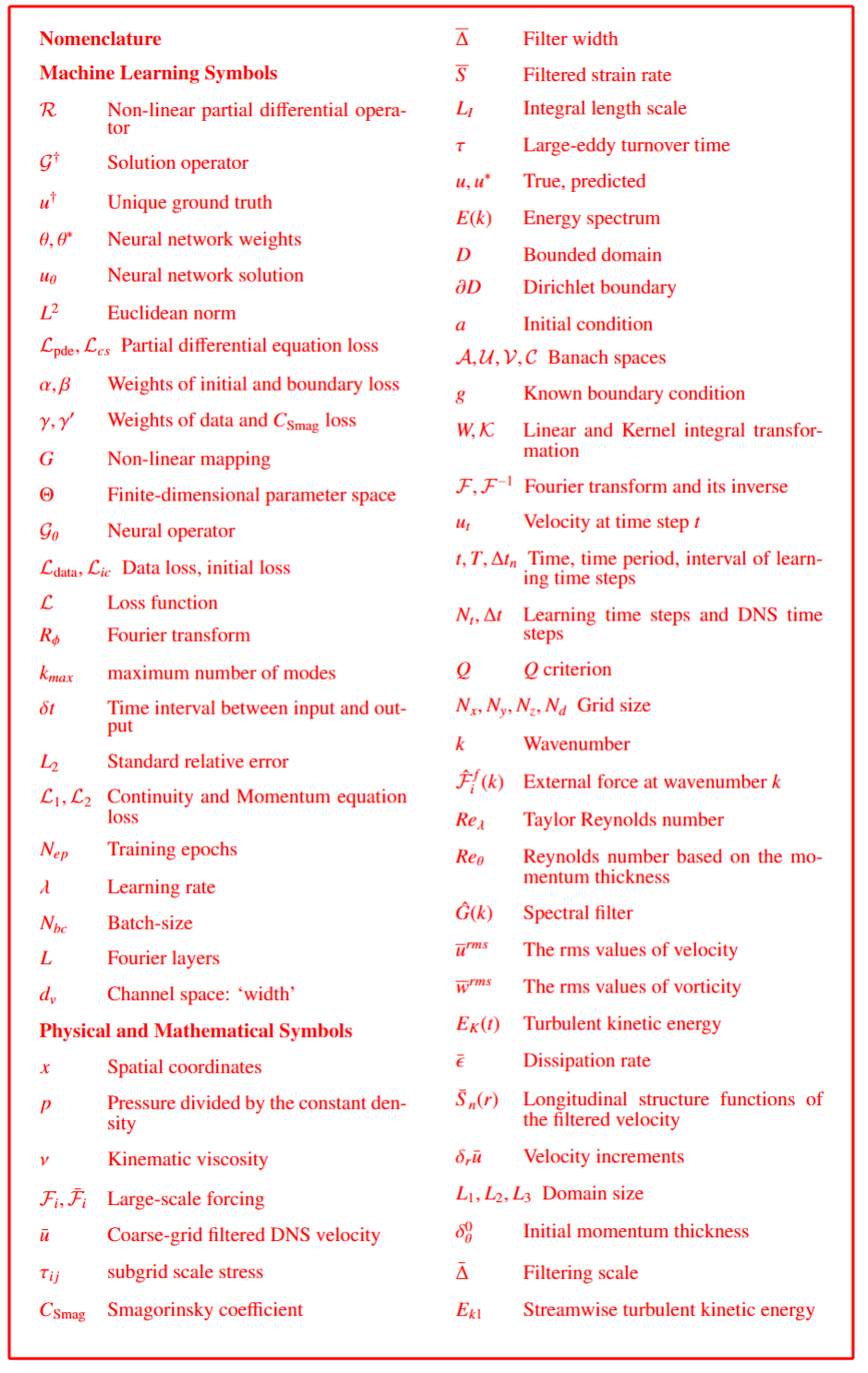}}
\end{minipage}

\end{figure}

\bibliographystyle{elsarticle-num} 
\bibliography{bibtxt}

\end{document}